\documentclass[%
 nofootinbib,reprint,aps,pra,superscriptaddress
]{revtex4-2}

\usepackage{amsmath}
\usepackage{amssymb}
\usepackage{amsfonts}
\usepackage{dsfont}
\usepackage{mathrsfs}
\usepackage{amsthm}
\usepackage{mathtools}
\usepackage{braket}
\usepackage{physics}
\usepackage{nicematrix}
\usepackage{bm} %negrito em equações
\usepackage{MnSymbol}
\usepackage{hyperref}

\DeclarePairedDelimiter{\dbra}{\llangle}{\rvert}
\DeclarePairedDelimiter{\dket}{\lvert}{\rrangle}
\newcommand{\dbraket}[2]{\llangle#1|#2\rrangle}
\newcommand{\dketbra}[2]{|#1\rrangle\llangle#2|}

\usepackage{tikz}
\usetikzlibrary{shapes, positioning, calc, decorations.pathreplacing}

\tikzset{
  tensor/.style={draw, minimum size=1cm, thick},
  indexlabel/.style={above, inner sep=1pt, font=\footnotesize},
  smalltensor/.style={draw, minimum size=0.65cm, thick},
  1inleg/.style={circle, draw=black, fill=white, inner sep=0pt, minimum size=5pt},
  1outleg/.style={circle, draw=black, fill=black, inner sep=0pt, minimum size=5pt},
  2inleg/.style={draw=black, fill=white, inner sep=0pt, minimum size=5pt},
  2outleg/.style={draw=black, fill=black, inner sep=0pt, minimum size=5pt},
  0inleg/.style={diamond, draw=black, fill=white, inner sep=0pt, minimum size=6pt},
  0outleg/.style={diamond, draw=black, fill=black, inner sep=0pt, minimum size=6pt},
  wire/.style={thick},
}

\theoremstyle{definition}

\usepackage{graphicx}
\usepackage{dcolumn}
\usepackage{bm}
\usepackage{appendix}
\usepackage{amsfonts}
\usepackage[normalem]{ulem}
\usepackage{xcolor}
\usepackage{booktabs}
\usepackage{comment}
\usepackage{afterpage}
\usepackage[export]{adjustbox}

\begin{document}

\title{A derivation of the late-time volume law for local operator entanglement {in chaotic many-body systems}}
\author{Guilherme Ilário Correr}
\email{guilherme.correr@helsinki.fi}
\thanks{corresponding author.}
\affiliation{Department of Physics, University of Helsinki, FIN-00014 Helsinki, Finland}
\author{John Goold}
\email{GOOLDJ@tcd.ie}
\affiliation{School of Physics, Trinity College Dublin, Dublin 2, Ireland}
\affiliation{Trinity Quantum Alliance, Unit 16, Trinity Technology and Enterprise Centre, Pearse Street, D02 YN67, Dublin 2, Ireland}
\affiliation{Algorithmiq Ltd, Kanavakatu 3C 00160, Helsinki, Finland}

\author{Marco Cattaneo}
\email{marco.cattaneo@helsinki.fi}
\affiliation{Department of Physics, University of Helsinki, FIN-00014 Helsinki, Finland}

\begin{abstract}Local Operator Entanglement (LOE) has emerged an indicator of quantum chaos in many-body systems. Numerical studies have shown that, in chaotic systems, LOE grows linearly in time and displays a volume-law behavior at late times, scaling proportionally with the number of local degrees of freedom. Despite extensive numerical evidence, complemented by analytical studies in integrable systems, a fully analytical understanding of the emergence of the volume law remains incomplete.
In this paper, we contribute toward this goal by deriving a late-time expression for LOE in chaotic systems that exhibits volume-law scaling. Our derivation proceeds by expressing the late-time LOE in the Liouville eigenstate basis and relies on three main assumptions: a higher-order non-resonance condition for the Hamiltonian eigenenergies, the Eigenstate Thermalization Hypothesis (ETH) ansatz for the matrix elements of the initial local operator, and the replacement of Hamiltonian eigenstates with random states in the final expression for LOE. Under these assumptions, we obtain an explicit formula displaying volume-law scaling.
Finally, we complement our analytical derivation with numerical simulations of the 1D mixed-field Ising model, testing the resulting formula and exploring the regime of validity of our assumptions.
\end{abstract}

\maketitle

\section{\label{sec:introduction}Introduction}

Characterizing chaotic {quantum} systems through their spectral properties has been a persistent effort since the proposal of the Bohigas-Giannoni-Schmit and Berry-Tabor conjectures{~\cite{BGS_conjecture, berry_tabor_conjecture} {and subsequent foundational work in quantum chaos} \cite{haake_quantum_signatures_chaos, Glushkov2021, Casati1989}}.
Where dynamics is concerned, different figures of merit have been introduced to diagnose chaotic phenomenology in many-body systems. The most successful ones are probably the out-of-equilibrium autocorrelation functions~\cite{ballistic_entanglement_chaotic}, entanglement dynamics in state space~\cite{entanglement_dynamics_manybody, entanglement_MBL, ballistic_entanglement_chaotic, complexity_integrability_prosen}, complexity quantifiers~\cite{krylov_complexity_saturation__chaos, linear_growth_complexity_proof, brown_susskind_conjecture}, out-of-time-ordered-correlators~\cite{otocs_quantum_mech, otocs_stanford, OTOCS_and_chaos}, entanglement of quantum evolutions~\cite{Zanardi2001,Styliaris2021}, and the entanglement of time-evolved local operators~\cite{Prosen_ising_chain_localoperatorentanglement, complexity_integrability_prosen, vincenzo_alba_LOE_54chain, vincenzo_alba_LOE_more, dowling_modi_scrambling_chaos_localoperatorentanglement}.

Despite the ability of these quantities to capture different features of quantum chaos, such as diffusive transport, scrambling, and exponentially complex dynamics, quite often the same behavior can be found in some families of integrable systems. Among the existing quantifiers, entanglement in operator space, usually termed as the Local Operator Entanglement (LOE)~\cite{Prosen_ising_chain_localoperatorentanglement}, clearly tells chaotic and integrable behavior apart for short-range Hamiltonians. Since its introduction, it has been thoroughly studied for integrable free~\cite{Prosen_ising_chain_localoperatorentanglement, spin_chain_localisation_localoperatorentanglement, prl_LOE_integrable_models, loe_cft_free_fermions}, integrable interacting~\cite{bertini_prosen_LOE_dualunitary, vincenzo_alba_LOE_54chain, vincenzo_alba_LOE_more} and chaotic systems~\cite{vincenzo_alba_LOE_more, bertini_prosen_LOE_dualunitary, coarsegrained_localoperatorentanglement, spin_chain_localisation_localoperatorentanglement}.

The LOE is defined as the entanglement entropy of the reduced density matrix of a vectorized local operator. It has been shown to scale, in the thermodynamic limit, at most logarithmically with time for all integrable systems, increasing indefinitely for interacting systems~\cite{vincenzo_alba_LOE_54chain, bertini_prosen_LOE_dualunitary} and saturating for some free integrable cases~\cite{Prosen_ising_chain_localoperatorentanglement, prl_LOE_integrable_models}.
This quantity is not only connected to the complexity scaling of the classical simulation of the dynamics using 
density matrix renormalization group~\cite{Prosen_ising_chain_localoperatorentanglement, XY_integrable_localoperatorentanglement, complexity_integrability_prosen, vincenzo_alba_LOE_54chain, bertini_prosen_LOE_dualunitary, vincenzo_alba_LOE_more}, but also constitutes an interesting probe of chaos, yielding a reliable quantity to study the peculiar features of generic dynamics in operator space.

Moreover, in a recent work it has been proven through analytical methods that the LOE is capable of distinguishing scrambling integrable systems from scrambling chaotic systems~\cite{dowling_modi_scrambling_chaos_localoperatorentanglement}. In addition to analytical evidence for integrable models, numerical studies on chaotic systems~\cite{coarsegrained_localoperatorentanglement, vincenzo_alba_LOE_more} have shown a characteristic linear increase with time, suggesting a \textit{volume law} for entanglement at late times, analogous to the Page curve~\cite{page_entanglement_subsystem, lubkin_entanglement_subsystem}. Throughout this work, ``volume law" refers to scaling proportional to the number of local degrees of freedom, equivalently logarithmic in the Hilbert-space dimension of the subsystem.

%Volume laws for entanglement, i.e., entanglement entropy of a subsystem that scales logarithmically in its dimension, typically emerge for the dynamics of randomly sampled unitaries. In this direction, a volume law for pseudorandom quantum circuits was found by considering the LOE obtained with dynamics generated by Haar random unitaries~\cite{butterfly_effect_localoperatorsentanglement}, which has deep connections with the late-time behavior of correlation functions obtained through chaotic Hamiltonian dynamics~\cite{chaos_and_complexity_design_yoshida, k-eth_kaneko-sagawa, designs_free_probability_pappalardi}.

The aim of this work is to provide further analytical foundations for the late-time behavior of LOE in chaotic systems. In particular, we present an analytical derivation of the volume law for late-time LOE of an initially local operator for chaotic dynamics. Our result relies on three central assumptions. The first one is the ``4 non-resonance condition'' for the eigenvalues of the system Hamiltonian, which is widely accepted for chaotic systems~\cite{eth_Srednicki1994, eth_Srednicki1999, ETH_review}. The second  is the Eigenstate Thermalization Hypothesis (ETH) applied to expectation values of local operators{~\cite{eth_Srednicki1999,ETH_review,ETH_Deutsch_2018, Gogolin_eisert_equilibration_thermalisation, Polkovnikov2011}}, which is also well established in the chaotic regime. The third assumption is that, for the purpose of computing the total LOE, which we will show can be written as a highly nontrivial expression involving multiple summations over system eigenvectors, the latter can be replaced by Haar-distributed random vectors, provided the system dimension is large. 

Admittedly, the third assumption is the most delicate. It can be fully justified only when restricting the analysis to an energy shell of eigenstates, a point that we corroborate numerically through simulations of the 1D Mixed Field Ising Model (MFIM)~\cite{chaos_distributions_entanglement_khemani}. We therefore acknowledge that our analytical approach relies on an assumption that may not be justified in full generality. Even so, since it holds at the level of an energy shell, we believe that the calculation remains useful, as it provides hints about the mechanism underlying the emergence of the volume law for late-time LOE in chaotic systems.

%{Starting from a} basis of the Krylov subspace consisting of eigenvectors of the Liouvillian that generates the dynamics, we obtain an analytical formula for the average late-time value of the LOE, expressed in terms of the total Hilbert space dimension and the dimension of the subsystem over which the entanglement entropy is computed. The formula exhibits a volume law with respect to the subsystem dimension. The results are numerically verified using exact diagonalization, and the interplay between the two approaches allows to understand the role and amplitude of the generic eigenstates hypothesis to the LOE volume law. These findings are connected to the computation of late time correlation functions in many-body systems, such as applying $k$-ETH~\cite{k-eth_kaneko-sagawa, designs_free_probability_pappalardi} or through the correspondence between late time chaotic dynamics and unitary designs~\cite{chaos_in_quantum_yoshida, chaos_and_complexity_design_yoshida, butterfly_effect_localoperatorsentanglement}, providing a different perspective to explore the properties of chaotic dynamics.

The rest of this manuscript is structured as follows. The LOE is introduced in Section~\ref{sec:LOE}, while in Section~\ref{sec:hypothesis} we carefully explain the assumptions at the basis of our calculation. Then, in Section~\ref{sec:liouville_basis} we introduce the Liouvillian basis we will use in the computation of LOE. Next, the analytical derivation of the volume law is presented in Section~\ref{sec:results}, while Section~\ref{sec:numerics} is devoted to the numerical validations of the analytical results.  Finally, we draw some concluding remarks in Section~\ref{sec:conclusion}.

%=======================================================================================================================================
\section{\label{sec:LOE}Local Operator Entanglement and volume law}

Our aim is to study the dynamics of a traceless\footnote{This condition was not strictly required in the first definition of the LOE~\cite{Prosen_ising_chain_localoperatorentanglement}. However, it was shown to be necessary in order for the evolution of the LOE to be as fast as possible~\cite{coarsegrained_localoperatorentanglement}, and is an important ally in analytical computations~\cite{dowling_modi_scrambling_chaos_localoperatorentanglement}. Furthermore, the traceful part of the operator was shown to give trivial contributions when discussing operator growth computed via OTOCs~\cite{dowling_modi_scrambling_chaos_localoperatorentanglement}.} local hermitian operator $\mathcal{O}$ acting on the Hilbert space $\mathcal{H}$ of dimension $\mathrm{d}$. Its evolution in Heisenberg picture is generated by the many-body Hamiltonian $H$: 
{\begin{equation}
    \mathcal{O}(t)=U^{\dagger}(t)\mathcal{O}U(t),\quad U(t) \equiv e^{-iHt}.
\end{equation}}
The LOE is defined as the entanglement $2-$Rényi entropy of the vectorization    of $\mathcal{O}(t)$ given a bipartition of $\mathcal{H}$ ~\cite{Prosen_ising_chain_localoperatorentanglement}.  To vectorize $\mathcal{O}$ we work in the doubled Hilbert space, $\mathcal{H}\otimes\mathcal{H}'$, which can be obtained through the isomorphism 
\begin{equation}
    \label{eqn:Isom_opTOvec}
    \Omega:\ketbra{a}{b}\mapsto\ket{a}\otimes\ket{b^*}
\end{equation} that takes operators acting on $\mathcal{H}$ to vectors in the doubled space~\cite{qmechanics_liouville_space}:

\begin{equation}
    A=\sum_{a,b=1}^{{d}} A_{ab}\ketbra{a}{b}\mapsto\dket{A}=\frac{1}{\sqrt{\mathrm{d}}}\sum_{a,b=1}^{d}A_{ab}\ket{a}\otimes\ket{b^*},
\end{equation}
where 
{\begin{equation}
    \dket{A} = \frac{1}{\sqrt{\mathrm{d}}}\Omega[A].
\end{equation}}
%we applied the isomorphism $\Omega$ in the definition of $A$.
The additional normalization factor comes from the definition of the infinite-temperature inner product~\cite{krylov_review}:
\begin{equation} 
    \dbraket{A}{B}:=\Tr(A^{\dagger} B)/\mathrm{d},
\end{equation} {corresponding} to the correlation function between $A$ and $B$ at infinite temperature. 

{In the study of LOE, we assume the local operator $\mathcal{O}$} to be normalized {according to the infinite-temperature inner product}, 
\begin{equation}\label{eqn:innerproduct}
    \dbraket{\mathcal{O}}{\mathcal{O}}=\Tr(\mathcal{O}^{\dagger}\mathcal{O})/\mathrm{d}=1.
\end{equation}
Therefore, the density matrix \begin{equation}\rho^{(\mathcal{O})}(t)=\dketbra{\mathcal{O}(t)}{\mathcal{O}(t)}\end{equation} is a well-defined pure state, and we can compute entropies of entanglement for the vectorized time-evolved operator. In the context of this work, we will consider uniquely the $2-$Rényi entropy of entanglement~\cite{renyi_entropies}: given a bipartition $\mathcal{H}_{\mathscr{A}}\otimes\mathcal{H}_{\mathscr{B}}$, we define the LOE as
\begin{equation}
    S_2\Big[\rho^{(\mathcal{O})}_{\mathscr{A}}(t)\Big]:=-\ln\Big[\Tr\big(\rho^{(\mathcal{O})}_{\mathscr{A}}(t)\big)^2\Big],
\end{equation}
{where we apply the partial trace: Given an operator decomposed in a bipartite basis of a space $\mathcal{H}_{\mathscr{A}}\otimes\mathcal{H}_{\mathscr{B}}$,
\begin{equation}
    O = \sum_{i,j = 1}^{\mathrm{d}_\mathscr{A}}\sum_{k,l = 1}^{\mathrm{d}_\mathscr{B}}A^{ij}_{kl}\dket{i}_{\mathscr{A}}\otimes\dket{k}_{\mathscr{B}}\dbra{j}_{\mathscr{A}}\otimes\dbra{l}_{\mathscr{B}},
\end{equation}
}{the reduced operator obtained with the partial trace is defined as
\begin{equation}
    O_{\mathscr{A}}= \Tr_\mathscr{B}(O)\coloneq\sum_{i=1}^{\mathrm{d}_\mathscr{A}}\sum_{k,l = 1}^{\mathrm{d}_\mathscr{B}}O^{ii}_{kl}\dket{k}_{\mathscr{B}}\dbra{l}_{\mathscr{B}}.
\end{equation}
}

\begin{figure}
    \centering
    \includegraphics[width=0.7\columnwidth]{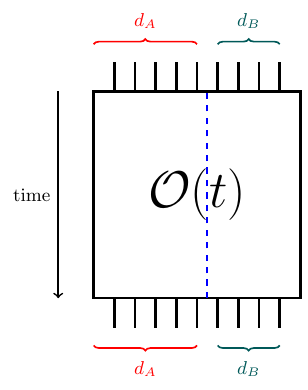}
    \caption{A pictorial representation of the bipartition considered throughout this work for the computation of entanglement in the extended space, $\mathcal{H}\otimes\mathcal{H}'$. We consider the case for which the bipartition is induced from a bipartition within the physical Hilbert space, $\mathcal{H}=\mathcal{H}_A\otimes\mathcal{H}_B$, with dimensions $\mathrm{d}_A$ and $\mathrm{d}_B$, respectively, such that $\mathrm{d}_A\mathrm{d}_B=\mathrm{d}$. This way, each part of the doubled space is divided between $\mathcal{H}\otimes\mathcal{H}'\equiv\Big(\mathcal{H}_{A}\otimes\mathcal{H}'_{A}\Big)\otimes\Big(\mathcal{H}_{B}\otimes\mathcal{H}'_{B}\Big)$.}
    \label{fig:bipartition_LOE}
\end{figure}

The choice of bipartition is in principle arbitrary. However, it is useful to first introduce a bipartition in the physical Hilbert space $\mathcal{H}$, and then extend it naturally to the doubled space. In this way, the operator entanglement is directly related to entanglement in the physical space. {After introducing a bipartition in the physical space,} $\mathcal{H}=\mathcal{H}_A\otimes\mathcal{H}_B$, we can write the doubled space as 
\begin{equation}
    \mathcal{H}\otimes\mathcal{H}'=\mathcal{H}_A\otimes\mathcal{H}_B\otimes\mathcal{H}_{A'}\otimes\mathcal{H}_{B'}.  
\end{equation}
Through a reshuffling of the partition order, we obtain $\mathcal{H}_{\mathscr{A}}=\mathcal{H}_A\otimes\mathcal{H}_A'$ and $\mathcal{H}_{\mathscr{B}}=\mathcal{H}_B\otimes\mathcal{H}_B'$. Figure~\ref{fig:bipartition_LOE} presents an illustration of the bipartition considering the tensor representation of the vectorized operator.

The LOE is a quantity that {grows} at most linearly {in time} for chaotic systems and logarithmically for integrable ones~\cite{Prosen_ising_chain_localoperatorentanglement, XY_integrable_localoperatorentanglement, complexity_integrability_prosen, vincenzo_alba_LOE_54chain, bertini_prosen_LOE_dualunitary, vincenzo_alba_LOE_more}. This scaling is observed for both finite systems and systems in the thermodynamic limit. {In particular,} when we consider finite {chaotic} systems, the entanglement has to grow linearly with time and then saturate at a particular value. This value will depend on the size of the bipartition, and {it is believed to satisfy a} \textit{volume law} {as a function of the dimension of the smallest subsystem, which in this work we assume to be subsystem $\mathscr{A}$. In other words, $S_2\Big[\rho^{(\mathcal{O})}_{\mathscr{A}}(t)\Big]$ should be proportional to the number of local degrees of freedom in $\mathscr{A}$ (its ``volume'' $\abs{\mathscr{A}}$).}

{If we assume that there are $n_A$ ``sites'' in the subsystem $A$ of the physical Hilbert space, then its dimension is given by
\begin{equation}
    \mathrm{d}_A = q^{n_A},
\end{equation}
where $q$ is the dimension of each site. For instance, if we focus on many-body qubit systems, $q=2$. Then, the volume of the subsystem $\mathscr{A}$ in the bipartition of the doubled Hilbert space is $\abs{\mathscr{A}}= 2 n_A$, and the LOE satisfies a volume law if 
\begin{equation}
    S_2\Big[\rho^{(\mathcal{O})}_{\mathscr{A}}(t)\Big]\sim 2 n_A \sim  \ln(\mathrm{d}_A^2),
\end{equation}
where we assumed $n_A<n_B$. Accordingly, throughout we paper we set 
\begin{equation}
    \dim(\mathcal{H}_A)=\mathrm{d}_A, \; \dim(\mathcal{H}_B)=\mathrm{d}_B,\; \dim(\mathcal{H})=\mathrm{d}= \mathrm{d}_A \times \mathrm{d}_B.
\end{equation}
}

A volume law for late-time LOE was found for arbitraty bipartitions and pseudorandom circuits, i.e., random circuits that {have enough depth\footnote{I.e., the number of discrete time steps needed to execute the gates in a quantum circuit.}} in order to replicate the Haar measure~\cite{butterfly_effect_localoperatorsentanglement}. In particular, the authors found
\begin{equation}\label{eq:volume_law_butterfly_effect}
   \langle\Tr\big(\rho^{(\mathcal{O})}_{\mathscr{A}}(t)\big)^2\rangle_\text{Haar} =\frac{\mathrm{d}_A^2+\mathrm{d}_B^2}{\mathrm{d}^2}+\frac{1}{\mathrm{d}}\Tr(\mathcal{O}^{\dagger}\mathcal{O}\mathcal{O}^{\dagger}\mathcal{O}),
\end{equation}
where $ \langle\Tr\big(\rho^{(\mathcal{O})}_{\mathscr{A}}(t)\big)^2\rangle_\text{Haar}$ is the Haar averaged purity of the vectorized operator $\rho^{(\mathcal{O})}_{\mathscr{A}}(t)$ evolved through a pseudorandom circuit, and
for the same bipartitions considered in our work, i.e., $\dim(\mathcal{H}_{\mathscr{A}})=\mathrm{d}_A^2$ and $\dim(\mathcal{H}_{\mathscr{B}})=\mathrm{d}_B^2$. 
Equation~\eqref{eq:volume_law_butterfly_effect} is a volume law that contains an additional correction {proportional to $\Tr(\mathcal{O}^{\dagger}\mathcal{O}\mathcal{O}^{\dagger}\mathcal{O})$}. If $\mathrm{d}_A=\mathrm{d}_B$, {it can be shown that this term brings a contribution of the same order as the first term on the RHS of Equation~\eqref{eq:volume_law_butterfly_effect}}. 

An analytical derivation of the volume law for  continuous-time dynamics generated by chaotic Hamiltonians is still missing. This is what we aim to address in this work, focusing on the late-time behaviour of LOE.

%In the context of the equivalence between unitary designs and late-time chaotic dynamics through the $k-$ETH (Appendix~\ref{app:ETH_designs}), a missing piece is to obtain the same behavior regarding only arguments about the continuous time Hamiltonian dynamics. The main idea of this equivalence is that when an ETH physical operator (Appendix~\ref{app:ETH_designs}) is concerned, the late-time values of correlation functions of the observable under chaotic Hamiltonian dynamics should be the same as those obtained under $k-$designs of a particular order that depends on the function. In the same manner, this should be equivalent to the assumptions of non-degeneracy of the eigenvalues and eigenstates of Hamiltonian being uniformly distributed states. We address the computation following the second approach in the following section applying the Liouvillian basis in operator space.

%=======================================================================================================================================

\section{\label{sec:hypothesis}Hypotheses and reasoning}

In this section we present and discuss the key hypotheses we use for the derivation of the volume law. We start by introducing the eigenvalues and eigenvectors of the Hamiltonian $H$ we aim to study: 
\begin{equation}
    \label{eqn:eigenvalues_eigenvectors}
    H \ket{E_a}= E_a \ket{E_a}, \quad a=1,\ldots,\mathrm{d},
\end{equation}
where $\{\ket{E_a}\}_{a=1}^\mathrm{d}$ is an orthonormal basis of eigenvectors.
Then, our three main hypotheses are:

\begin{enumerate}
    \item \textbf{4 non-resonance}~\cite{non_resonance_bertini, non_resonance_Riddell, non_resonance_thermalization_A_Winter}: Given two arbitrary sets containing $k$ of the eigenstates of $H$, say $\Lambda_k$ and $\Lambda^{'}_k$,  the Hamiltonian $H$ satisfies the $k$ non-resonance if, $\forall\Lambda_k,\;\Lambda^{'}_k$, the equality $\sum_{E_a\in\Lambda_k}E_a=\sum_{E_a\in\Lambda^{'}_k}E_a$ implies $\Lambda_k=\Lambda^{'}_k$. In our calculations, we will use up to the $k=4$ non-resonance condition for the eigenenergies of $H$.
    \item \textbf{ETH ansatz}: The matrix elements of the local operator $\mathcal{O}$ in the basis of the Hamiltonian eigenstates can be expressed through the ETH ansatz~\cite{ETH_review, ETH_Deutsch_2018}:
\begin{equation}\label{eq:standard_eth}
    \mathcal{O}_{ab} = \mathcal{O}_{\text{microc}}(\overline{E})\delta_{ab}+e^{-S(\overline{E})/2}f_{\mathcal{O}}(\overline{E},\omega_{ab})\Delta_{ab},
\end{equation}
where $\overline{E}=(E_a+E_b)/2$, while $\mathcal{O}_{\text{microc}}(E_a)$ and $f_{\mathcal{O}}(\overline{E},\omega_{ab})$ are order-one smooth functions whose behavior depend on the physical model. Moreover, $S(\overline{E})$ is the thermodynamic entropy that is an extensive function, so $e^{-S(\overline{E})/2}$ scales as $1/\sqrt{\mathrm{d}}$. Finally, $\Delta_{ab}$ are random real or complex numbers with zero mean and unit variance.
    \item \textbf{Eigenstates behaving as random states:} In the final expression of the late-time LOE as a function of the system eigenstates, the latter can be replaced by Haar random states, provided the system dimension is sufficiently large. 
\end{enumerate}

Next, we discuss each hypothesis and their justification. The first hypothesis, $k$ non-resonance,  is a generalization of the standard assumption on the non-degeneracy of the Hamiltonian eigenenergies to their differences, and to the differences of their differences. It is a common assumption appearing in late-time calculations of correlation functions for chaotic systems~\cite{patrycja_thermalization_non_resonance, non_resonance_bertini, non_resonance_Riddell, non_resonance_thermalization_A_Winter}, including the convergence of average values to the statistical mechanics prediction and decay of fluctuations. It is usually justified by the emergence of the Wigner-Dyson distributions and level repulsion~\cite{ETH_review,silvia_pappalardi_notes}. %An essential consequence of this assumption for thermalization in closed quantum systems is that no matter how we split our system, there will always exist interaction between the parts~\cite{non_resonance_thermalization_A_Winter}. In other words, there is no preference on which part we define to be the subsystem or the bath. In the case of non-interacting Hamiltonians this will not be generally true, unless the model is specially suited to satisfy rational independence of the spectrum~\cite{non_resonance_bertini}.  
The value $k$ that is typically required depends on the order of the correlation function we are interested in studying. It is common to assume 2 non-resonance to compute correlation functions with fourth order on the unitary time evolution, see for instance~\cite{OTOCS_scaling_brandao_huang, patrycja_thermalization_non_resonance, silvia_pappalardi_notes}. Generic assumptions of non-resonance are also assumed when computing higher-order correlation functions~\cite{designs_free_probability_pappalardi, k-eth_kaneko-sagawa}, and are believed to hold in chaotic systems~\cite{non_resonance_thermalization_A_Winter, eth_Srednicki1999}. Throughout this work, 4 non-resonance was assumed. 

Secondly, the ETH ansatz is commonly employed to study the matrix elements of local operators in chaotic systems~\cite{ETH_review,silvia_pappalardi_notes}. It is a fundamental assumption in the explanation of the emergence of thermalization in quantum systems, and is considered a solid signature of chaotic Hamiltonians.

The third assumption is motivated by early studies on quantum thermalization, in which Hamiltonian eigenstates are often replaced by Haar-random states to compute matrix elements of local operators~\cite{eth_Deutsch1991,eth_Srednicki1994}. We acknowledge that this assumption is strictly justified only within a narrow energy window and in the bulk of the spectrum.  Nevertheless, deriving analytical results without it proved intractable. We therefore adopt it as an assumption that is rigorously valid only when restricted to a small energy window. We emphasize this point, and the contrast with the full Hilbert space case, through numerical simulations in Section~\ref{sec:numerics}. Extending the assumption to the full eigenstate spectrum, however, still provides valuable insight into the origin of the volume-law behavior observed in the late-time LOE. {The limitations of this sort of hypothesis for the physical model studied in this work are investigated further from the entanglement perspective in~\cite{chaos_distributions_entanglement_khemani}, where the authors compare the eigenstate entanglement with Haar random state entanglement and, taking into account possible conservation laws, with the entanglement displayed by states obtained under the ``Bianchi-Donà distribution'' \cite{bianchi_dona}.}

Moreover, we note that assuming an initially local operator implies that 
$\mathcal{O}$ has support over a linear superposition of most system eigenstates for a chaotic Hamiltonian with delocalized eigenvectors. Therefore, focusing on the bulk is reasonable, since the majority of relevant eigenvectors lie there.
This reasoning is in consonance with calculations comparing the state entanglement obtained by Haar averaging with infinite-time averages for chaotic Hamiltonians, where the initial state is usually assumed to be given by an infinite temperature state~\cite{designs_free_probability_pappalardi, k-eth_kaneko-sagawa}.

%This correspondence is done here by considering the bulk of the many-body spectrum (the ``infinite temperature'' eigenstates), however, for large dimension and a tuned observable, this can be valid for most of the eigenstates. Given that the Hamiltonian eigenstates are a completely delocalized basis, in order for an operator to be localized, it needs to be defined as a superposition of many of these eigenstates, which justifies the assumption that most of the contributions are coming from the eigenstates behaving as random states. This reasoning is in consonance with calculations comparing the state entanglement obtained by Haar averaging with infinite-time averages for chaotic Hamiltonians, where the initial state is usually assumed to be given by an infinite temperature state~\cite{designs_free_probability_pappalardi, k-eth_kaneko-sagawa}. This idea is further discussed in Appendix~\ref{app:volume_law_states}.

Finally, our third assumption is also inspired by the result that, under the restriction to small enough energy windows, eigenstates of chaotic Hamiltonians satisfy entanglement volume laws for the von-Neumann entropy corresponding to the Page curve, obtained for Haar random states~\cite{volume_law_eigenstates}. Similar arguments were applied to obtain the entanglement Rényi entropies of Hamiltonian eigenstates~\cite{renyi_entropy_chaotic_eigenstates}. In our calculation, we try to show this equivalence also for entanglement in Heisenberg picture, i.e., LOE.  In addition, a similar assumption is also at the basis of ``$k-$ETH'', i.e., the extension of the ETH ansatz to $k$-order correlations~\cite{k-eth_kaneko-sagawa}, or the generalized ETH equivalence with unitary $k$-designs~\cite{correlations_ETH_free-prob}. We stress that these results should be considered as an inspiration for our calculation rather than a rigorous justification of our third assumption. 

%It is important to point out that Hypothesis $2$ should be taken with a grain of salt, despite being reasonable as justified with numerical studies (\cite{k-eth_kaneko-sagawa}, Section~\ref{sec:numerics} and Appendix~\ref{app:additional_numerics}). First, some entanglement correlation functions of systems without conserved charges were shown to not correspond exactly to what is expected from uniform random pure states~\cite{entanglement_midspectrum_states_deviation, renyi_entropy_chaotic_eigenstates, universal_eigenstate_entanglement_huang}, due to constraints on the locality of the Hamiltonian and orthogonality of the eigenstates. Another important role is played by conservation laws, for instance magnetization or even energy in Hamiltonian systems, leading to corrections on the probability distribution that reproduces the eigenstates~\cite{bianchi_dona, chaos_distributions_entanglement_khemani, entanglement_u1_charge_langlett_nieva, OTOCS_scaling_brandao_huang}. Secondly, this should not be true for any observable considered, otherwise the late-time Hamiltonian ensemble would correspond to a Haar random $k-$design, which is not true~\ref{app:ETH_designs}.  Still, this hypothesis is well justified as described above and provides analytical techniques to handle the case when the energy window is small, i.e., within a microcanonical ensemble~\cite{canonical_typicality_ginibre_ensemble, OTOCS_scaling_brandao_huang}.

\section{\label{sec:liouville_basis}Liouvillian basis for time averages}

%One novel way of quantifying and thinking about complexity in many-body systems comes from the Krylov basis~\cite{universaloperatorgrowth, krylov_review}. This basis is obtained through a Gram-Schmidt orthogonalization process applied to the set of vectors obtained by the nested action of the generator of the dynamics on the initial state. This procedure generates a basis of the subspace of the complete Hilbert space where the dynamics is effectively happening, referred to as the Krylov subspace. The evolution of an initially local operator can be interpreted through this basis, which not only offers a conceptual understanding of complexity growth in Heisenberg picture, but also provides a framework for defining a measure of complexity, called \textit{Krylov complexity}, in many-body systems.~\cite{krylov_complexity_saturation__chaos, krylov_localization_supression_complexity, krylovcomplexity_integrability_to_chaos, multiseed_krylov_complexity, krylov_circuits, Krylov_quantum_circuits_localization}.

%Both the LOE and Krylov complexity for a given Hamiltonian depend on the choice of the initial operator to be evolved, which is usually taken as local. For this reason, characterizing the decomposition of the initial operator in the basis of eigenstates of the system Hamiltonian is of particular importance.

Our goal is to compute the late-time value of LOE for the dynamics generated by the chaotic Hamiltonian $H$ in Equation~\eqref{eqn:eigenvalues_eigenvectors}. In particular, to simplify the calculations we will compute the long-time average of the exponential of the LOE, defined through
\begin{equation}\label{eqn:timeAvLOEfirst}
    \overline{\Tr\left[\big(\rho^{(\mathcal{O})}_{\mathscr{A}}\big)^2\right]}=\lim_{t_{\text{max}}\rightarrow\infty}\frac{1}{t_{\text{max}}}\int_0^{t_{\text{max}}}dt\Tr\left[\big(\rho^{(\mathcal{O})}_{\mathscr{A}}(t)\big)^2\right].
\end{equation}
From now on we will refer to this quantity as the late-time averaged purity or simply purity, in connection to the related quantity in quantum information~\cite{nielsen_chuang_quantum_computing}. After the average, we compute the LOE by applying the logarithm function, which is shown to be exponentially close to calculating the average of the Rényi entropy~\cite{renyi_entropy_chaotic_eigenstates}. Note that we are not considering the LOE of the average state, but rather the average value of the LOE at late times.

For the sake of computing Equation~\eqref{eqn:timeAvLOEfirst}, we need to evolve the local operator $\mathcal{O}$ until time $t$, properly vectorized. To do so, let us first introduce the Liouvillian   $\mathcal{L}[\,\cdot\,]:=[H,\,\cdot\,]$, defined in the operator space as a matrix through \begin{equation}\mathbb{L}\dket{\mathcal{O}}:=\dket{\mathcal{L}[\mathcal{O}]}.\end{equation}
Using the Liouvillian, the Heisenberg picture evolution of $\mathcal{O}$ is written as
\begin{equation}
    \dket{\mathcal{O}(t)}=e^{it\mathbb{L}}\dket{\mathcal{O}}.
\end{equation}

From the form of the Liouvillian as a super-operator, one can easily check that its eigenvectors satisfy
\begin{equation}\label{eqn:LiouvillianEigenvectors}
    \begin{split}
        \mathbb{L}\dket{\omega_{ab}}&=\omega_{ab}\dket{\omega_{ab}},\\
        \dket{\omega_{ab}}&\equiv\frac{1}{\sqrt{\mathrm{d}}}\Omega[\ketbra{E_a}{E_b}],
    \end{split}
\end{equation}
given the eigenvalues and eigenvectors of the Hamiltonian in Equation~\eqref{eqn:eigenvalues_eigenvectors}, and the Bohr frequencies 
\begin{equation}
    \omega_{ab}\equiv E_a-E_b.
\end{equation}
Since $\ket{E_a}$ are normalized states, we observe $\dbraket{\omega_{ab}}{\omega_{ab}}=1/\mathrm{d}$.

{The computation of similar time averages for the late-time evolution of states in Schrödinger picture are typically performed in the basis of eigenvectors of the system Hamiltonian $H$.
This choice is convenient because, if the eigenenergies are nondegenerate, such as for the spectrum of chaotic systems, the oscillating phases between distinct eigenenergies cancel under long-time averaging, so that only the diagonal contributions remain~\cite{ETH_review}.
}

{For chaotic systems, the Liouvillian eigenvectors $\dket{\omega_{ab}}$ are nondegenerate only if $\omega_{ab}\neq 0$, while the eigenvector associated with $\omega_{ab}=0$ is always at least $\mathrm{d}$-times degenerate. %Oscillating phases within this degenerate subspace do not vanish in the long-time average. 
To work with a single vector in this degenerate subspace, we introduce an effective basis of lower dimension defined by}
\begin{equation}\label{eq:omega_vectors_normalized}
    \dket{\omega_m}:=\frac{1}{\mathcal{N}_m}\sum_{(a,b)\in\mathcal{I}_m}\mathcal{O}_{ab}\dket{\omega_{ab}},\quad {m=0,\ldots,\mathcal{K}-1.}
\end{equation}
 $\mathcal{I}_m$ is the set of all $(a,b)$ such that $\omega_{ab}=\omega_m$, while 
\begin{equation}\label{eqn:normalization_constant}
    \mathcal{N}_m=\sqrt{\frac{1}{\mathrm{d}}\sum_{(a,b)\in\mathcal{I}_m}|\mathcal{O}_{ab}|^2} = \dbraket{\omega_m}{\mathcal{O}}
\end{equation} is a normalization {constant}.  Moreover, 
\begin{equation}
\mathcal{O}_{ab}\equiv\bra{E_a}\mathcal{O}\ket{E_b}.    
\end{equation} 

{Note that the dimension of this basis is equivalent to the number $\mathcal{K}$ of different non-trivial subspaces related to resolving the initial operator in the Liouvillian eigenbasis.} Moreover, $\dket{\omega_0}$ is the projection of $\mathcal{O}$ over the degenerate subspace with $\omega_{ab}=0$. $\mathcal{K}$ corresponds to the number of non-vanishing projections of $\mathcal{O}$ over the Liouvillian eigenstates, and it is bounded from above by $\mathrm{d}^2-\mathrm{d}+1$~\cite{edge_Krylov_space,krylov_localization_supression_complexity}.

This basis 
was introduced in the context of quantum Krylov methods~\cite{krylov_review} as another possible basis for the Krylov subspace~\cite{edge_Krylov_space,krylovcomplexity_integrability_to_chaos,krylov_localization_supression_complexity}. In this context, $\mathcal{K}$ %expressing the number of non-zero contributions to the operator resolved Liouvillian basis 
is the dimension of the Krylov subspace. Therefore, the set of vectors obtained by resolving the matrix elements of the operator $\mathcal{O}$ in the Liouvillian basis can be said to be a minimal effective basis that describes the dynamics in operator space. 

We take advantage of this basis to write the Heisenberg evolution of the local operator $\mathcal{O}$ in the simple form
\begin{equation}
\begin{split}
    \dket{\mathcal{O}(t)}&=\sum_{m=0}^{\mathcal{K}-1}e^{i\omega_mt}\dbraket{\omega_m}{\mathcal{O}}\dket{\omega_m}\\
    &=\sum_{m=0}^{\mathcal{K}-1}e^{i\omega_mt}\mathcal{N}_m\dket{\omega_m}.
\end{split}
\end{equation}
 {Starting from this expression}, the hypotheses we are going to consider can be reasonably applied, simplifying the form of the infinite-time integrals and providing a clear late-time value in terms of the basis $\dket{\omega_m}$.

\section{\label{sec:results}Analytical Results}

\subsection{\label{subsec:computation}Long-time average}
Given the assumption of normalized initial operator according to the infinite temperature inner product, we proceed with obtaining the density matrix related to the vectorized operator as
\begin{equation}
    \begin{split}
        \rho^{\mathcal{O}}(t) &=\dketbra{\mathcal{O}(t)}{\mathcal{O}(t)}\\
        &=\sum^{\mathcal{K}-1}_{m,n=0}e^{i(\omega_m-\omega_n)t}\mathcal{N}_m\mathcal{N}_n\dketbra{\omega_m}{\omega_n}.
    \end{split}
\end{equation}
Considering the bipartition defined in Sec.~\ref{sec:liouville_basis}, $\mathcal{H}_{\mathscr{A}}\otimes\mathcal{H}_{\mathscr{B}}$, with bipartite basis vectors given by $\dket{j,\alpha}\equiv\dket{j}_{\mathscr{A}}\otimes\dket{\alpha}_{\mathscr{B}}$, we will define the density matrix in a bipartite basis of $\mathcal{H}\otimes\mathcal{H}'$. From now on, we will use indices $m$, $n$, $p$ and $q$ for the basis of the Liouvillian, $\dket{\omega_m}$, other Latin letters for the basis of subspace $\mathcal{H}_{\mathscr{A}}$ (summing from $1$ to $\mathrm{d}_A^2$), and Greek letters for the basis of subspace $\mathcal{H}_{\mathscr{B}}$ (summing from $1$ to $\mathrm{d}_B^2$). {Moreover, to simplify the notation we adopt the Einstein summation convention for these indices, i.e., summation over repeated indices is implied.} 

In the basis {introduced above, the density matrix is written as}
\begin{equation}
    \begin{split}
        \rho^{\mathcal{O}}(t) =& e^{i(\omega_m-\omega_n)t}\mathcal{N}_m\mathcal{N}_n\\
        &\times \dbraket{j,\alpha}{\omega_m}\dbraket{\omega_n}{k, \beta}\dketbra{j, \alpha}{k, \beta}.
    \end{split}
\end{equation}
For convenience, we define {a family of} matrices {labeled by the Liouvillian eigenvalue $m$}: %corresponding to the decomposition of the eigenvectors of the Liouvillian on the bipartite basis chosen
\begin{equation}\omega^{(m)}_{j\alpha} := \dbraket{j,\alpha}{\omega_m},\end{equation} 
{This matrix is analogous to the one arising during the Schmidt Decomposition of pure states.} 
Tracing out subsystem $\mathcal{H}_{\mathscr{B}}$,
{\begin{equation}
    \begin{split}
         \rho^{\mathcal{O}}_{\mathscr{A}}(t) =& e^{i(\omega_m-\omega_n)t}\mathcal{N}_m\mathcal{N}_n\omega^{(m)}_{j\alpha}\omega^{(n)*}_{k\beta}\dketbra{j}{k}_{\mathscr{A}}\delta_{\alpha\beta}\\
        %\rho^{\mathcal{O}}_{\mathscr{A}}(t)
        =& e^{i(\omega_m-\omega_n)t}\mathcal{N}_m\mathcal{N}_n\omega^{(m)}_{j\alpha}\omega^{(n)\dagger}_{\alpha k}\dketbra{j}{k}_{\mathscr{A}}.
    \end{split}
\end{equation}}

{Next}, to compute the $2-$Rényi entropy, we calculate the purity of the reduced density matrix, {which reads}
\begin{equation}\label{eq:trace_rho_sqr_timedependent}
    \begin{split}
    &\Tr\left[\big(\rho_{\mathscr{A}}^{({\mathcal{O}})}(t)\big)^2\right] = e^{i(\omega_m+\omega_p-\omega_n-\omega_q)t}\mathcal{N}_m\mathcal{N}_n\mathcal{N}_p\mathcal{N}_q \\
        &\times\Tr_{\mathscr{A}}(\omega^{(m)}\omega^{(n)\dagger}\omega^{(p)}\omega^{(q)\dagger}).
    \end{split}
\end{equation}

The trace of these $\omega$ matrices is used here to simplify the notation. We should always be aware that what we are actually calculating is the decomposition of the Liouvillian eigenstates in the bipartite basis. Explicitly,
\begin{equation}
    \begin{split}
        &\Tr_{\mathscr{A}}(\omega^{(m)}\omega^{(n)\dagger}\omega^{(p)}\omega^{(q)\dagger})=\Tr_{\mathscr{B}}(\omega^{(q)\dagger}\omega^{(m)}\omega^{(n)\dagger}\omega^{(p)}) \\
        &=\dbraket{j,\alpha}{\omega_m}\dbraket{\omega_n}{k,\alpha}\dbraket{k,\beta}{\omega_p}\dbraket{\omega_q}{j,\beta}.
    \end{split}
\end{equation}

We notice that in Equation~\eqref{eq:trace_rho_sqr_timedependent} all time dependencies are encoded in the phases. 
%We can then take advantage of the fact that all the degenerate eigenfrequencies are grouped within the same eigenstate, therefore it simplifies the infinite time average. Considering that the exponentials are rapidly oscillating, the only remaining terms of the integration will be those for which the argument of the exponential is zero. From construction, all $\omega_m$ are different from each other, therefore there are no degeneracies for the eigenvalues appearing in the sums. This is not implied from the non-degeneracy of the eigenvalues of the Hamiltonian, but is due to the way we constructed the Liouvillian basis. For instance, one could have $E_3-E_2 = E_2-E_1$ and still have non-degenerate energies. However, this is not enough to obtain the simplifications desired, as it would simplify the long-time integrals only when we have exponentials of two frequencies. We need to apply a higher order condition of non-resonance, in this case, $k=2$ non-resonance for the frequencies, implying that we need a $k=4$ non-resonance for the energies of the Hamiltonian~\cite{non_resonance_thermalization_A_Winter, non_resonance_bertini, non_resonance_Riddell}, as mentioned in Subsection~\ref{sec:hypothesis}. 
Next, we take the long-time average defined in Equation~\eqref{eqn:timeAvLOEfirst} and make use of the 4 non-resonance condition introduced in Sec.~\ref{sec:hypothesis}. In the long-time average  only the non-oscillating terms survive~\cite{non_resonance_thermalization_A_Winter, non_resonance_bertini, non_resonance_Riddell}. From this consideration, we have only three remaining terms: ${\omega_m=\omega_n=\omega_p=\omega_q}$; $\omega_m=\omega_n$, $\omega_p=\omega_q$, $\omega_m\neq\omega_p$; and $\omega_m=\omega_q$, $\omega_p=\omega_n$, $\omega_m\neq\omega_p$. As a result, we get the long-time average in terms of the traces of the $\omega$ matrices:
\begin{equation}\label{eq:ergodic_avg_open_terms}
    \begin{split}
\overline{\Tr[\big(\rho^{(\mathcal{O})}_{\mathscr{A}}\big)^2]}=\sum_{m=0}^{\mathcal{K}-1}\mathcal{N}_m^4\Tr_{\mathscr{A}}(\omega^{(m)\dagger}\omega^{(m)}\omega^{(m)\dagger}\omega^{(m)})\\
        +\sum_{m\neq p=0}^{\mathcal{K}-1}\mathcal{N}_m^2\mathcal{N}_p^2\Tr_{\mathscr{A}}(\omega^{(m)}\omega^{(m)\dagger}\omega^{(p)}\omega^{(p)\dagger})\\
        +\sum_{m\neq p=0}^{\mathcal{K}-1}\mathcal{N}_m^2\mathcal{N}_p^2\Tr_{\mathscr{B}}(\omega^{(m)\dagger}\omega^{(m)}\omega^{(p)\dagger}\omega^{(p)}).
    \end{split}
\end{equation}
Note that deriving Equation~\eqref{eq:ergodic_avg_open_terms} only required the 4 non-resonance condition. 
A physical intuition on the late-time purity is discussed in Appendix~\ref{app:physical_intuition_purity}.

\subsection{\label{subsec:volume_law_operators}Expressing the time-averaged purity in terms of the system eigenstates}

%We now get to the step where we explore the late time Equation~\eqref{eq:ergodic_avg_open_terms} to obtain the volume laws. To proceed further, we take advantage of the $4$ non-resonance condition again, this time after introducing the vectorization map explicitly in Equation~\eqref{eq:ergodic_avg_open_terms} to obtain a form in terms of the eigenstates of the Hamiltonian.
Next, write Equation~\eqref{eq:ergodic_avg_open_terms} as a function of the Hamiltonian eigenstates. For simplicity, in this section we perform the calculations explicitly only for the first term in the RHS of Equation~\eqref{eq:ergodic_avg_open_terms}, and we refer to Appendix~\ref{app:additional_calculations_summations} for the analysis of the remaining terms. 

First of all, recalling the results in Eqs.~\eqref{eqn:LiouvillianEigenvectors} and~\eqref{eqn:normalization_constant}, we observe
\begin{equation}
\begin{split}
    &\Big(\dbra{j,j'}_{\mathscr{A}}\otimes\dbra{\alpha, \alpha'}_{\mathscr{B}}\Big)\dket{\omega_m}\\
    &=\frac{1}{\mathcal{N}_m}\sum_{(a,b)\in\mathcal{I}_m}\frac{1}{\sqrt{\mathrm{d}}}\mathcal{O}_{ab}E^{a}_{j\alpha}E^{b*}_{j'\alpha'}.
\end{split}
\end{equation}
We have introduced the notation 
\begin{equation}
    \dket{j,j'}_{\mathscr{A}}\equiv\ket{j}_A\otimes\ket{j'}_{A'},
\end{equation}
and analogously for subspace $\mathscr{B}$. The prime symbol on indexes means that the index refers to the output space in the vectorized representation of operators. %Notice that if the sum over $j$ defined for $\mathscr{A}$ is from $1$ to $d_A^2$, the sums for $j$ in $A$ and $j'$ in $A'$ are from $1$ to $d_A$ for each of them. 
Furthermore, 
\begin{equation}
    E^{a}_{j\alpha}\equiv\braket{j,\alpha}{E_a}.
\end{equation}
We then get, for the first term in the RHS of ~\eqref{eq:ergodic_avg_open_terms},
\begin{equation}\label{eqn:intermediateStep}
    \begin{split}
        &\sum_{m=0}^{\mathcal{K}-1}\mathcal{N}_m^4\Tr_{\mathscr{A}}(\omega^{(m)}\omega^{(m)\dagger}\omega^{(m)}\omega^{(m)\dagger})\\
        &=\frac{1}{d^2}\sum_{m=0}^{\mathcal{K}-1}\sum_{(a,b),(c,d)\in\mathcal{I}_m}\sum_{(e,f),(g,h)\in\mathcal{I}_m}\mathcal{O}_{ab}\mathcal{O}^*_{cd}\mathcal{O}_{ef}\mathcal{O}^*_{gh}\\
        &\hspace{6em}\times E^a_{j\alpha}E^{b*}_{j'\alpha'}E^{c*}_{k\alpha}E^{d}_{k'\alpha'}E^e_{k\beta}E^{f*}_{k'\beta'}E^{g*}_{j\beta}E^{h}_{j'\beta'}.
    \end{split}
\end{equation}
Here the sums over the indexes of the eigenstates are shown explicitly to stress that this set does not cover the whole spectrum, but only those eigenstates satisfying $E_a-E_b=\omega_m\Rightarrow(a,b)\in\mathcal{I}_m$. 

{Due to non-resonance assumption, only the $\omega_m=0$ frequency will be degenerate, and we can then split Equation~\eqref{eqn:intermediateStep} into a ``diagonal'' part collecting terms related to $\omega_m=0$, which we name $\mathcal{D}$, and a ``off-diagonal'' part ($\omega_m\neq 0$), which we name $\mathcal{P}$. 
For the diagonal part we obtain
\begin{equation}
    \begin{split}
        \mathcal{D}=\sum_{a, b, c, d=1}^{\mathrm{d}}\mathcal{O}_{aa}\mathcal{O}^*_{bb}\mathcal{O}_{cc}\mathcal{O}^*_{dd}E^a_{j\alpha}E^{a*}_{j'\alpha'}E^{b*}_{k\alpha}E^{b}_{k'\alpha'}\\
        \times E^c_{k\beta}E^{c*}_{k'\beta'} E^{d*}_{j\beta}E^{d}_{j'\beta'}.
    \end{split}
\end{equation}
The off-diagonal terms are all non-degenerate, implying that there is only one pair $(a,b)\in\mathcal{I}_m$ with $a\neq b$ for each of the $\omega_m$. In this sense, repeated sums over the same $m$ reduce to the multiplication of the only possible case:
\begin{equation}
    \begin{split}
        &\mathcal{P}=\frac{1}{\mathrm{d}^2}\sum_{m=1}^{\mathcal{K}-1}\sum_{(a,b),(c,d)\in\mathcal{I}_m}\sum_{(e,f),(g,h)\in\mathcal{I}_m}\\
        &\mathcal{O}_{ab}\mathcal{O}^*_{cd}\mathcal{O}_{ef}\mathcal{O}^*_{gh} E^a_{j\alpha}E^{b*}_{j'\alpha'}E^{c*}_{k\alpha}E^{d}_{k'\alpha'}E^e_{k\beta}E^{f*}_{k'\beta'}E^{g*}_{j\beta}E^{h}_{j'\beta'}\\
        &=\frac{1}{\mathrm{d}^2}\sum_{a\neq b=1}^{\mathrm{d}}|\mathcal{O}_{ab}|^4E^a_{j\alpha}E^{b*}_{j'\alpha'}E^{a*}_{k\alpha}E^{b}_{k'\alpha'}E^a_{k\beta}E^{b*}_{k'\beta'}E^{a*}_{j\beta}E^{b}_{j'\beta'}.
    \end{split}
\end{equation}

Therefore, the first term in Equation~\eqref{eq:ergodic_avg_open_terms} reads

\begin{equation}
    \begin{split}
        &\sum_{m=0}^{\mathcal{K}-1}\mathcal{N}_m^4\Tr_{\mathscr{A}}(\omega^{(m)}\omega^{(m)\dagger}\omega^{(m)}\omega^{(m)\dagger})\\
        &=\frac{1}{\mathrm{d}^2}\sum_{a, b, c, d=1}^{\mathrm{d}}\mathcal{O}_{aa}\mathcal{O}^*_{bb}\mathcal{O}_{cc}\mathcal{O}^*_{dd}E^a_{j\alpha}E^{a*}_{j'\alpha'}E^{b*}_{k\alpha}E^{b}_{k'\alpha'}\\
        &\times E^c_{k\beta}E^{c*}_{k'\beta'} E^{d*}_{j\beta}E^{d}_{j'\beta'}\\
        &+\frac{1}{\mathrm{d}^2}\sum_{a\neq b=1}^{\mathrm{d}}|\mathcal{O}_{ab}|^4 E^a_{j\alpha}E^{b*}_{j'\alpha'}E^{a*}_{k\alpha}E^{b}_{k'\alpha'}E^a_{k\beta}E^{b*}_{k'\beta'}E^{a*}_{j\beta}E^{b}_{j'\beta'}.
    \end{split}
\end{equation}

Summing up the different terms coming from Equation~\eqref{eq:ergodic_avg_open_terms}, we have $6$ remaining contributions (see Appendix~\ref{app:additional_calculations_summations}). The final sum is given by
\begin{widetext}
    \begin{equation}\label{eq:final_purity_eigenstates}
        \begin{split}
            \overline{\Tr[\big(\rho^{(\mathcal{O})}_{\mathscr{A}}\big)^2]}=\frac{1}{\mathrm{d}^2}\sum_{a, b, c, d=1}^{\mathrm{d}}\mathcal{O}_{aa}\mathcal{O}^*_{bb}\mathcal{O}_{cc}\mathcal{O}^*_{dd}E^a_{j\alpha}E^{a*}_{j'\alpha'}E^{b*}_{k\alpha}E^{b}_{k'\alpha'}E^c_{k\beta}E^{c*}_{k'\beta'}E^{d*}_{j\beta}E^{d}_{j'\beta'}\\
            +\frac{2}{\mathrm{d}^2}\sum_{a,c=1}^{\mathrm{d}}\sum_{e\neq f=1}^{\mathrm{d}}\mathcal{O}_{aa}\mathcal{O}^*_{cc}|\mathcal{O}_{ef}|^2E^a_{j\alpha}E^{a*}_{j'\alpha'}E^{c*}_{k\alpha}E^{c}_{k'\alpha'}E^{e}_{k\beta}E^{f*}_{k'\beta'}E^{e*}_{j\beta}E^{f}_{j'\beta'}\\
             +\frac{1}{\mathrm{d}^2}\sum_{a \neq b = 1}^{\mathrm{d}}\sum_{c \neq d=1}^{\mathrm{d}}\mathcal{O}_{ab}\mathcal{O}^*_{cd}\mathcal{O}_{ab}\mathcal{O}^*_{cd}E^a_{j\alpha}E^{b*}_{j'\alpha'}E^{c*}_{k\alpha}E^{d}_{k'\alpha'}E^{a}_{k\beta}E^{b*}_{k'\beta'}E^{c*}_{j\beta}E^{d}_{j'\beta'}\\
             +\frac{2}{\mathrm{d}^2}\sum_{a,g=1}^{\mathrm{d}}\sum_{c\neq d=1}^{\mathrm{d}}\mathcal{O}_{aa}|\mathcal{O}_{cd}|^2\mathcal{O}^*_{gg}E^a_{j\alpha}E^{a*}_{j'\alpha'}E^{c*}_{k\alpha}E^{d}_{k'\alpha'}E^{c}_{k\beta}E^{d*}_{k'\beta'}E^{g*}_{j\beta}E^{g}_{j'\beta'}\\
             +\frac{1}{\mathrm{d}^2}\sum_{a \neq b = 1}^{\mathrm{d}}\sum_{c \neq d=1}^{\mathrm{d}}|\mathcal{O}_{ab}|^2|\mathcal{O}_{cd}|^2E^a_{j\alpha}E^{b*}_{j'\alpha'}E^{c*}_{k\alpha}E^{d}_{k'\alpha'}E^{c}_{k\beta}E^{d*}_{k'\beta'}E^{a*}_{j\beta}E^{b}_{j'\beta'}\\
             -\frac{1}{\mathrm{d}^2}\sum_{a \neq b = 1}^{\mathrm{d}}|\mathcal{O}_{ab}|^4 E^a_{j\alpha}E^{b*}_{j'\alpha'}E^{a*}_{k\alpha}E^{b}_{k'\alpha'}E^{a}_{k\beta}E^{b*}_{k'\beta'}E^{a*}_{j\beta}E^{b}_{j'\beta'}.
         \end{split}
    \end{equation}
\end{widetext}

{We pause to interpret the structure of the above equation, resulting from our first two hypotheses. As shown in Appendix~\ref{app:physical_intuition_purity}, the late-time formula takes the form of a mutual information between the two subsystems, expressed in terms of traces of squared reduced density matrices of a diagonal ensemble defined by the operator's decomposition in the Liouvillian basis. Equation (41) shows how this mutual information can be computed via eigenstate correlation functions of the observable, where index contractions over eigenstates determine the contributions of diagonal and off-diagonal operator elements, weighted according to their respective ETH scalings with system dimension. This reasoning connects to a broader framework in which chaos diagnostics such as ETH, the Spectral Form Factor, Krylov complexity \cite{krylovcomplexity_integrability_to_chaos}, and OTOCs \cite{OTOCS_scaling_brandao_huang}, all reflect, in different ways, the consequences of eigenvalue and eigenvector correlations, both in their early-time dynamics and/or late-time behavior, under the chaos assumption. The eigenstate correlations relevant to our computation will be evaluated, under our third working hypothesis, using Random Tensor Networks, introduced in the following section.}

\subsection{Replacing Hamiltonian eigenstates with random eigenstates}

%As discussed in Subsection~\ref{sec:hypothesis}, we are going to regard the eigenstates of the chaotic Hamiltonian as generated according to unitary designs of a particular order to be defined soon. A Haar random vector satisfies the invariance property of the measure, i.e., when integrating over Haar random vectors, the distributions $V\ket{\psi}\sim \ket{\psi}$ are equivalent, being $V$ any unitary. This same statement defines the Haar measure in the unitary group manifold, $VU\sim U,\;\forall V$. If we take any column vector of $U$, let us say $U\ket{a}$, then $V(U\ket{a})\sim U\ket{a}$. Therefore, the columns of $U$ are invariant vectors under unitaries, being then distributed according to the Haar measure due to the uniqueness. This way, if we write each eigenvector of the Hamiltonian as one of the columns of a random unitary $U$, $U\ket{a}=\ket{E_a}$, we can argue that the integration over the elements of the unitaries for a particular design order~\cite{introductiontohaar, randomtensornetworks_graphical_haar, RTNI_haar}

%\begin{equation}
%    \begin{split}
    %    &\int_{U\sim \varepsilon}U_{a_1,i_1}\cdots U_{a_k,i_k} U^{*}_{b_1,j_1}\cdots U^{*}_{b_k,j_k}d\mu_{\varepsilon}(U)\\
    %    &=\int_{E\sim \varepsilon}E^{a_1}_{i_1}\cdots E^{a_k}_{i_k} E^{b_1*}_{j_1}\cdots E^{b_k*}_{j_k}d\mu_{\varepsilon}(E),
   % \end{split}
%\end{equation}
Next, according to our third assumption in Section~\ref{sec:hypothesis}, we replace the Hamiltonian eigenstates with Haar-random eigenstates. We can thus apply the notions of Weingarten calculus to Equation~\eqref{eq:final_purity_eigenstates}. In particular, due to concentration of measure in the large-$\mathrm{d}$ limit, we can also replace the product of correlated matrix elements of the random states with their Haar-average. Representing the Haar random ensemble with $\text{H}$, we obtain
\begin{widetext}
    \begin{equation}\label{eq:final_purity_unitaries}
        \begin{split}
            \overline{\Tr[\big(\rho^{(\mathcal{O})}_{\mathscr{A}}\big)^2]}=\frac{1}{\mathrm{d}^2}\sum_{a, b, c, d=1}^{\mathrm{d}}\mathcal{O}_{aa}\mathcal{O}^*_{bb}\mathcal{O}_{cc}\mathcal{O}^*_{dd}\int_{U\sim \text{H}} U_{a,j\alpha}U^{*}_{a,j'\alpha'}U^{*}_{b,k\alpha}U_{b,k'\alpha'}U_{c,k\beta}U^{*}_{c,k'\beta'}U^{*}_{d,j\beta}U_{d,j'\beta'}d\mu_{\text{H}}(U)\\
            +\frac{2}{\mathrm{d}^2}\sum_{a,c=1}^{\mathrm{d}}\sum_{e\neq f=1}^{\mathrm{d}}\mathcal{O}_{aa}\mathcal{O}^*_{cc}\mathcal{O}_{ef}\mathcal{O}^*_{ef}\int_{U\sim \text{H}} U_{a,j\alpha}U^{*}_{a,j'\alpha'}U^{*}_{c,k\alpha}U_{c,k'\alpha'}U_{e,k\beta}U^{*}_{f,k'\beta'}U^{*}_{e,j\beta}U_{f,j'\beta'}d\mu_{\text{H}}(U)\\
             +\frac{1}{\mathrm{d}^2}\sum_{a \neq b = 1}^{\mathrm{d}}\sum_{c \neq d=1}^{\mathrm{d}}\mathcal{O}_{ab}\mathcal{O}^*_{cd}\mathcal{O}_{ab}\mathcal{O}^*_{cd}\int_{U\sim \text{H}} U_{a,j\alpha}U^{*}_{b,j'\alpha'}U^{*}_{c,k\alpha}U_{d,k'\alpha'}U_{a,k\beta}U^{*}_{b,k'\beta'}U^{*}_{c,j\beta}U_{d,j'\beta'}d\mu_{\text{H}}(U)\\
             +\frac{2}{\mathrm{d}^2}\sum_{a,g=1}^{\mathrm{d}}\sum_{c\neq d=1}^{\mathrm{d}}\mathcal{O}_{aa}\mathcal{O}^*_{cd}\mathcal{O}_{cd}\mathcal{O}^*_{gg}\int_{U\sim \text{H}} U_{a,j\alpha}U^{*}_{a,j'\alpha'}U^{*}_{c,k\alpha}U_{d,k'\alpha'}U_{c,k\beta}U^{*}_{d,k'\beta'}U^{*}_{g,j\beta}U_{g,j'\beta'}d\mu_{\text{H}}(U)\\
             +\frac{1}{\mathrm{d}^2}\sum_{a \neq b = 1}^{\mathrm{d}}\sum_{c \neq d=1}^{\mathrm{d}}\mathcal{O}_{ab}\mathcal{O}^*_{cd}\mathcal{O}_{cd}\mathcal{O}^*_{ab}\int_{U\sim \text{H}} U_{a,j\alpha}U^{*}_{b,j'\alpha'}U^{*}_{c,k\alpha}U_{d,k'\alpha'}U_{c,k\beta}U^{*}_{d,k'\beta'}U^{*}_{a,j\beta}U_{b,j'\beta'}d\mu_{\text{H}}(U)\\
             -\frac{1}{\mathrm{d}^2}\sum_{a \neq b = 1}^{\mathrm{d}}|\mathcal{O}_{ab}|^4 \int_{U\sim \text{H}} U_{a,j\alpha}U^{*}_{b,j'\alpha'}U^{*}_{a,k\alpha}U_{b,k'\alpha'}U_{a,k\beta}U^{*}_{b,k'\beta'}U^{*}_{a,j\beta}U_{b,j'\beta'}d\mu_{\text{H}}(U).
        \end{split}
    \end{equation}
\end{widetext}
It is essential to point out that the ETH hypothesis is also appearing implicitly in the assumption behind Equation~\eqref{eq:final_purity_unitaries}, as we regarded the elements of the matrix of $\mathcal{O}$ in the energy eigenbasis to be independent of the structure of the eigenstates, while depending on smooth functions of the eigenenergies only.

Each of the integrals in Equation~\eqref{eq:final_purity_unitaries} has $(4!)^2$ terms, due to their decomposition in terms of Weingarten functions with two sums running over permutations in $S_4$. On the other hand, if we decided to use the approximation where the identity is the dominant term appearing in the Weingarten sum~\cite{butterfly_effect_localoperatorsentanglement}, we would have $24$ terms in each integral. However, the column indexes appearing in the unitaries will be defined in terms of two indexes, expressively increasing the number of Kronecker deltas. We can, instead of performing these calculations explicitly, rewrite each term within the integral as a random tensor network, where the $U$ are the random tensors distributed according to the unitary Haar measure, enabling the use of graphical calculus for random tensor networks~\cite{randomtensornetworks_graphical_haar, RTNI_haar}. Doing so, we can reduce the number of terms drastically, ending up with $13$ terms with different weights depending on the dimensions $\mathrm{d}_A$, $\mathrm{d}_B$ and $\mathrm{d}$.  A detailed discussion of the steps involved in this calculation is presented in Appendix~\ref{app:random_tensor_networks}.

\subsection{Final expressions for LOE}
\label{sec:final:results}
Finally, after performing the computations, we end up with $13$ different terms, each with a corresponding weight $w_n$. The details of the calculation and specific formulas for the weights in terms of $\mathrm{d}_A$, $\mathrm{d}_B$ and $\mathrm{d}$ are discussed in Appendix~\ref{app:random_tensor_networks}. Here, we skip the details and present the final result by restricting ourselves to two different limits: ${\mathrm{d}_A=\text{O}(1),\, \mathrm{d}\gg 1}$, and {${\mathrm{d}_A=\sqrt{\mathrm{d}}} \gg 1$}.

\subsubsection{$\mathrm{d}_A=\text{O}(1),\, \mathrm{d}\gg 1$}

In this limit, we obtain the late-time formula for LOE as

\begin{flalign}\label{eq:dA_O1_big_windows}
    &\Big\langle\overline{S_2(\rho_{\mathscr{A}}^{\mathcal{O}}(t))}\Big\rangle\approx 2\ln(\mathrm{d}_A)&&\\
    &\hspace{6em}-\ln\left\{1 +(\mathrm{d}-1)\sigma_{\text{off-diag}}^2\left[\sigma^2_{\text{diag}}-1\right]\right\}.\nonumber&&
\end{flalign}

We introduce the averages over diagonal and off-diagonal values of the operator matrix elements in the energy basis ~\cite{ETH_review}:
\begin{equation}\label{eq:diagonal_average}
    \langle \mathcal{O}_{aa} \rangle_{\text{diag}}:= \frac{1}{\mathrm{d}}\sum_{b=1}^{\mathrm{d}}\mathcal{O}_{bb}
\end{equation}
and
\begin{equation}\label{eq:off_diagonal_average}
    \langle \mathcal{O}_{ab} \rangle_{\text{off-diag}}:= \frac{1}{\mathrm{d}(\mathrm{d}-1)}\sum_{c\neq d=1}^{\mathrm{d}}\mathcal{O}_{cd},
\end{equation}
respectively. The quantities $\sigma^2$  appearing in Equation~\eqref{eq:dA_O1_big_windows} are the variances over the same ensembles:
\begin{equation}\label{eqn:variances}
\begin{split}
    \sigma_{\text{diag}}^2 &= \frac{1}{\mathrm{d}}\sum_{b=1}^{\mathrm{d}}\mathcal{O}^2_{bb}-\langle \mathcal{O}_{aa} \rangle_{\text{diag}}^2,\\
    \sigma_{\text{off-diag}}^2 &= \frac{1}{\mathrm{d}(\mathrm{d}-1)}\sum_{c\neq d=1}^{\mathrm{d}}\mathcal{O}_{cd}^2-\langle \mathcal{O}_{ab} \rangle_{\text{off-diag}}^2.
\end{split}
\end{equation}

\subsubsection{${\mathrm{d}_A=\sqrt{\mathrm{d}}} \gg 1$}

In the case ${d_A=\sqrt{\mathrm{d}}\gg 1}$, we obtain
 
\begin{flalign}\label{eq:dA_big_big_windows}
    &\Big\langle\overline{S_2(\rho_{\mathscr{A}}^{\mathcal{O}}(t))}\Big\rangle\approx \ln(\mathrm{d})&\\
    &\hspace{5.5em}-\ln\left\{1 + \sigma_{\text{diag}}^2\left[1+(\mathrm{d}-1)\sigma_{\text{off-diag}}^2\right]\right\},\nonumber&&
\end{flalign}
for the same quantities defined in Equation~\eqref{eqn:variances}. One can notice that this equation does not involve the Page correction that appears for $\mathrm{d}_A=\sqrt{\mathrm{d}}$~\cite{volume_law_eigenstates, page_entanglement_subsystem}. The reason behind this is that we are only considering the leading order contributions given by the scales of ETH magnitudes. Page corrections should appear in the next leading order. Eqs.~\eqref{eq:dA_O1_big_windows} and~\eqref{eq:dA_big_big_windows} predict a volume law plus an additional term of same order that depends on the initial operator $\mathcal{O}$ for late-time LOE.

The fluctuations (variances) of the matrix elements scale as the inverse of the dimension of the Hilbert space dimension, $1/\mathrm{d}$, decaying to zero for the off-diagonal elements, while achieving $\text{O}(1)$ values for the diagonal elements (see Appendix~\ref{app:additional_numerics} for a study of the numerical scaling of this quantity as a function of $\mathrm{d}$), as expected from ETH \cite{mfim_cirac_thermalization, master_equations_ETH_Donovan_Mitchison}. Therefore, for small system sizes this correction should become more significant, shifting the volume law by a constant depending on $\mathrm{d}$. When we consider large dimensional systems, the term depending on the fluctuations will be of order
\begin{equation}
\begin{split}
   &\ln\left\{1 +(\mathrm{d}-1)\sigma_{\text{off-diag}}^2\left[\sigma^2_{\text{diag}}-1\right]\right\}\\
   &=\ln\left[1 +\text{O}(\mathrm{d})\text{O}(1/\mathrm{d})\text{O}(1)\right]=\ln\left[\text{O}(1)\right] 
\end{split}
\end{equation}
for small $d_A$, and
\begin{equation}
\begin{split}
    &\ln\left\{1 + \sigma_{\text{diag}}^2\left[1+(\mathrm{d}-1)\sigma_{\text{off-diag}}^2\right]\right\}\\
    &=\ln\left\{1 +\text{O}(1)\left[1 + \text{O}(\mathrm{d})\right]\text{O}(1/\mathrm{d})\right\}=\ln\left[\text{O}(1)\right]
\end{split}
\end{equation}
for $\mathrm{d}_A\approx \mathrm{d}$. Therefore we obtain a volume law corrected by at most the logarithm of an order $1$ number. Comparing this behavior with the result obtained for the state space evolution (see Appendix~\ref{app:volume_law_states}), we can argue that the late-time behavior of the LOE obtained through our calculation has an intrinsic dependency on the initial operator, in contrast with the universal behavior given by infinite temperature initial states on the Page value case for Schrödinger evolutions.

The result is consistent with the analytical formula for late-time LOE in random circuits derived in Ref.~\cite{butterfly_effect_localoperatorsentanglement}. There, the authors neglect the term proportional to $\Tr(\mathcal{O}^{\dagger}\mathcal{O}\mathcal{O}^{\dagger}\mathcal{O})/\mathrm{d}$, by using the argument that its order is subleading. This term is analogous to the additional terms depending on $\sigma^2$ in Eqs.~\eqref{eq:dA_O1_big_windows} and~\eqref{eq:dA_big_big_windows}. In our case this term cannot be discarded, as explained above.

The results found here in the regime of large system sizes are closely connected with what is expected from these quantities, both from the consideration of a linear increase in time of the LOE for chaotic systems~\cite{dowling_modi_scrambling_chaos_localoperatorentanglement, coarsegrained_localoperatorentanglement, butterfly_effect_localoperatorsentanglement}, and with the results obtained through Haar designs~\cite{butterfly_effect_localoperatorsentanglement, designs_free_probability_pappalardi, k-eth_kaneko-sagawa} and numerics~\cite{coarsegrained_localoperatorentanglement, vincenzo_alba_diffusion_operator_entanglement_spreading}. The additional dependence on the initial operator coming from $\sigma^2_{\text{diag}}$ and $\sigma^2_{\text{off-diag}}$, which has the same order of the leading contribution when regarding ETH physical observables, comes from generic physical operators and should provide the same contributions regarding the behavior as a function of the dimension.

\section{\label{sec:numerics}Numerical benchmarks}

As a benchmark for our analytical results and focusing on the consequences of the third hypothesis when applied to energy windows or the whole Hilbert space, we consider the maximally chaotic set of parameters for the Mixed Field Ising Model (MFIM) in a 1D chain~\cite{chaos_distributions_entanglement_khemani} with broken reflection symmetry. Its Hamiltonian is given by
\begin{align}\label{eq:mixed_field_Ising}
    H=J\sum_{j=0}^{L-2}\sigma^z_j\sigma^z_{j+1}+\sum_{j=0}^{L-1}\Big(h_z\sigma^z_j+h_x\sigma^x_j\Big)+g_0\sigma_0^z+g_l\sigma^z_{L-1}.
\end{align}
{From now on, we will fix the parameters $(J, h_x, h_z, g_0, g_l) = (1.0, 1.1, 0.3, 0.25, -0.25)$.} 

{The above Hamiltonian models a chain of interacting spin-$1/2$ particles, or a chain of hardcore bosons. The parameter $J$ characterizes the $Z$-coupling strength between first neighbors, while $h_z$ and $h_x$ capture the strength of additional magnetic fields respectively in the $z$ and $x$ direction. When $h_z=0$, this model is known as the integrable transverse-field Ising model, which is mappable to free fermions. In order to break all the symmetries of the model, we introduce the reflection symmetry breaking magnetic fields, with strengths $g_0$ and $g_l$ at sites $0$ and $L$, respectively.}

{In the regime of parameters chosen above for the set $J$, $h_z$, and $h_x$, the MFIM is chaotic and satisfies conventional ETH for local observables~\cite{master_equations_ETH_Donovan_Mitchison}. Moreover, it can be shown that its eigenstates display a volume law entanglement similar to the behavior presented by ensembles of uniformly distributed states (see Appendix~\ref{app:additional_numerics} and~\cite{chaos_distributions_entanglement_khemani}), therefore it is suitable to validate our analytical results. }

As an initial operator, we consider the normalized $\sigma_x$ operator at the center of the chain. Specifically, we choose it on site $L/2$ for even $L$ and $(L-1)/2$ for   odd $L$. This operator is local, therefore satisfies ETH, has zero trace, and is normalized to $1$ according to the inner product in Equation~\eqref{eqn:innerproduct}. {Figure~\ref{fig:diagonal_terms_ETH} presents the behavior of the diagonal elements of the chosen operator in the eigenbasis of the Hamiltonian. We can observe a generic decay of fluctuations for increasing system dimension.} %, saturating at values of order $\text{O}(1)$.}

\begin{figure}[t]
    \centering
    \includegraphics[width=\columnwidth]{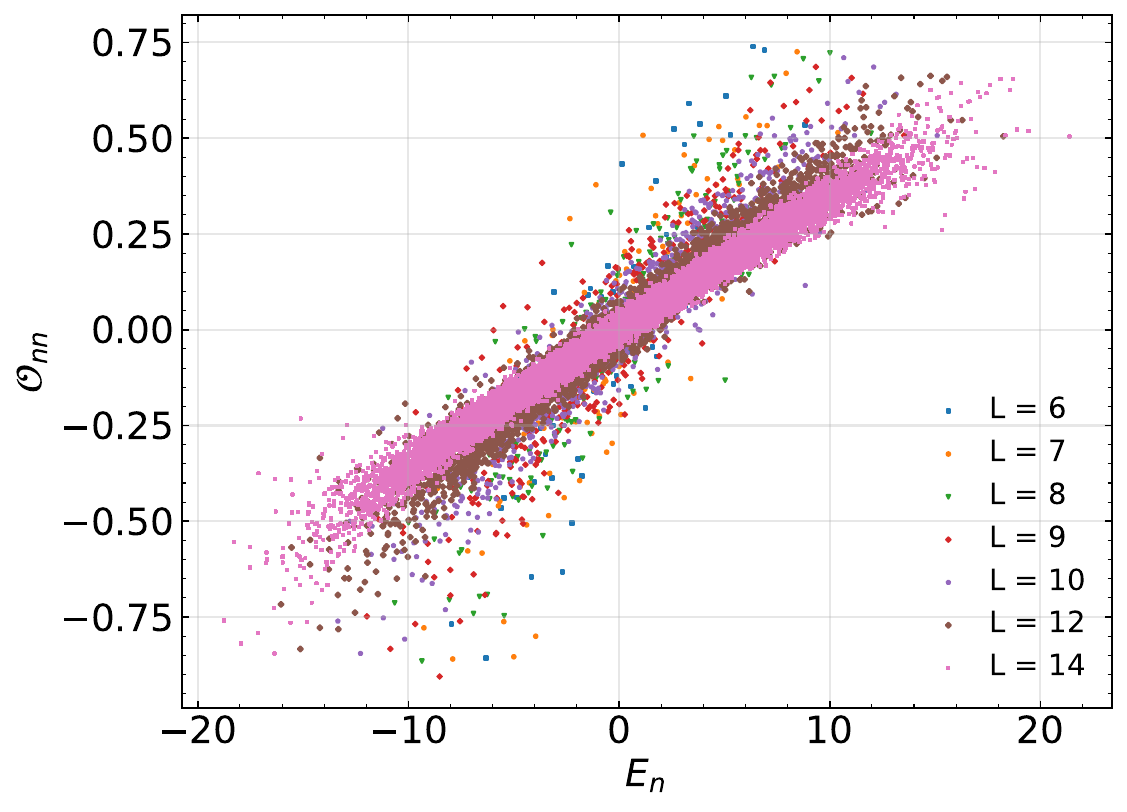}
    \caption{Diagonal matrix elements of the operator $\sigma_x$ at the center of the chain (at $L/2$) in the energy eigenbasis of the maximally chaotic mixed field Ising model~\eqref{eq:mixed_field_Ising} as a function of the energy for different Hilbert Space dimensions. Different symbols and colors represent different dimensions, from $L=6$ to $L=14$.}
    \label{fig:diagonal_terms_ETH}
\end{figure}

Throughout this section, we are going to compare numerical simulations of Equation~\eqref{eq:final_purity_eigenstates}, where eigenstates are computed through exact diagonalization, with the corresponding formula after replacing the eigenstates with Haar-random states, Equation~$\eqref{eq:final_purity_unitaries}$, which is computed fully analytically using Weingarten calculus (we refer the readers to Appendix~\ref{app:random_tensor_networks} and Equation~\eqref{eq:app_purity_weightedum} therein for further details). We point out that we are computing Equation~\eqref{eq:final_purity_unitaries} exactly, while the formulas presented in Section~\ref{sec:final:results} are an approximation thereof. {As we employ exact diagonalization, we cannot access large system sizes beyond $L=15$, and our results cannot be directly extrapolated to the thermodynamic limit. Nevertheless, the system sizes considered are sufficient to gather reliable spectral statistics and check the validity of our hypotheses.}

\subsection{Full LOE for the total Hilbert space}

\begin{figure}[t]
    \centering
    \includegraphics[width=\columnwidth]{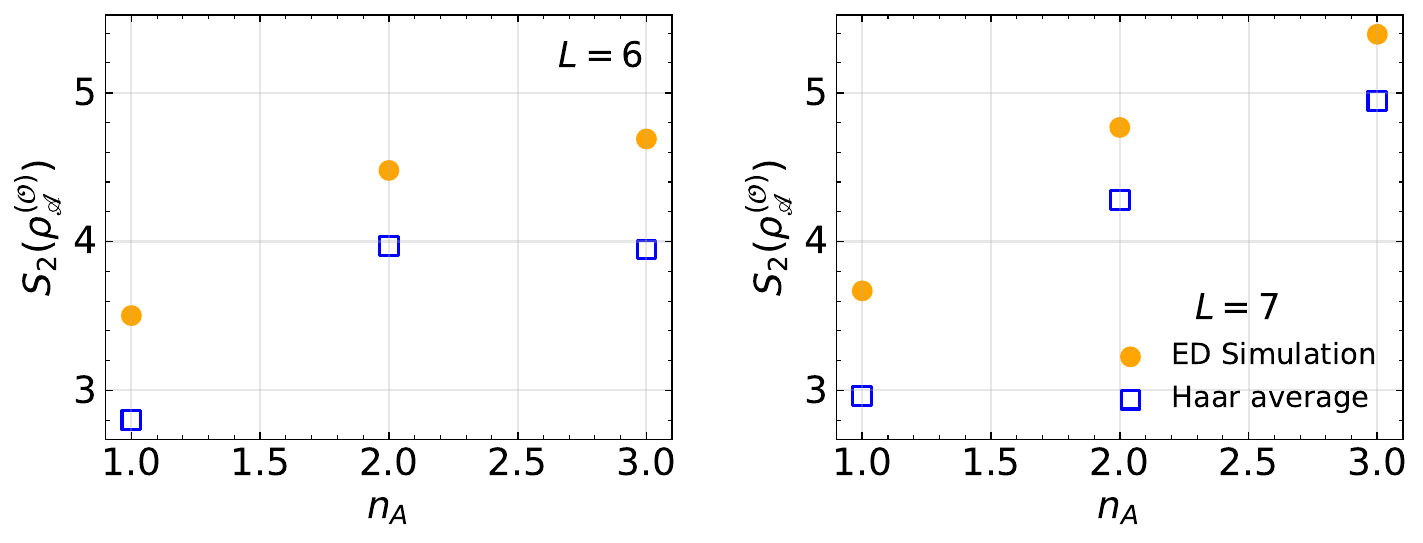}
    \caption{Late-time LOE of the full Hilbert space as a function of the size of subsystem A for $L = 6$ (left) and $L=7$ (right), computed through both numerical exact diagonalization (orange full circles) and analytical Haar average (blue empty squares).}
    \label{fig:6_7_complete_LOE}
\end{figure}

\begin{figure}[t]
    \centering
    \includegraphics[width=\columnwidth]{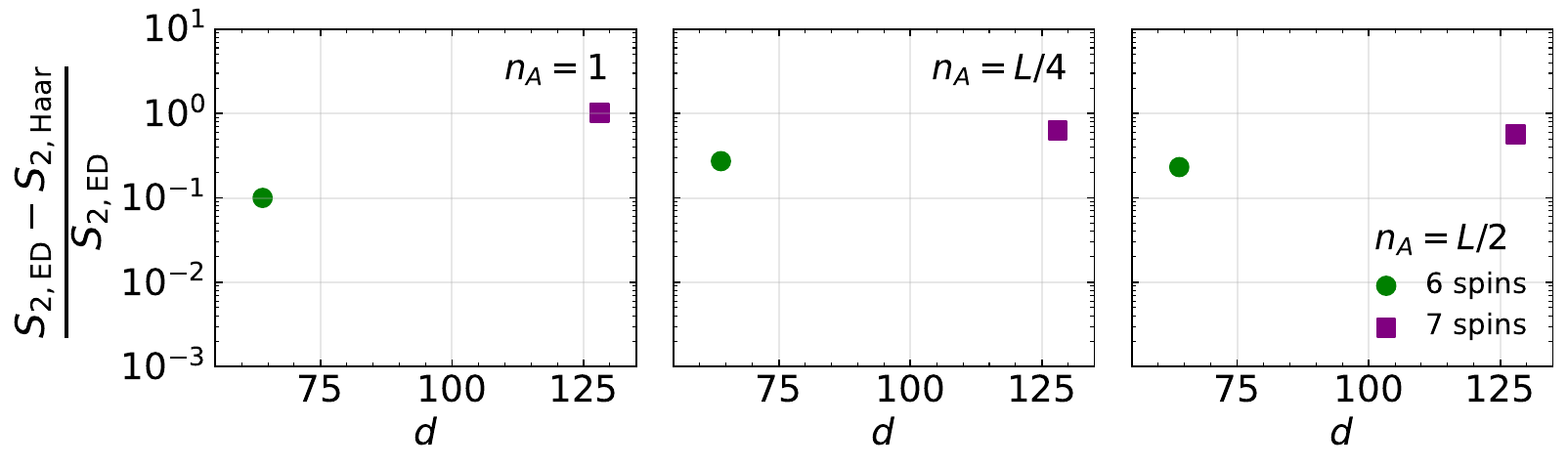}
    \caption{Error function comparing the simulations based on exact diagonalization with Haar random averaged values of late-time LOE calculation after time average for the complete spectrum of the cases $L=6$ (green circles) and $L=7$ (purple squares). Each panel shows the values at different bipartition sizes, including system A with $1$, $L/4$ and $L/2$ spins.}
    \label{fig:6_7_error_completeLOE}
\end{figure}

We first consider the late-time LOE computed on the full Hilbert space of the system. In this scenario, we do not expect our third assumption in Section~\ref{sec:hypothesis} to hold exactly. 

The simulations were carried out for small system sizes ($L=6$ and $L=7$) due to the high computational cost of computing the inter-dependent summations in Equation~\eqref{eq:final_purity_eigenstates}, in total running over $d^8$ indexes, which make this computation unfeasible for higher dimensions with standard algorithms.

A comparison between numerically exact diagonalization (ED) and Haar random average is shown in Figures~\ref{fig:6_7_complete_LOE} and~\ref{fig:6_7_error_completeLOE}. We observe that both curves display a similar scaling, but they do not exactly match, not even for small subsystems sizes. The relative errors between them is of the order of $10^{-1}$ or $10^0$. 

This result is expected for continuous dynamics even in the absence of additional symmetries, due to the energy conservation constraint in the sampling of the energy eigenstates. 

\subsection{Restriction to an energy window}

\begin{figure}[t]
    \centering
    \includegraphics[width=\columnwidth]{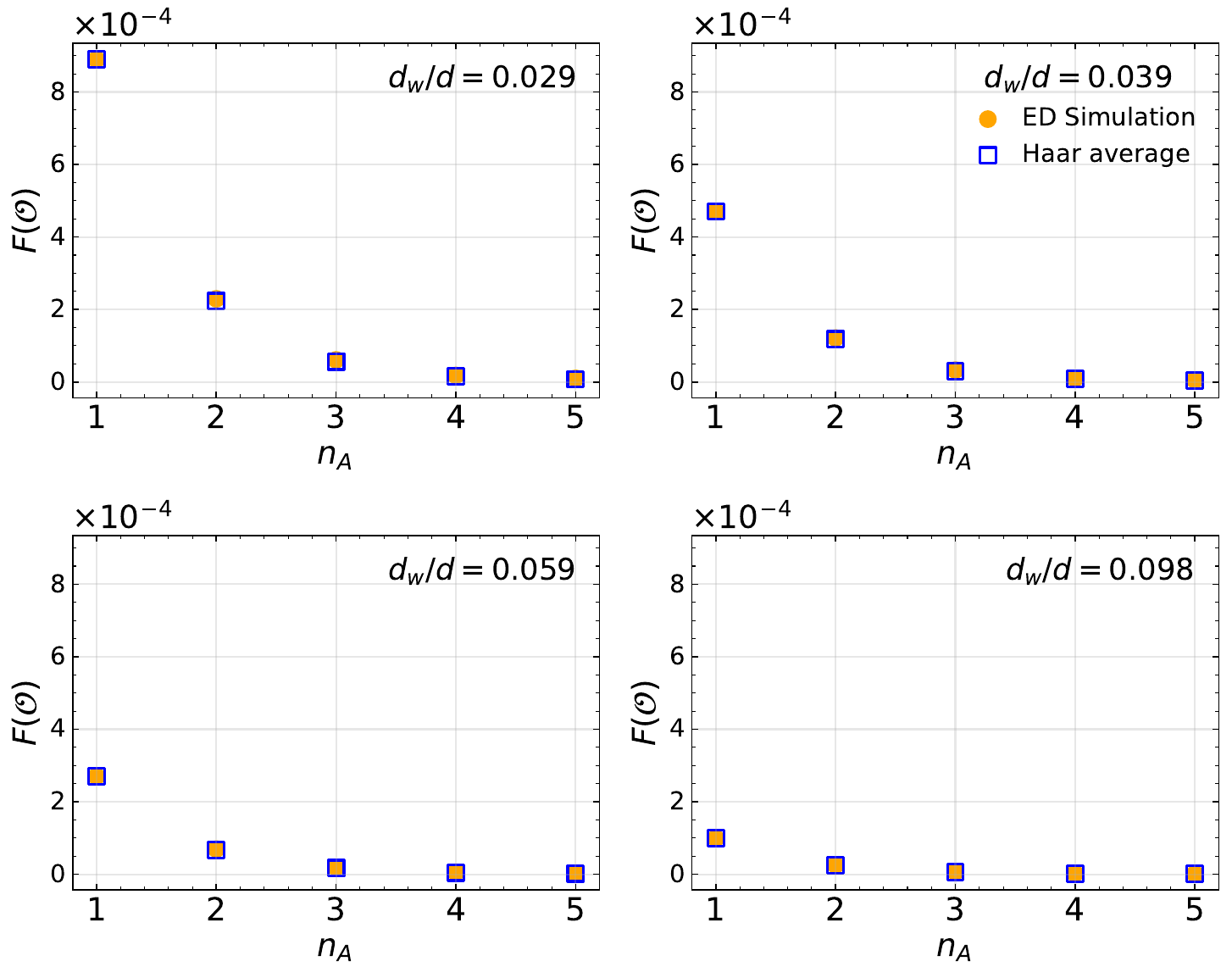}
    \caption{$F(\mathcal{O})$ (Equation~\eqref{eq:F_explicit_form})  as a function of the size of subsystem A, computed through both numerical exact diagonalization (orange full circles) and analytical Haar average (blue empty squares). We have set $L=10$, and each panel corresponds to a different size of the energy shell.}
    \label{fig:diag_10_window_dependent}
\end{figure}

\begin{figure}[t]
    \centering
    \includegraphics[width=\columnwidth]{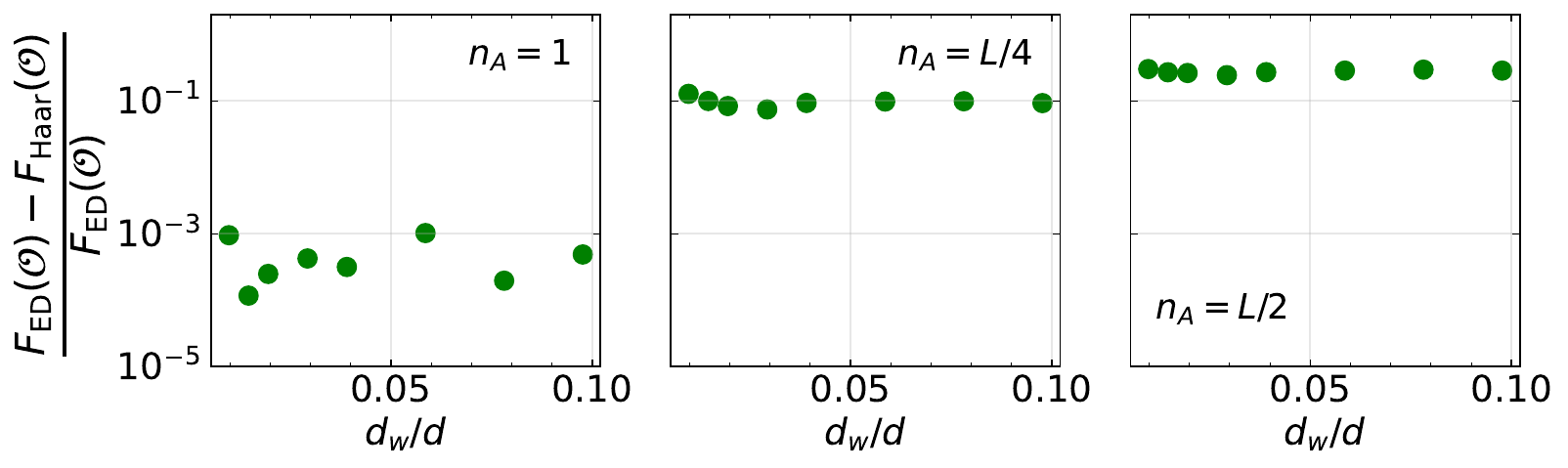}
    \caption{Relative error between the simulations based on ED and Haar random analytics for the term $F(\mathcal{O})$ (Equation~\eqref{eq:F_explicit_form}) as a function of the rescaled dimension of the energy shell $\mathrm{d}_w/\mathrm{d}$, with $L=10$. Each panel corresponds to a different bipartition size. From left to right: $n_A=1$, $L/4$ and $L/2$.}
    \label{fig:error_diag_10_window_dependent}
\end{figure}

\begin{figure}[t]
    \centering
    \includegraphics[width=\columnwidth]{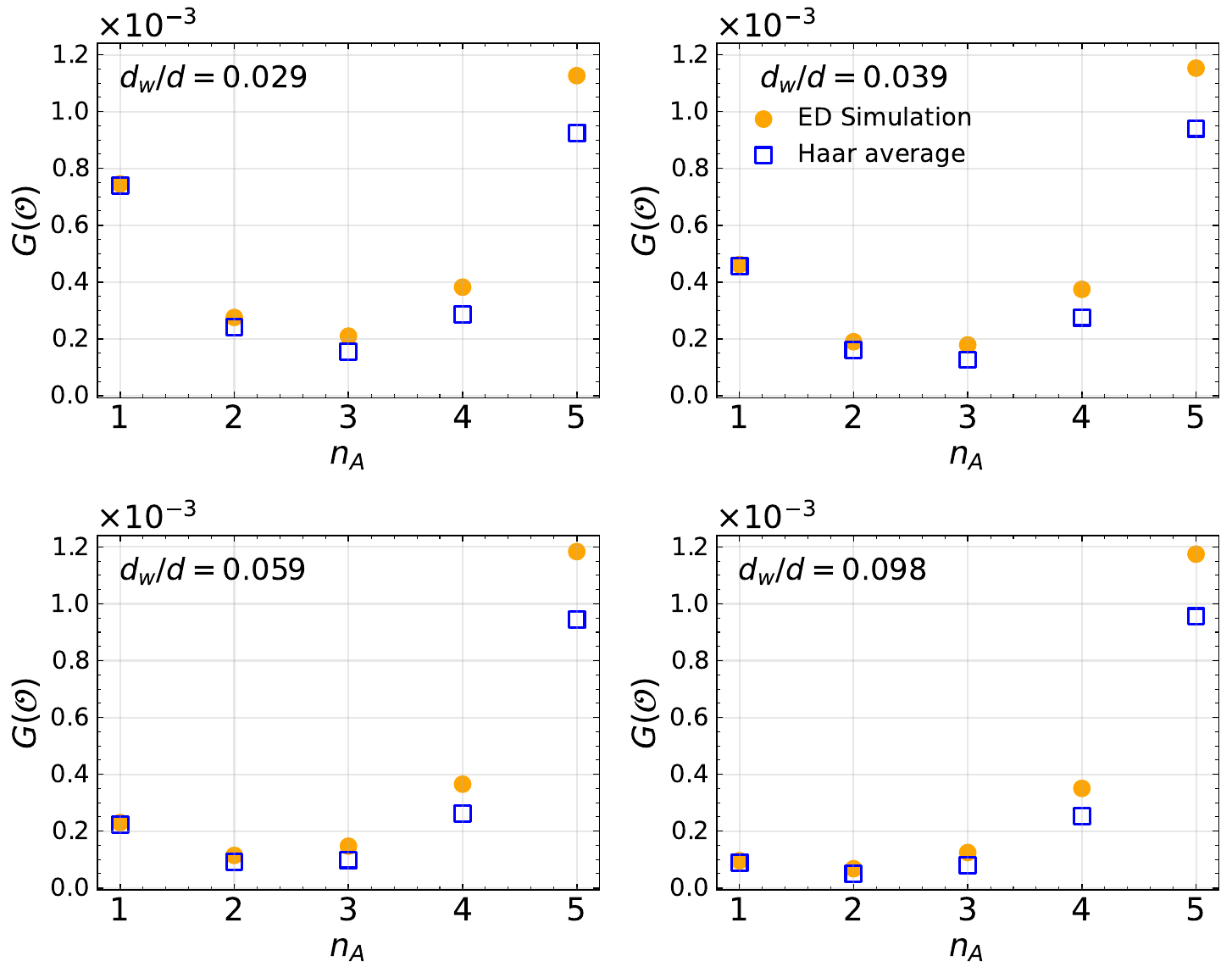}
    \caption{$G(\mathcal{O})$ (Equation~\eqref{eq:G_explicit_form}) as a function of the size of subsystem A, computed through both numerical exact diagonalization (orange full circles) and analytical Haar average (blue empty squares). We have set $L=10$, and each panel corresponds to a different size of the energy shell.}
    \label{fig:offdiag_10_window_dependent}
\end{figure}

\begin{figure}[t]
    \centering
    \includegraphics[width=\columnwidth]{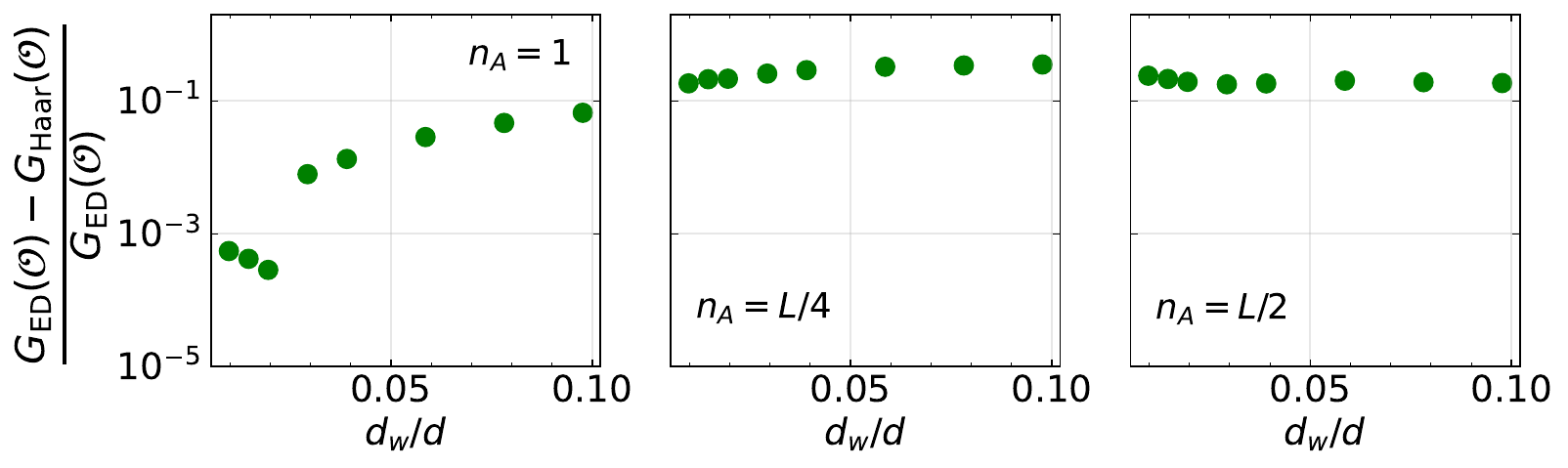}
    \caption{Relative error between the simulations based on ED and Haar random analytics for the term $G(\mathcal{O})$ (Equation~\eqref{eq:G_explicit_form}) as a function of the rescaled dimension of the energy shell $\mathrm{d}_w/\mathrm{d}$, with $L=10$. Each panel corresponds to a different bipartition size. From left to right: $n_A=1$, $L/4$ and $L/2$.}
    \label{fig:error_offdiag_10_window_dependent}
\end{figure}

\begin{figure}[t]
    \centering
    \includegraphics[width=\columnwidth]{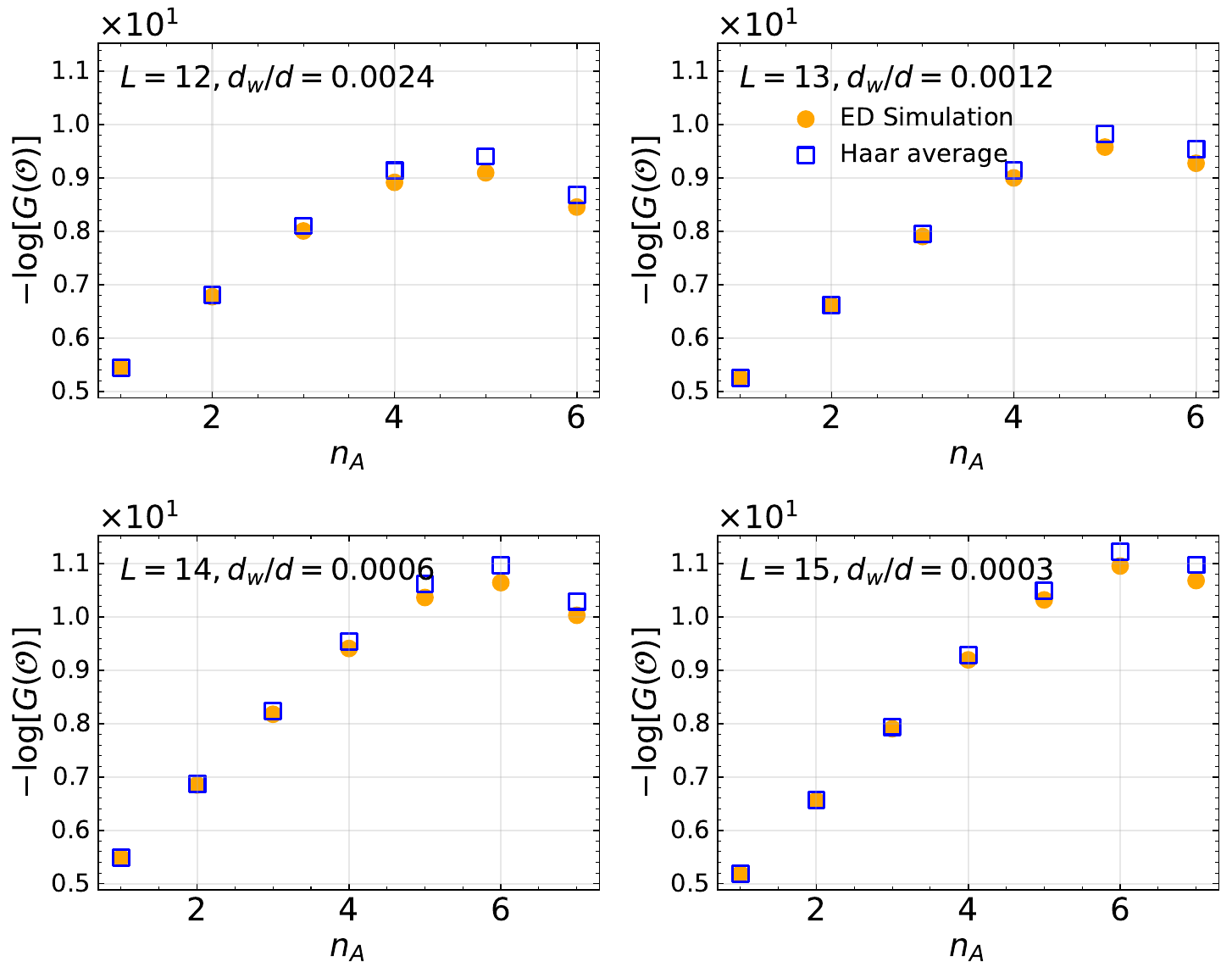}
    \caption{$-\ln G(\mathcal{O})$ (Equation~\eqref{eq:G_explicit_form}) as a function of the size of subsystem A, computed through both numerical exact diagonalization (orange full circles) and analytical Haar average (blue empty squares). The window dimension is set to $\mathrm{d}_w=10$, and each panel corresponds to a different system size $L$.}
    \label{fig:log_offdiag_different_n_spins}
\end{figure}

As computing the late-time LOE in the full Hilbert space is computationally prohibitive for larger systems, we perform numerical studies in which the LOE is restricted to an energy window of the Hilbert space. We point out that, by doing so, we are effectively projecting the initial operator $\mathcal{O}$ over the subspace of eigenstates within the chosen energy window. As a consequence, $\mathcal{O}$ loses its locality property, and this is why the LOE in the full Hilbert space is typically studied. We anyway investigate this quantity to show that our third assumption in Section~\ref{sec:hypothesis} holds in a small energy window.

We consider the case $L=10$, together with selected results for other numbers of spins. Further results for $12$, $13$, $14$, and $15$ spins are presented in Appendix~\ref{app:additional_numerics}. Within the plots, when a fixed system size is considered, we vary energy shell dimension $\mathrm{d}_w$ by changing the window size $\Delta E$. The different plots are presented in terms of the rescaled size of the shell with respect to the total dimension, $\mathrm{d}_w/d$. In all cases the window is at the center of the spectrum, i.e., given $E_0$ at the center of the spectrum and a window with size $\Delta E$, we assume that the window is defined by $\delta E= [E_0 + \Delta E/2, E_0 - \Delta E/2]$.

Given the very different nature of diagonal and off-diagonal terms of the operator in the energy basis, we first look at the terms of Equation~\eqref{eq:final_purity_eigenstates} where only diagonal terms of $\mathcal{O}$ appear or only off-diagonal terms appear, before studying the general behavior of the complete LOE. For simplicity, we define these two terms as
\begin{equation}\label{eq:F_explicit_form}
\begin{split}
    F(\mathcal{O})\equiv\sum_{a, b, c, d=1}^{\mathrm{d}}&\mathcal{O}_{aa}\mathcal{O}^*_{bb}\mathcal{O}_{cc}\mathcal{O}^*_{dd}\\
    &\times E^a_{j\alpha}E^{a*}_{j'\alpha'}E^{b*}_{k\alpha}E^{b}_{k'\alpha'}E^c_{k\beta}E^{c*}_{k'\beta'}E^{d*}_{j\beta}E^{d}_{j'\beta'}
\end{split}
\end{equation}
and
\begin{equation}\label{eq:G_explicit_form}
\begin{split}
    G(\mathcal{O})\equiv \sum_{a \neq b = 1}^{\mathrm{d}}&\sum_{c \neq d=1}^{\mathrm{d}}|\mathcal{O}_{ab}|^2|\mathcal{O}_{cd}|^2\\
    &\times E^a_{j\alpha}E^{b*}_{j'\alpha'}E^{c*}_{k\alpha}E^{d}_{k'\alpha'}E^{c}_{k\beta}E^{d*}_{k'\beta'}E^{a*}_{j\beta}E^{b}_{j'\beta'}.
\end{split}
\end{equation}

Figures~\ref{fig:diag_10_window_dependent} and~\ref{fig:error_diag_10_window_dependent} depict the value of the term $F(\mathcal{O})$ in Equation~\eqref{eq:F_explicit_form}, obtained numerically with ED and analytically with Haar averaging, and the normalized error when comparing the two values, as a function of the window dimension $\mathrm{d}_W$. In Figure~\ref{fig:diag_10_window_dependent},  we observe good agreement between the two quantities at all window sizes considered. Regarding the relative error shown in Figure~\ref{fig:error_diag_10_window_dependent}, we notice that for small subsystem sizes such as $n_A=1$  ($n_A=2$) the error is very small, with values of the order of $10^{-4}$ ($10^{-1}$). Moreover, we observe that for a fixed $n_A$ these values do not have a strong dependence on the window size. Finally, for $n_A=L/2=5$, we see that the relative error is roughly equal to $5\cross 10^{-1}$, confirming that the approximation based on Haar random states gets worse for a larger subsystem size. This can also be observed in our simulations regarding eigenstate entanglement (Appendix~\ref{app:additional_numerics}) and works related to the topic~\cite{volume_law_eigenstates}. A similar behavior is observed also for larger system sizes (see Appendix~\ref{app:additional_numerics}).

 %We attribute it to the fact that the expectation that the eigenstates should satisfy the behavior of RMT usually is concerned with the elements of local observables on this basis. We impose this constraint by taking ETH to be valid. However, we introduced the additional hypothesis that the components of the eigenstates on a local basis are given by RMT (the Haar integrations performed with random tensor networks, Appendix~\ref{app:random_tensor_networks}).
%When the locality of the subregion is removed, by looking at subsystem sizes that are half of the whole system, the state itself should not reproduce a ``thermal'' or random state as precisely, and the approximation should worsen. This can be interpreted both from the perspective of part of the system acting as a thermal bath (closed system thermalization) or by considering that the study of these components of the eigenstates on a bipartite basis to be given in terms of an observable, the partial swap operator. This first perspective is precisely one of the ways in which ETH was introduced in Srednicki's seminal papers~\cite{eth_Srednicki1994, eth_Srednicki1999}. The second perspective is discussed in greater detail in Appendix~\ref{app:volume_law_states} and in Ref.~\cite{k-eth_kaneko-sagawa}, where entanglement Rényi entropies are defined as observables.

We remark that in the final calculation for LOE, what we have to take into account is $-\ln F(\mathcal{O})$, $-\ln G(\mathcal{O})$. In Appendix~\ref{app:additional_numerics} we show this quantities displays a nice and smooth volume law.

Next, we focus on the term $G(\mathcal{O})$ (Equation~\eqref{eq:G_explicit_form}) that deals with the non-diagonal contributions. The results are shown in Figures~\ref{fig:offdiag_10_window_dependent} and~\ref{fig:error_offdiag_10_window_dependent}. Instead of the exponential decay observed for $F(\mathcal{O})$ that implies a volume law for $-\ln F(\mathcal{O})$, $-\ln G(\mathcal{O})$ in Figure~\ref{fig:offdiag_10_window_dependent} initially decays as a function of the subsystem size and then becomes an increasing function, both for the case of ED and for Haar random states. The relative error in Figure~\ref{fig:error_offdiag_10_window_dependent} also shows that the Haar-random approximation does not work as effectively for this term, as it can be around $10^{-1}$ even for smaller subsystem sizes and increased window dimension.

The behavior of $-\ln G(\mathcal{O})$ is shown in Figure~\ref{fig:log_offdiag_different_n_spins} (see also Appendix~\ref{app:additional_numerics} for further results). We observe that this term presents a behavior similar to a volume law at smaller $n_A$ that dips down at larger $n_A$. On the other hand, this effect at large $n_A$ is alleviated by increasing the total number of spins, and the behavior approaches a regular volume law, as illustrated in Figure~\ref{fig:log_offdiag_different_n_spins}. We can then argue that the results imply a stronger dependence of the behavior of the off-diagonal terms at $n_A\approx L/2$ on the system sizes.

Finally, we focus on the complete LOE that takes into account both contributions from $F(\mathcal{O})$ and $G(\mathcal{O})$. The results are depicted in Figure~\ref{fig:complete_10_window_dependent}. We observe the same behavior as for $G(\mathcal{O})$, which is a smooth volume law for small $n_A$ that is again followed by a dip. The total relative error between ED and Haar-random states (Figure~\ref{fig:error_complete_10_window_dependent}) follows a similar behavior as for the $F$ and $G$ terms, being reasonably small for smaller dimensions (around $10^{-3}$) and increasing as a function of $n_A$. 
For $10$ spins and $n_A=L/4$, it reaches its maximal value of around $5\cross 10^{-1}$ given the energy windows considered in this section. These error values around $10^{-1}$ are only observed for larger values $n_A$, around $L/4$, still.

\begin{figure}[t]
    \centering
    \includegraphics[width=\columnwidth]{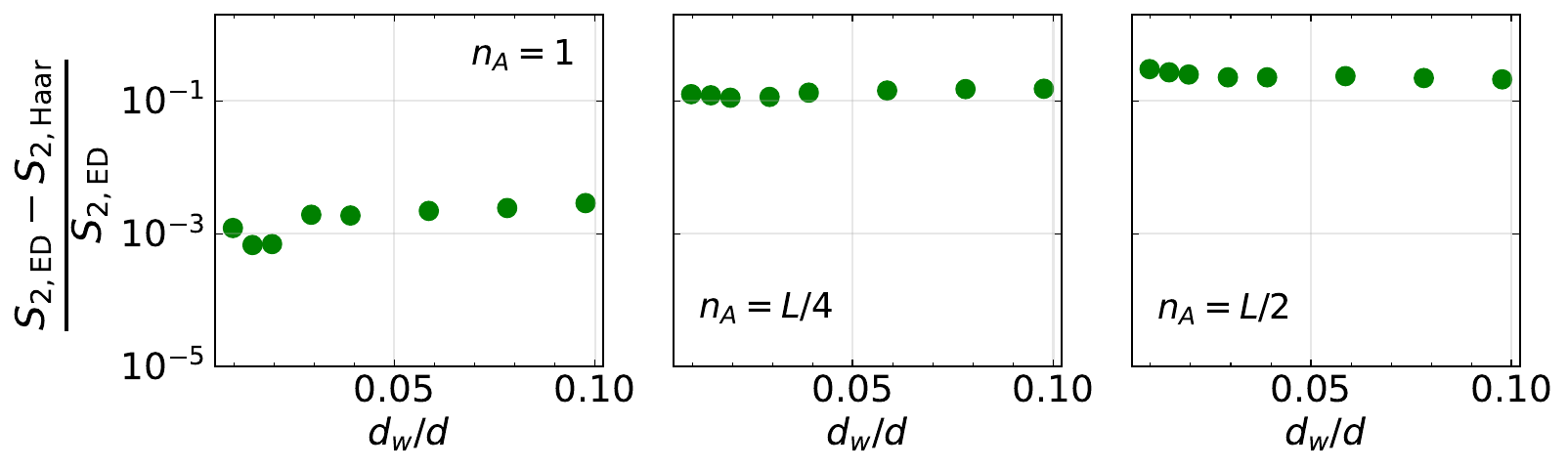}
    \caption{Relative error between the simulations based on ED and Haar random analytics for the total LOE as a function of the rescaled dimension of the energy shell $\mathrm{d}_w/\mathrm{d}$, with $L=10$. Each panel corresponds to a different bipartition size. From left to right: $n_A=1$, $L/4$ and $L/2$.}
    \label{fig:error_complete_10_window_dependent}
\end{figure}

\begin{figure}[t]
    \centering
    \includegraphics[width=\columnwidth]{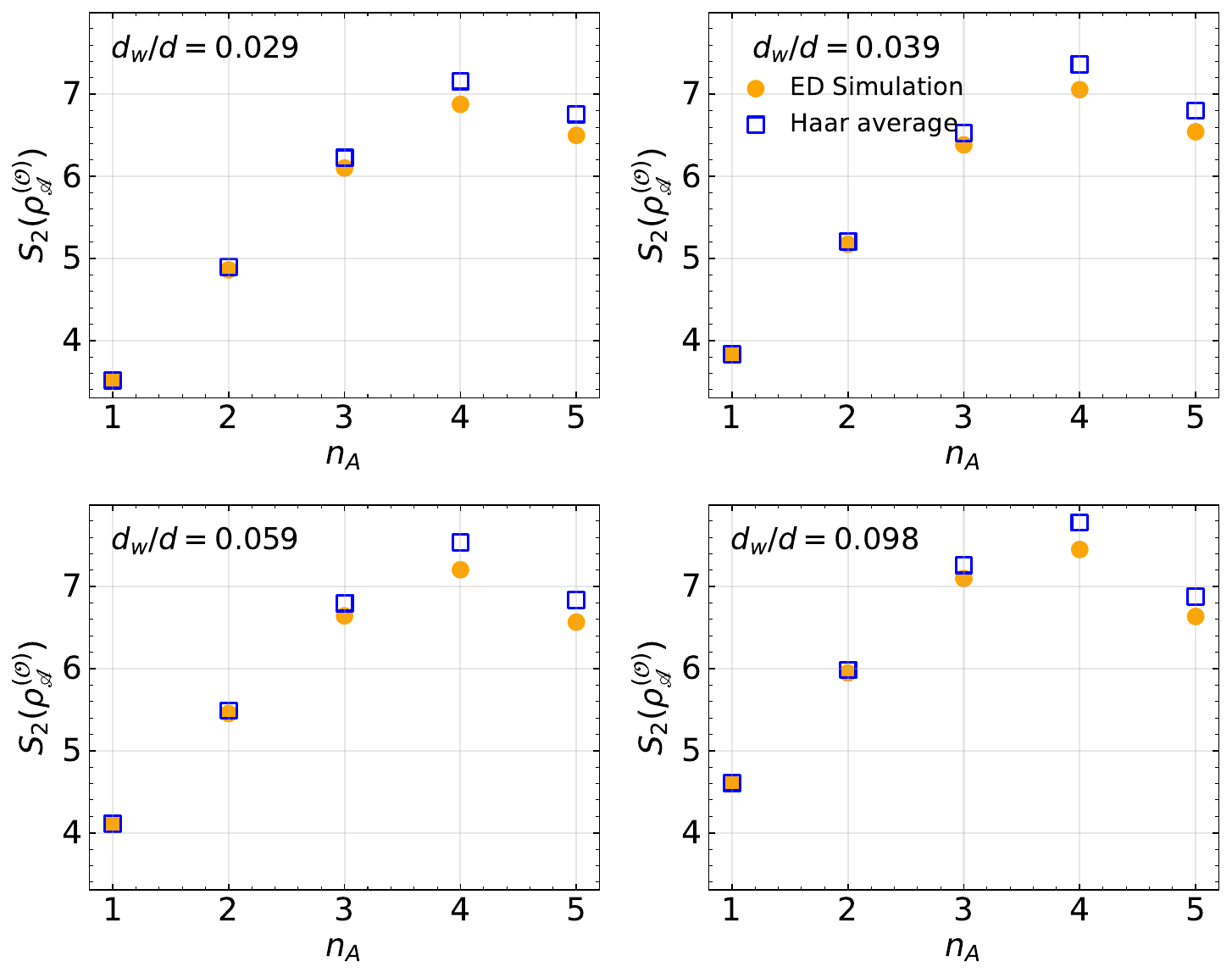}
    \caption{Total LOE as a function of the size of subsystem A, computed through both numerical exact diagonalization (orange full circles) and analytical Haar average (blue empty squares). We have set $L=10$, and each panel corresponds to a different size of the energy shell.}
    \label{fig:complete_10_window_dependent}
\end{figure}

\section{\label{sec:conclusion}Conclusions}
In this paper, we have studied the analytical foundations of late-time local operator entanglement in chaotic systems. By using a Liouvillian basis of the Krylov subspace spanned by the evolution, introduced in Section~\ref{sec:liouville_basis}, and assuming the 4 non-resonance condition for the eigenenergies of the chaotic Hamiltonian, we have derived an exact formula for the long time average of LOE as a function of the Hamiltonian eigenstates and matrix elements of the initial local operator $\mathcal{O}$, expressed in Equation~\eqref{eq:final_purity_eigenstates}.

Next, we have  applied two fundamental assumptions: ETH for the matrix elements of the initial local operator, and the replacement of system eigenstates with random states for the sake of calculating LOE. Then, we were able to transform Equation~\eqref{eq:final_purity_eigenstates} into Equation~\eqref{eq:final_purity_unitaries}, which was a function of the matrix elements of the local operator and Haar averages of correlated elements of random vectors, which could be computed exactly and analytically through Weingarten calculus (see the details in Appendix~\ref{app:random_tensor_networks}). Using this analytical result and the ETH assumption, we have then derived an approximated expression for late-time LOE both in the case of small subsystem size and for LOE computed through a bipartition of the system with equal number of sites. These formulas are expressed in Equations~\eqref{eq:dA_O1_big_windows} and~\eqref{eq:dA_big_big_windows}, and display a volume law plus an additional term of the same order.  

A more detailed explanation of the assumptions we made and a discussion on their validity can be found in Section~\ref{sec:hypothesis}. The 4 non-resonance condition and the ETH assumption are well established and corroborated in a plethora of numerical studies. Replacing Hamiltonian eigenstates with Haar random states in the final expression for late-time LOE is instead fully justified only when restricting ourselves in a small energy window. We checked this through numerical simulations of the Mixed Field Ising Model in Section~\ref{sec:numerics}: the equivalence between Equation~\eqref{eq:final_purity_eigenstates} and~\eqref{eq:final_purity_unitaries} emerges for small energy windows, while these formulas do not exactly match but display the same behavior as a function of the subsystem dimension on the full Hilbert space. 

Although our third assumption cannot be rigorously justified over the full Hilbert space, it is necessary to obtain a closed analytical expression for the late-time LOE. We believe it still provides useful insight into the origin of the volume-law behavior of late-time LOE, starting from the structure of the system eigenstates and the decomposition of LOE in the Liouvillian basis. Future work may consider more refined assumptions on the distribution of eigenstates, but obtaining analytical results under them is likely to be very challenging.

\begin{acknowledgments}
The authors wish to thank the Finnish Computing Competence Infrastructure (FCCI) for supporting this project with computational and data storage resources. G.I.C. acknowledges the financial support of the Finnish Ministry of Education and Culture. G.I.C. and M.C. acknowledge the financial support of the Research Council of Finland through the Finnish Quantum Flagship project (358878, UH). J.G. thanks the Royal Society and Research Ireland for financial support. 
\end{acknowledgments}

\bibliography{refs}
\bibliographystyle{apsrev4-2}

\onecolumngrid
\appendix

%=======================================================================================================================================

\section{\label{app:volume_law_states}Volume law for state space}

The purity equation~\eqref{eq:ergodic_avg_open_terms} was obtained by considering just the $k=2$ non-resonance hypothesis for the eigenvalues of the generator of the time evolution operator, which is valid both for operator or state space. For instance, if, instead of the Liouvillian, we considered the Hamiltonian as the generator, we would just need to require a $k=2$ non-resonance for the energies. In this sense, we can affirm that this form is also valid for states in the following way
\begin{equation}\label{eq:state_ergodic_avg}
    \begin{split}
        \overline{\Tr[\big(\rho_{{A}}\big)^2]}&=\sum_{m=1}^{\mathrm{d}}|\braket{\psi}{E_m}|^4\Tr_{A}(|E^{(m)}|^2|E^{(m)}|^2)\\
        &+\sum_{m\neq p=1}^{\mathrm{d}}|\braket{\psi}{E_m}|^2|\braket{\psi}{E_p}|^2\Tr_{{A}}(E^{(m)}E^{(m)\dagger}E^{(p)}E^{(p)\dagger})\\
        &+\sum_{m\neq p=1}^{\mathrm{d}}|\braket{\psi}{E_m}|^2|\braket{\psi}{E_p}|^2\Tr_{{B}}(E^{(m)\dagger}E^{(m)}E^{(p)\dagger}E^{(p)}),
    \end{split}
\end{equation}
being $d$ the dimension of the  Hilbert space and considering a bipartition $A-B$, with subsystem dimensions $\mathrm{d}_A$ and $\mathrm{d}_B$. The initial state is simply given by $\ket{\psi}$. Here, the $E^{(m)}$ matrices are defined in the same way as $\omega^{(m)}$. If we take into account an initial state $\ket{\psi}$ characterized by uniform probabilities in the energy eigenbasis (sometimes refered as an infinite temperature state \cite{silvia_pappalardi_notes, designs_free_probability_pappalardi}) for simplicity, it is reasonable to assume the full delocalization of the diagonal ensemble, implying weights given by $|\braket{\psi}{E_m}|^2=1/\mathrm{d},\,\forall m$. This statement is strictly different from the operator case, as here we only have one order of magnitude, given by the dimension of the window. In the operator case, we have to take into account the two different orders of magnitude (diagonal terms and off-diagonal terms) coming from the ETH. We then get
\begin{equation}
    \begin{split}
        \overline{\Tr[\big(\rho_{{A}}\big)^2]}&=\frac{1}{\mathrm{d}^2}\sum_{m=1}^{\mathrm{d}}\Tr_{A}(|E^{(m)}|^2|E^{(m)}|^2)\\
        &+\frac{1}{\mathrm{d}^2}\sum_{m\neq p=1}^{\mathrm{d}}\Tr_{{A}}(E^{(m)}E^{(m)\dagger}E^{(p)}E^{(p)\dagger})\\
        &+\frac{1}{\mathrm{d}^2}\sum_{m\neq p=1}^{\mathrm{d}}\Tr_{{B}}(E^{(m)\dagger}E^{(m)}E^{(p)\dagger}E^{(p)}).
    \end{split}
\end{equation}
By comparing these traces of the eigenvector matrices decomposed in the bipartition with the partial swapping operator $S_X$ defined on~\cite{k-eth_kaneko-sagawa}, we can identify that these are basically the matrix elements of $S_X$ (here denoted as $S_A$ because of the naming of our bipartitions) in the energy eigenbasis,
\begin{equation}\label{eq:purity_function_partialswap}
    \overline{\Tr[\big(\rho_{{A}}\big)^2]}=\frac{1}{\mathrm{d}^2}\Bigg[\sum_{m=1}^{\mathrm{d}}\bra{E_m E_m}S_A\ket{E_m E_m}\nonumber\\
    +\sum_{m\neq p=1}^{\mathrm{d}}\bra{E_m E_p}S_A\ket{E_m E_p}\nonumber\\
    +\sum_{m\neq p=1}^{\mathrm{d}}\bra{E_m E_p}S_A\ket{E_p E_m}\Bigg],
\end{equation}
again we use the simpler nomenclature DIA, SEMI and PERM. These can be averaged over uniformly distributed eigenstates over the Hilbert space, leading to weights depending on $\mathrm{d}$ and $\mathrm{d}_A$ of the form
\begin{equation}
    \begin{split}
        \text{DIA}\equiv\bra{E_m E_m}S_A\ket{E_m E_m}=\frac{\mathrm{d}/\mathrm{d}_A + \mathrm{d}_A}{\mathrm{d} + 1}\\
        \text{SEMI}\equiv\bra{E_m E_p}S_A\ket{E_m E_p}=\frac{\mathrm{d}^2/\mathrm{d}_A - \mathrm{d}_A}{\mathrm{d}^2 - 1}\\
        \text{PERM}\equiv\bra{E_m E_p}S_A\ket{E_p E_m}=\frac{\mathrm{d} \mathrm{d}_A - \mathrm{d}/\mathrm{d}_A}{\mathrm{d}^2 - 1}.
    \end{split}
\end{equation}
In this simple example of entanglement of infinite temperature time evolved states, we can obtain a closed analytical formula without any further approximation regarding the dimensions of $\mathrm{d}_A$ and $\mathrm{d}$. By substituting the averaged values of the elements of $S_A$ within the purity, we obtain the entanglement
\begin{equation}
    \Big\langle\overline{S_2(\rho_A(t))}\Big\rangle=-\ln\left(\Big\langle\overline{\Tr[\big(\rho_{{A}}\big)^2]}\Big\rangle\right)\simeq n_A\ln(2)-\ln(\frac{\mathrm{d}+\mathrm{d}_A^2}{\mathrm{d}+1}),
\end{equation}
which does not depend on the window and is exactly the equation obtained for the entanglement according to the $2-$Rényi entropy of entanglement using only the completely diagonal elements of $S_A$ (DIA) in Ref.~\cite{k-eth_kaneko-sagawa}. We can notice that, even though the terms SEMI and PERM appear within the late-time purity, they compensate each other and give a net behavior that equals a traditional Page volume law. This result comes from the fact that infinite temperature states always have a uniform probability distribution over the eigenstates, which is not what would be expected for the probabilities of operators, $|\dbraket{\mathcal{O}}{\omega_m}|^2$ due to ETH, generating the unbalances appearing in the main text. We can see a similar result by taking into account infinite temperature states which can select between the DIA, SEMI and PERM weights, generating entanglement dependencies that do not follow a volume law at any $\mathrm{d}_A$ as illustrated in Figure~\ref{fig:states_weights}. The behavior displayed without considering the three parts together presents dependencies on the dimension of the window considered.

\begin{figure}[htb]
    \centering
    \includegraphics[width=0.5\columnwidth]{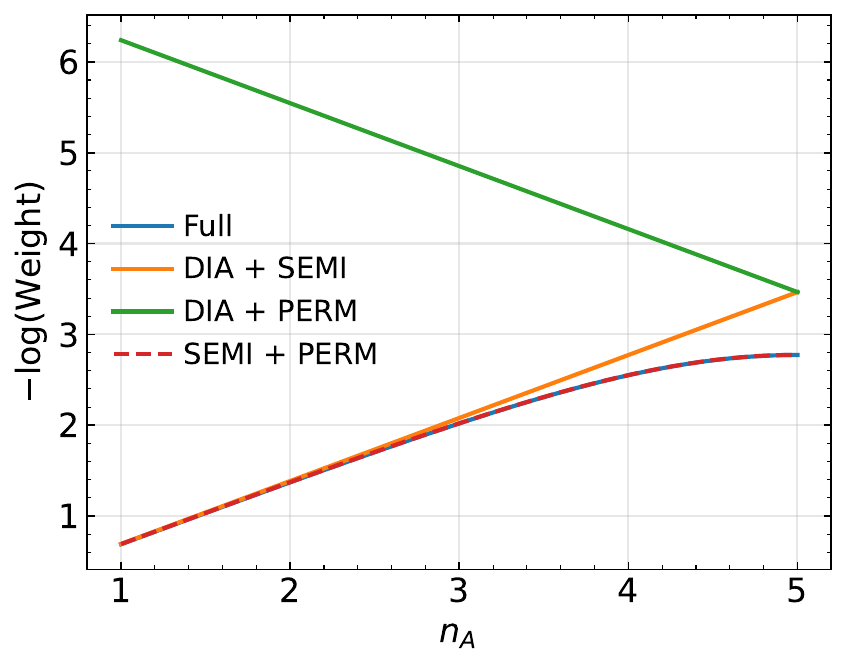}
    \caption{Late-time average entanglement according to Equation~\eqref{eq:purity_function_partialswap}, after averaging over random eigenstates, for $L=10$ spins and energy window dimension $d=40$. The different plots represent different combinations of the weights, which give a different overall behavior as a function of $n_A$. Notice that the result presented in the text is given by the ``Full'' case.}
    \label{fig:states_weights}
\end{figure}

\section{\label{app:physical_intuition_purity}Late-time purity in a mutual information-like form}

We define the diagonal ensemble for the initial operator $\mathcal{O}$ in the Liouvillian basis as
\begin{equation}
    \rho^{(\mathcal{O})}_{\omega}:=\sum_{m=0}^{\mathcal{K}-1}|\dbraket{\mathcal{O}}{\omega_m}|^2\dketbra{\omega_m}{\omega_m},
\end{equation}
Notice that throughout this section we will refer to the partial trace over subsystem $\mathscr{B}$ for the bipartition $\mathcal{H}_\mathscr{A}\otimes \mathcal{H}_\mathscr{B}$ calculated for the diagonal ensemble above, i.e.,
\begin{equation}
    \rho^{(\mathcal{O})}_{\omega, \mathscr{A}}:=\Tr_{\mathscr{B}}(\rho^{(\mathcal{O})}_{\omega})
\end{equation}
Using these definitions, we can identify the terms appearing in Equation~\eqref{eq:ergodic_avg_open_terms}
\begin{equation}\label{eqn:rho_omega_A_squared}
    \begin{split}
        \Tr[\big(\rho_{\omega,\mathscr{A}}^{(\mathcal{O})}\big)^2]&=\sum_{mp=0}^{\mathcal{K}-1}|\dbraket{\mathcal{O}}{\omega_m}|^2|\dbraket{\mathcal{O}}{\omega_p}|^2\dbraket{j,\alpha}{\omega_m}\dbraket{\omega_m}{k,\alpha}\dbraket{k,\beta}{\omega_p}\dbraket{\omega_p}{j,\beta}\\
        &=\sum_{m,p=0}^{\mathcal{K}-1}|\dbraket{\mathcal{O}}{\omega_m}|^2|\dbraket{\mathcal{O}}{\omega_p}|^2\Tr_{\mathscr{A}}(\omega^{(m)}\omega^{(m)\dagger}\omega^{(p)}\omega^{(p)\dagger})\\
        &=\sum_{m=0}^{\mathcal{K}-1}|\dbraket{\mathcal{O}}{\omega_m}|^4\Tr_{\mathscr{A}}(|\omega^{(m)}|^2|\omega^{(m)}|^2)\\
        &+\sum_{m\neq p=0}^{\mathcal{K}-1}|\dbraket{\mathcal{O}}{\omega_m}|^2|\dbraket{\mathcal{O}}{\omega_p}|^2\Tr_{\mathscr{A}}(\omega^{(m)}\omega^{(m)\dagger}\omega^{(p)}\omega^{(p)\dagger}).
    \end{split}
\end{equation}
An analogous relation is valid when tracing out $\mathscr{A}$ instead of $\mathscr{B}$. However, the order of the matrices appearing in the trace in Equation~\eqref{eqn:rho_omega_A_squared} is inverted:
\begin{equation}
    \begin{split}
        \Tr[\big(\rho_{\omega,\mathscr{B}}^{(\mathcal{O})}\big)^2]&=\sum_{m,p=0}^{\mathcal{K}-1}|\dbraket{\mathcal{O}}{\omega_m}|^2|\dbraket{\mathcal{O}}{\omega_p}|^2\Tr_{\mathscr{B}}(\omega^{(m)\dagger}\omega^{(m)}\omega^{(p)\dagger}\omega^{(p)})\\
        &=\sum_{m=0}^{\mathcal{K}-1}|\dbraket{\mathcal{O}}{\omega_m}|^4\Tr_{\mathscr{B}}(|\omega^{(m)}|^2|\omega^{(m)}|^2)\\
        &+\sum_{m\neq p=0}^{\mathcal{K}-1}|\dbraket{\mathcal{O}}{\omega_m}|^2|\dbraket{\mathcal{O}}{\omega_p}|^2\Tr_{\mathscr{B}}(\omega^{(m)\dagger}\omega^{(m)}\omega^{(p)\dagger}\omega^{(p)})
    \end{split}
\end{equation}
To rewrite the purity in a more intuitive form, we define the non-renormalized projected state of the initial density matrix over the subspace related to $\omega_m$, 
\begin{equation}
    \rho_m^{(\mathcal{O})}=\dketbra{\omega_m}{\omega_m}\Big(\dketbra{\mathcal{O}}{\mathcal{O}}\Big)\dketbra{\omega_m}{\omega_m}.
\end{equation}
With this, we have that
\begin{equation}
    \sum_{m=0}^{\mathcal{K}-1}|\dbraket{\mathcal{O}}{\omega_m}|^4\Tr_{\mathscr{B}}(|\omega^{(m)}|^2|\omega^{(m)}|^2)=\sum_{m=0}^{\mathcal{K}-1}\Tr[\big(\rho_{m,\mathscr{B}}^{(\mathcal{O})}\big)^2]
\end{equation}
We then get to the final expression
\begin{align}\label{eq:decomposed_ergodic_average_tracesqr}
    \overline{\Tr[\big(\rho^{(\mathcal{O})}_{\mathscr{A}}\big)^2]}=\Tr[\big(\rho_{\omega,\mathscr{A}}^{(\mathcal{O})}\big)^2]+\Tr[\big(\rho_{\omega,\mathscr{B}}^{(\mathcal{O})}\big)^2]-\sum_{m=0}^{\mathcal{K}-1}\Tr[\big(\rho_{m,\mathscr{B}}^{(\mathcal{O})}\big)^2],
\end{align}
therefore the last term in the RHS is the sum of the purities for subsystem $\mathscr{B}$ coming from the projection of the initial operator on the basis of the Liouvillian.

\section{\label{app:additional_calculations_summations}Obtaining the terms in Equation~\eqref{eq:final_purity_eigenstates}}

The strategy we use to simplify the summations over the eigenfrequencies of the Liouvillian will be the same as for the first term; therefore we again consider separately the diagonal (index $m=0$) and off-diagonal (index $m\neq0$) contributions. Here we present all the summations explictly for clarity. For the second term in the RHS of Equation~\eqref{eq:ergodic_avg_open_terms}, the split between diagonal and off-diagonal terms reads
\begin{equation}
    \begin{split}
        \sum_{m\neq p=0}^{\mathcal{K}-1}|\dbraket{\mathcal{O}}{\omega_m}|^2|\dbraket{\mathcal{O}}{\omega_p}|^2\Tr_{\mathscr{A}}(\omega^m\omega^{m\dagger}\omega^p\omega^{p\dagger})&=2\sum_{p=1}^{\mathcal{K}-1}|\dbraket{\mathcal{O}}{\omega_0}|^2|\dbraket{\mathcal{O}}{\omega_p}|^2\Tr_{\mathscr{A}}(\omega^0\omega^{0\dagger}\omega^p\omega^{p\dagger})\\
        &+\sum_{m\neq p=1}^{\mathcal{K}-1}|\dbraket{\mathcal{O}}{\omega_m}|^2|\dbraket{\mathcal{O}}{\omega_p}|^2\Tr_{\mathscr{A}}(\omega^m\omega^{m\dagger}\omega^p\omega^{p\dagger}),
    \end{split}
\end{equation}
which, after the introduction of the vectorized form of the vectors $\dket{\omega_m}$, becomes
\begin{equation}
    \begin{split}
        &\sum_{m\neq p=0}^{\mathcal{K}-1}|\dbraket{\mathcal{O}}{\omega_m}|^2|\dbraket{\mathcal{O}}{\omega_p}|^2\Tr_{\mathscr{A}}(\omega^m\omega^{m\dagger}\omega^p\omega^{p\dagger})\\
        &=\frac{2}{\mathrm{d}^2}\sum_{p=1}^{\mathcal{K}-1}\sum_{a,c=1}^{\mathrm{d}}\sum_{(ef),(gh)\in\mathcal{I}_p}\sum_{jj'kk'=1}^{\mathrm{d}_A}\sum_{\alpha\alpha'\beta\beta'=1}^{\mathrm{d}_B}\mathcal{O}_{aa}\mathcal{O}^*_{cc}\mathcal{O}_{ef}\mathcal{O}^*_{gh}E^a_{j\alpha}E^{a*}_{j'\alpha'}E^{c*}_{k\alpha}E^{c}_{k'\alpha'}E^{e}_{k\beta}E^{f*}_{k'\beta'}E^{g*}_{j\beta}E^{h}_{j'\beta'}\\
        &+\frac{1}{\mathrm{d}^2}\sum_{m\neq p=1}^{\mathcal{K}-1}\sum_{(ab),(ef)\in\mathcal{I}_m}\sum_{(cd),(gh)\in\mathcal{I}_p}\sum_{jj'kk'=1}^{\mathrm{d}_A}\sum_{\alpha\alpha'\beta\beta'=1}^{\mathrm{d}_B}\mathcal{O}_{ab}\mathcal{O}^*_{cd}\mathcal{O}_{ef}\mathcal{O}^*_{gh}E^a_{j\alpha}E^{b*}_{j'\alpha'}E^{c*}_{k\alpha}E^{d}_{k'\alpha'}E^{e}_{k\beta}E^{f*}_{k'\beta'}E^{g*}_{j\beta}E^{h}_{j'\beta'}\\
        &=\frac{2}{\mathrm{d}^2}\sum_{a,c=1}^{\mathrm{d}}\sum_{e\neq f=1}\sum_{jj'kk'=1}^{\mathrm{d}_A}\sum_{\alpha\alpha'\beta\beta'=1}^{\mathrm{d}_B}\mathcal{O}_{aa}\mathcal{O}^*_{cc}\mathcal{O}_{ef}\mathcal{O}^*_{ef}E^a_{j\alpha}E^{a*}_{j'\alpha'}E^{c*}_{k\alpha}E^{c}_{k'\alpha'}E^{e}_{k\beta}E^{f*}_{k'\beta'}E^{e*}_{j\beta}E^{f}_{j'\beta'}\\
        &+\frac{1}{\mathrm{d}^2}\sum_{m\neq p=1}^{\mathcal{K}-1}\sum_{(ab)\in\mathcal{I}_m}\sum_{(cd)\in\mathcal{I}_p}\sum_{jj'kk'=1}^{\mathrm{d}_A}\sum_{\alpha\alpha'\beta\beta'=1}^{\mathrm{d}_B}\mathcal{O}_{ab}\mathcal{O}^*_{cd}\mathcal{O}_{ab}\mathcal{O}^*_{cd}E^a_{j\alpha}E^{b*}_{j'\alpha'}E^{c*}_{k\alpha}E^{d}_{k'\alpha'}E^{a}_{k\beta}E^{b*}_{k'\beta'}E^{c*}_{j\beta}E^{d}_{j'\beta'},
    \end{split}
\end{equation}
where we applied the hypothesis that the off-diagonal terms are always non-degenerate due to the non-resonance condition, therefore in order for $(ab),(ef)\in\mathcal{I}_m$, $(ab)=(ef)$ and $a\neq b$. Now, we want to remove all the sets $\mathcal{I}_m$ and $\mathcal{I}_p$ in order to simplify the summations. To do so, we observe
\begin{equation}
    \sum_{m\neq p=1}^{\mathcal{K}-1}\sum_{(ab)\in\mathcal{I}_m}\sum_{(cd)\in\mathcal{I}_p}=\sum_{m,p=1}^{\mathcal{K}-1}\sum_{(ab)\in\mathcal{I}_m}\sum_{(cd)\in\mathcal{I}_p}-\sum_{m=p=1}^{\mathcal{K}-1}\sum_{(ab)\in\mathcal{I}_m}\sum_{(cd)\in\mathcal{I}_p}.
\end{equation}
The last two terms are easier to simplify, and are going to give rise to the sums $\sum_{a\neq b=1}^{\mathrm{d}}\sum_{c\neq d=1}^{\mathrm{d}}$ and $\sum_{a\neq b=1}^{\mathrm{d}}$, respectively. Therefore, the final result for this term is
\begin{equation}
    \begin{split}
         &\sum_{m\neq p=0}^{\mathcal{K}-1}|\dbraket{\mathcal{O}}{\omega_m}|^2|\dbraket{\mathcal{O}}{\omega_p}|^2\Tr_{\mathscr{A}}(\omega^m\omega^{m\dagger}\omega^p\omega^{p\dagger})\\
         &=\frac{2}{\mathrm{d}^2}\sum_{a,c=1}^{\mathrm{d}}\sum_{e\neq f=1}\sum_{jj'kk'=1}^{\mathrm{d}_A}\sum_{\alpha\alpha'\beta\beta'=1}^{\mathrm{d}_B}\mathcal{O}_{aa}\mathcal{O}^*_{cc}\mathcal{O}_{ef}\mathcal{O}^*_{ef}E^a_{j\alpha}E^{a*}_{j'\alpha'}E^{c*}_{k\alpha}E^{c}_{k'\alpha'}E^{e}_{k\beta}E^{f*}_{k'\beta'}E^{e*}_{j\beta}E^{f}_{j'\beta'}\\
         &+\frac{1}{\mathrm{d}^2}\sum_{a \neq b = 1}^{\mathrm{d}}\sum_{c \neq d=1}^{\mathrm{d}}\sum_{jj'kk'=1}^{\mathrm{d}_A}\sum_{\alpha\alpha'\beta\beta'=1}^{\mathrm{d}_B}\mathcal{O}_{ab}\mathcal{O}^*_{cd}\mathcal{O}_{ab}\mathcal{O}^*_{cd}E^a_{j\alpha}E^{b*}_{j'\alpha'}E^{c*}_{k\alpha}E^{d}_{k'\alpha'}E^{a}_{k\beta}E^{b*}_{k'\beta'}E^{c*}_{j\beta}E^{d}_{j'\beta'}\\
         &-\frac{1}{\mathrm{d}^2}\sum_{a \neq b = 1}^{\mathrm{d}}\sum_{jj'kk'=1}^{\mathrm{d}_A}\sum_{\alpha\alpha'\beta\beta'=1}^{\mathrm{d}_B}|\mathcal{O}_{ab}|^2|\mathcal{O}_{ab}|^2 E^a_{j\alpha}E^{b*}_{j'\alpha'}E^{a*}_{k\alpha}E^{b}_{k'\alpha'}E^{a}_{k\beta}E^{b*}_{k'\beta'}E^{a*}_{j\beta}E^{b}_{j'\beta'}.
     \end{split}
\end{equation}

Finally, for the last term,

\begin{equation}
    \begin{split}
        \sum_{m\neq p=0}^{\mathcal{K}-1}|\dbraket{\mathcal{O}}{\omega_m}|^2|\dbraket{\mathcal{O}}{\omega_p}|^2\Tr_{\mathscr{A}}(\omega^m\omega^{p\dagger}\omega^p\omega^{m\dagger})&=2\sum_{p=1}^{\mathcal{K}-1}|\dbraket{\mathcal{O}}{\omega_0}|^2|\dbraket{\mathcal{O}}{\omega_p}|^2\Tr_{\mathscr{A}}(\omega^0\omega^{p\dagger}\omega^p\omega^{0\dagger})\\
        &+\sum_{m\neq p=1}^{\mathcal{K}-1}|\dbraket{\mathcal{O}}{\omega_m}|^2|\dbraket{\mathcal{O}}{\omega_p}|^2\Tr_{\mathscr{A}}(\omega^m\omega^{p\dagger}\omega^p\omega^{m\dagger})
    \end{split}
\end{equation}
for which we will apply a very similar procedure when compared to the previous one, only permuting the indexes within the trace in the first equation,
\begin{equation}
    \begin{split}
        &\sum_{m\neq p=0}^{\mathcal{K}-1}|\dbraket{\mathcal{O}}{\omega_m}|^2|\dbraket{\mathcal{O}}{\omega_p}|^2\Tr_{\mathscr{A}}(\omega^m\omega^{p\dagger}\omega^p\omega^{m\dagger})\\
        &=\frac{2}{\mathrm{d}^2}\sum_{p=1}^{\mathcal{K}-1}\sum_{a,g=1}^{\mathrm{d}}\sum_{(cd),(ef)\in\mathcal{I}_p}\sum_{jj'kk'=1}^{\mathrm{d}_A}\sum_{\alpha\alpha'\beta\beta'=1}^{\mathrm{d}_B}\mathcal{O}_{aa}\mathcal{O}^*_{cd}\mathcal{O}_{ef}\mathcal{O}^*_{gg}E^a_{j\alpha}E^{a*}_{j'\alpha'}E^{c*}_{k\alpha}E^{d}_{k'\alpha'}E^{e}_{k\beta}E^{f*}_{k'\beta'}E^{g*}_{j\beta}E^{g}_{j'\beta'}\\
        &+\frac{1}{\mathrm{d}^2}\sum_{m\neq p=1}^{\mathcal{K}-1}\sum_{(ab),(gh)\in\mathcal{I}_m}\sum_{(cd),(ef)\in\mathcal{I}_p}\sum_{jj'kk'=1}^{\mathrm{d}_A}\sum_{\alpha\alpha'\beta\beta'=1}^{\mathrm{d}_B}\mathcal{O}_{ab}\mathcal{O}^*_{cd}\mathcal{O}_{ef}\mathcal{O}^*_{gh}E^a_{j\alpha}E^{b*}_{j'\alpha'}E^{c*}_{k\alpha}E^{d}_{k'\alpha'}E^{e}_{k\beta}E^{f*}_{k'\beta'}E^{g*}_{j\beta}E^{h}_{j'\beta'}\\
        &=\frac{2}{\mathrm{d}^2}\sum_{a,g=1}^{\mathrm{d}}\sum_{c\neq d=1}^{d}\sum_{jj'kk'=1}^{\mathrm{d}_A}\sum_{\alpha\alpha'\beta\beta'=1}^{\mathrm{d}_B}\mathcal{O}_{aa}\mathcal{O}^*_{cd}\mathcal{O}_{cd}\mathcal{O}^*_{gg}E^a_{j\alpha}E^{a*}_{j'\alpha'}E^{c*}_{k\alpha}E^{d}_{k'\alpha'}E^{c}_{k\beta}E^{d*}_{k'\beta'}E^{g*}_{j\beta}E^{g}_{j'\beta'}\\
        &+\frac{1}{\mathrm{d}^2}\sum_{m\neq p=1}^{\mathcal{K}-1}\sum_{(ab)\in\mathcal{I}_m}\sum_{(cd)\in\mathcal{I}_p}\sum_{jj'kk'=1}^{\mathrm{d}_A}\sum_{\alpha\alpha'\beta\beta'=1}^{\mathrm{d}_B}\mathcal{O}_{ab}\mathcal{O}^*_{cd}\mathcal{O}_{cd}\mathcal{O}^*_{ab}E^a_{j\alpha}E^{b*}_{j'\alpha'}E^{c*}_{k\alpha}E^{d}_{k'\alpha'}E^{c}_{k\beta}E^{d*}_{k'\beta'}E^{a*}_{j\beta}E^{b}_{j'\beta'}.
    \end{split}
\end{equation}

Next, we apply the same  simplifications for the summations over the sets $\mathcal{I}_m$ and $\mathcal{I}_p$ as before to obtain the final result
\begin{equation}
    \begin{split}
         &\sum_{m\neq p=0}^{\mathcal{K}-1}|\dbraket{\mathcal{O}}{\omega_m}|^2|\dbraket{\mathcal{O}}{\omega_p}|^2\Tr_{\mathscr{A}}(\omega^m\omega^{p\dagger}\omega^p\omega^{m\dagger})\\
         &=\frac{2}{\mathrm{d}^2}\sum_{a,g=1}^{\mathrm{d}}\sum_{c\neq d=1}^{d}\sum_{jj'kk'=1}^{\mathrm{d}_A}\sum_{\alpha\alpha'\beta\beta'=1}^{\mathrm{d}_B}\mathcal{O}_{aa}\mathcal{O}^*_{cd}\mathcal{O}_{cd}\mathcal{O}^*_{gg}E^a_{j\alpha}E^{a*}_{j'\alpha'}E^{c*}_{k\alpha}E^{d}_{k'\alpha'}E^{c}_{k\beta}E^{d*}_{k'\beta'}E^{g*}_{j\beta}E^{g}_{j'\beta'}\\
         &+\frac{1}{\mathrm{d}^2}\sum_{a \neq b = 1}^{d}\sum_{c \neq d=1}^{d}\sum_{jj'kk'=1}^{\mathrm{d}_A}\sum_{\alpha\alpha'\beta\beta'=1}^{\mathrm{d}_B}\mathcal{O}_{ab}\mathcal{O}^*_{cd}\mathcal{O}_{cd}\mathcal{O}^*_{ab}E^a_{j\alpha}E^{b*}_{j'\alpha'}E^{c*}_{k\alpha}E^{d}_{k'\alpha'}E^{c}_{k\beta}E^{d*}_{k'\beta'}E^{a*}_{j\beta}E^{b}_{j'\beta'}\\
         &-\frac{1}{\mathrm{d}^2}\sum_{a \neq b = 1}^{\mathrm{d}}\sum_{jj'kk'=1}^{\mathrm{d}_A}\sum_{\alpha\alpha'\beta\beta'=1}^{\mathrm{d}_B}|\mathcal{O}_{ab}|^2|\mathcal{O}_{ab}|^2 E^a_{j\alpha}E^{b*}_{j'\alpha'}E^{a*}_{k\alpha}E^{b}_{k'\alpha'}E^{a}_{k\beta}E^{b*}_{k'\beta'}E^{a*}_{j\beta}E^{b}_{j'\beta'}.
    \end{split}
\end{equation}

Summing up these terms, we have the $6$ remaining contributions appearing in Equation~\eqref{eq:final_purity_eigenstates}, which reads (with all the summations written explicitly)
\begin{equation}
    \begin{split}
        &\overline{\Tr[\big(\rho^{(\mathcal{O})}_{\mathscr{A}}\big)^2]}\\
        &=\frac{1}{\mathrm{d}^2}\sum_{a, b, c, d=1}^{\mathrm{d}}\sum_{jj'kk'=1}^{\mathrm{d}_A}\sum_{\alpha\alpha'\beta\beta'=1}^{\mathrm{d}_B}\mathcal{O}_{aa}\mathcal{O}^*_{bb}\mathcal{O}_{cc}\mathcal{O}^*_{dd}E^a_{j\alpha}E^{a*}_{j'\alpha'}E^{b*}_{k\alpha}E^{b}_{k'\alpha'}E^c_{k\beta}E^{c*}_{k'\beta'}E^{d*}_{j\beta}E^{d}_{j'\beta'}\\
        &+\frac{2}{\mathrm{d}^2}\sum_{a,c=1}^{\mathrm{d}}\sum_{e\neq f=1}^{d}\sum_{jj'kk'=1}^{\mathrm{d}_A}\sum_{\alpha\alpha'\beta\beta'=1}^{\mathrm{d}_B}\mathcal{O}_{aa}\mathcal{O}^*_{cc}|\mathcal{O}_{ef}|^2E^a_{j\alpha}E^{a*}_{j'\alpha'}E^{c*}_{k\alpha}E^{c}_{k'\alpha'}E^{e}_{k\beta}E^{f*}_{k'\beta'}E^{e*}_{j\beta}E^{f}_{j'\beta'}\\
         &+\frac{1}{\mathrm{d}^2}\sum_{a \neq b = 1}^{\mathrm{d}}\sum_{c \neq d=1}^{d}\sum_{jj'kk'=1}^{\mathrm{d}_A}\sum_{\alpha\alpha'\beta\beta'=1}^{\mathrm{d}_B}\mathcal{O}_{ab}\mathcal{O}^*_{cd}\mathcal{O}_{ab}\mathcal{O}^*_{cd}E^a_{j\alpha}E^{b*}_{j'\alpha'}E^{c*}_{k\alpha}E^{d}_{k'\alpha'}E^{a}_{k\beta}E^{b*}_{k'\beta'}E^{c*}_{j\beta}E^{d}_{j'\beta'}\\
         &+\frac{2}{\mathrm{d}^2}\sum_{a,g=1}^{\mathrm{d}}\sum_{c\neq d=1}^{\mathrm{d}}\sum_{jj'kk'=1}^{\mathrm{d}_A}\sum_{\alpha\alpha'\beta\beta'=1}^{\mathrm{d}_B}\mathcal{O}_{aa}|\mathcal{O}_{cd}|^2\mathcal{O}^*_{gg}E^a_{j\alpha}E^{a*}_{j'\alpha'}E^{c*}_{k\alpha}E^{d}_{k'\alpha'}E^{c}_{k\beta}E^{d*}_{k'\beta'}E^{g*}_{j\beta}E^{g}_{j'\beta'}\\
         &+\frac{1}{\mathrm{d}^2}\sum_{a \neq b = 1}^{\mathrm{d}}\sum_{c \neq d=1}^{\mathrm{d}}\sum_{jj'kk'=1}^{\mathrm{d}_A}\sum_{\alpha\alpha'\beta\beta'=1}^{\mathrm{d}_B}|\mathcal{O}_{ab}|^2|\mathcal{O}_{cd}|^2E^a_{j\alpha}E^{b*}_{j'\alpha'}E^{c*}_{k\alpha}E^{d}_{k'\alpha'}E^{c}_{k\beta}E^{d*}_{k'\beta'}E^{a*}_{j\beta}E^{b}_{j'\beta'}\\
         &-\frac{1}{\mathrm{d}^2}\sum_{a \neq b = 1}^{\mathrm{d}}\sum_{jj'kk'=1}^{\mathrm{d}_A}\sum_{\alpha\alpha'\beta\beta'=1}^{\mathrm{d}_B}|\mathcal{O}_{ab}|^4 E^a_{j\alpha}E^{b*}_{j'\alpha'}E^{a*}_{k\alpha}E^{b}_{k'\alpha'}E^{a}_{k\beta}E^{b*}_{k'\beta'}E^{a*}_{j\beta}E^{b}_{j'\beta'}.
     \end{split}
\end{equation}

\section{\label{app:random_tensor_networks}Random tensor networks and ETH approximations}

In this appendix, we discuss the details of the calculations involved in computing the averages over the random tensor networks representing the integrals over the Haar measure of summations of the matrix elements of unitary matrices. In order to perform the graphical calculus, we apply the Wolfram Mathematica package for the integration of random tensor networks, RTNI~\cite{RTNI_haar}. The codes and further details mentioned here can be found in the notebook available at the github repository~\cite{github_correr}. 

Each of the integrals appearing in the sum~\eqref{eq:final_purity_unitaries} is represented as a tensor network in Figure~\ref{fig:integrals-tensornetworks-grid}. The convention used to represent the contractions is defined in Ref.~\cite{randomtensornetworks_graphical_haar} and consists of applying boxes to represent tensors and connected wires to represent index contractions. Each symbol, circle, square or diamond, represents a subspace with different dimension. In this case, the circle represents the space with dimension $\mathrm{d}_A$, the square the one with dimension $\mathrm{d}_B$, and the diamond the tensor product of these two spaces, resulting in dimension $\mathrm{d}$. If a symbol is filled (empty), it represents the output (input) of the tensor. For ease of understanding, readers may regard the rows of a unitary as the output and the columns as the input. When taking the adjoint, these roles are inverted. The smaller boxes represent kets or bras (white symbol for bra and filled symbol for ket), therefore a connection between a bra or ket to a unitary represents the selection of a column or a row.

\begin{figure*}[htbp]
    \includegraphics[width=0.7\columnwidth]{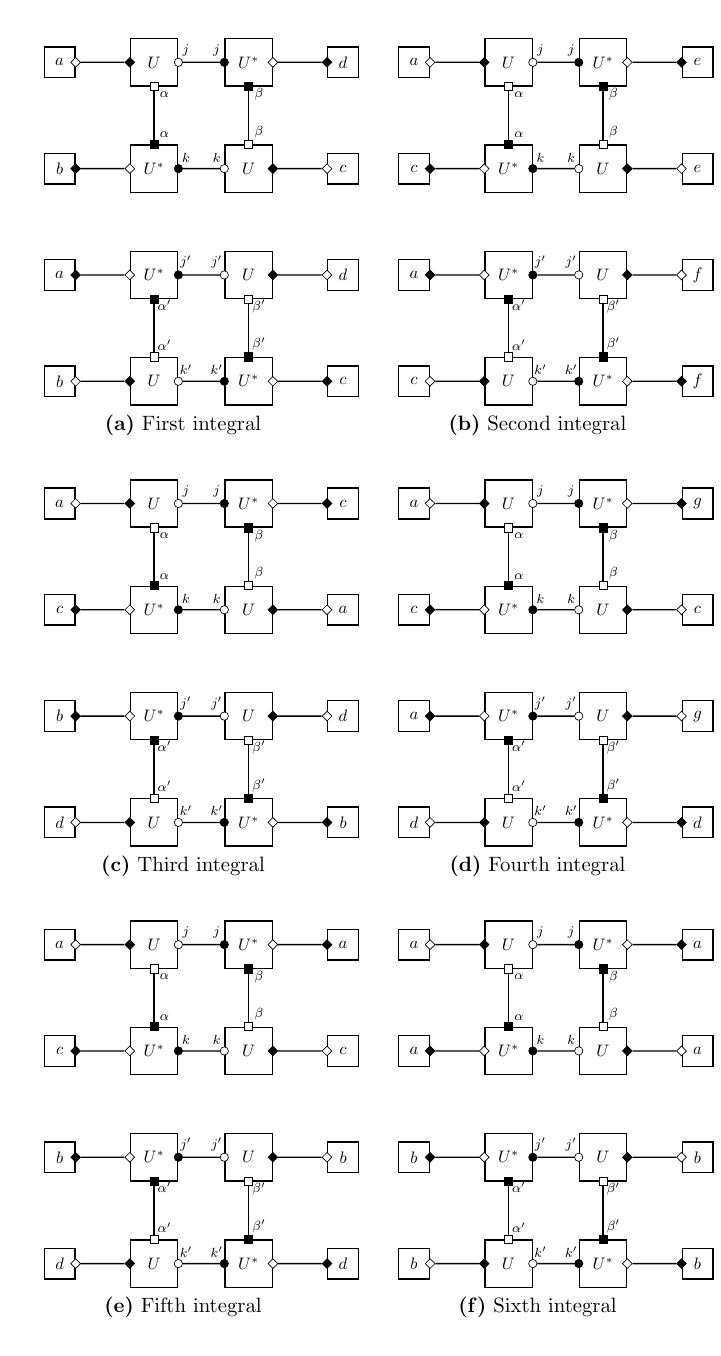}
    \centering
    \caption{Random tensor networks representing the different contractions over $d_A$ and $d_B$ indices appearing within the integrals in Equation~\eqref{eq:final_purity_unitaries}.}
    \label{fig:integrals-tensornetworks-grid}
\end{figure*}

After performing the graphical calculus for each of the tensor networks, the result is a sum of smaller tensor networks considering the index contractions given by summations over $d_A$ and $d_B$ with different weights. These smaller tensor networks, in fact, are simplified Kronecker deltas over the different indexes appearing in the matrix elements of the initial operator $\mathcal{O}$. Then, we simplify the indexes with the Kronecker deltas and get the final results only in terms of the matrix elements of $\mathcal{O}$ on the energy eigenbasis. Writing each of the weights individually here is unpractical. Therefore, we provide a supporting Wolfram Mathematica notebook \cite{github_correr} with the specific values.

We used the simplification that summations of Kronecker deltas over indexes that have to be different to simplify some of the terms to zero, i.e.,
\begin{equation}
    \sum_{a\neq b=1}^{\mathrm{d}}\delta_{ab}=0.
\end{equation}
The idea was to find, for each of the six terms in Equation~\eqref{eq:final_purity_unitaries}, the resulting non-trivial values after all the contractions. Equal contributions appearing in different terms of Equation~\eqref{eq:final_purity_unitaries} were summed up with the corresponding weights. For instance,
\begin{equation}
    z_n \left[\Tr^{\delta E}(\mathcal{O})\right]^4 + z_m \left[\Tr^{\delta E}(\mathcal{O})\right]^4 = (z_n + z_m)\left[\Tr^{\delta E}(\mathcal{O})\right]^4,
\end{equation}
where $z_n$ and $z_m$ are illustrative functions of $d_A$, $d_B$ and $d$. Given the resulting weights named according to the notation $w_n\equiv w_n(d_A,d_B,d)$, we get the $13$ terms

\begin{align}\label{eq:app_purity_weightedum}
\Big\langle\overline{\Tr[\big(\rho^{(\mathcal{O})}_{\mathscr{A}}\big)^2]}\Big\rangle
\begin{array}[t]{@{}l@{\hspace{0.1\linewidth}}l@{}}
\displaystyle
=\frac{w_0}{\mathrm{d}^2}\sum_{a=1}^{\mathrm{d}}\mathcal{O}_{aa}^4
&
\displaystyle
+\frac{w_1}{\mathrm{d}^2}\sum_{a=1}^{\mathrm{d}}\mathcal{O}_{aa}^3\Tr^{\delta E}(\mathcal{O})
\\[0.8ex]
\displaystyle
+\frac{w_2}{\mathrm{d}^2}\left[\sum_{a=1}^{\mathrm{d}}\mathcal{O}_{aa}^2\right]^2
&
\displaystyle
+\frac{w_3}{\mathrm{d}^2}\sum_{a=1}^{\mathrm{d}}\mathcal{O}_{aa}^2
\left[\Tr^{\delta E}(\mathcal{O})\right]^2
\\[0.8ex]
\displaystyle
+\frac{w_4}{\mathrm{d}^2}\left[\Tr^{\delta E}(\mathcal{O})\right]^4
&
\displaystyle
+\frac{w_5}{\mathrm{d}^2}\sum_{a \neq b=1}^{\mathrm{d}}
\mathcal{O}_{aa}\mathcal{O}_{cc}|\mathcal{O}_{ac}|^2
\\[0.8ex]
\displaystyle
+\frac{w_6}{\mathrm{d}^2}\sum_{a \neq b=1}^{\mathrm{d}}\mathcal{O}_{aa}^2|\mathcal{O}_{ac}|^2
&
\displaystyle
+\frac{w_7}{\mathrm{d}^2}\sum_{a=1}^{\mathrm{d}}\mathcal{O}_{aa}^2
\sum_{e \neq f=1}^{\mathrm{d}}|\mathcal{O}_{ef}|^2
\\[0.8ex]
\displaystyle
+\frac{w_8}{\mathrm{d}^2}\sum_{e \neq f = 1}^{\mathrm{d}}|\mathcal{O}_{ef}|^2
\left[\Tr^{\delta E}(\mathcal{O})\right]^2
&
\displaystyle
+\frac{w_9}{\mathrm{d}^2}\sum_{c \neq f = 1}^{\mathrm{d}}
\mathcal{O}_{cc}|O_{cf}|^2\Tr^{\delta E}(\mathcal{O})
\\[0.8ex]
\displaystyle
+\frac{w_{10}}{\mathrm{d}^2}\sum_{a \neq b = 1}^{\mathrm{d}}|\mathcal{O}_{ab}|^4
&
\displaystyle
+\frac{w_{11}}{\mathrm{d}^2}\sum_{a \neq b = 1}^{\mathrm{d}}|\mathcal{O}_{ab}|^2
\sum_{c \neq d = 1}^{\mathrm{d}}|\mathcal{O}_{cd}|^2
\\[0.8ex]
\displaystyle
+\frac{w_{12}}{\mathrm{d}^2}\sum_{a \neq b \neq c = 1}^{\mathrm{d}}
|\mathcal{O}_{ab}|^2|\mathcal{O}_{bc}|^2
&
\end{array}
\end{align}

This is the result applied to obtain the analytical values in Section~\ref{sec:results}. Details of the computation are shown in the notebook \cite{github_correr}. 

\subsection{Complete LOE\label{subapp:tensor_complete_LOE}}
In order to proceed further and obtain a more concise equation for the entanglement scaling as a function of $\mathrm{d}_A$ and $\mathrm{d}$, we study the LOE in the limits $\mathrm{d}_A=\mathrm{d}_B = \sqrt{\mathrm{d}}$, and $\mathrm{d}_A=\text{O}(1)$ and $\mathrm{d}\gg 1$. Using these conditions, we obtain the leading contribution to each of the weights $w_j$. After this, we study the order of magnitude of each of these terms depending on the elements of $\mathcal{O}$ imposing the constraints of ETH, which imply $\mathcal{O}_{aa}=\text{O}(1)$ and $\mathcal{O}_{ab}=\text{O}(1/\sqrt{\mathrm{d}})$, with $a\neq b$, and considering the number of different terms appearing in the sums with indexes varying from $1$ to $\mathrm{d}$. For instance, the first term in the equation above, $\frac{w_0}{\mathrm{d}^2}\sum_{a=1}^{\mathrm{d}}\mathcal{O}_{aa}^4$, will be of the order of $w_0/\mathrm{d}$, because $\mathcal{O}_{aa}=\text{O}(1)$.

\begin{comment}
\subsubsection{${d_A=\text{O}(1),\, d\gg 1}$ and ${d = \text{O}(1)}$}

In this case, the leading order term will be the one corresponding to $w_2$, with order $\text{O}(1)$, while the subleading order is the terms $w_0$ and $w_7$, with order $\text{O}(1/d)$. At leading order,

\begin{equation}
    \Big\langle\overline{\Tr[\big(\rho^{(\mathcal{O})}_{\mathscr{A}}\big)^2]}\Big\rangle=\frac{1}{d^2}\frac{\left(\sum_{a=1}^{d}\mathcal{O}_{aa}^2\right)^2}{d_A^2},
\end{equation}
therefore
\begin{equation}
    \begin{split}
        \Big\langle\overline{S_2(\rho^{(\mathcal{O})}(t)\big)}\Big\rangle&\approx2\ln(d_A)-2\ln(\frac{\sum_{a=1}^{d}\mathcal{O}_{aa}^2}{d})\\
        &=2\ln(d_A)-2\ln\left[\frac{\sum_{a=1}^{d}\mathcal{O}_{aa}^2}{d}+\left(\frac{\sum_{a=1}^{d}\mathcal{O}_{aa}}{d}\right)^2-\left(\frac{\sum_{a=1}^{d}\mathcal{O}_{aa}}{d}\right)^2\right]\\
        &=2\ln(d_A)-2\ln\left\{\sigma_{\text{diag}}^2 + \left[\frac{\Tr^{\delta E}(\mathcal{O})}{d}\right]^2\right\}.
    \end{split}
\end{equation}
\end{comment}

\subsubsection{${\mathrm{d}_A=\text{O}(1),\, \mathrm{d}\gg 1}$}

The terms composing the leading order will be $w_2$ and $w_7$, with order $\mathrm{d}^2/\mathrm{d}_A^2$. The average of the trace is then

\begin{equation}
    \begin{split}
        \Big\langle\overline{\Tr[\big(\rho^{(\mathcal{O})}_{\mathscr{A}}\big)^2]}\Big\rangle&=\frac{1}{\mathrm{d}_A^2 \mathrm{d}^2}\left[\left(\sum_{a=1}^{\mathrm{d}}\mathcal{O}_{aa}^2\right)^2+2\sum_{a=1}^{\mathrm{d}}\mathcal{O}_{aa}^2\sum_{e\neq f=1}^{\mathrm{d}}|\mathcal{O}_{ef}|^2\right]\\
        &=\frac{1}{\mathrm{d}_A^2 \mathrm{d}^2}\left[\sum_{a=1}^{\mathrm{d}}\mathcal{O}_{aa}^2\Tr^{\delta E}(\mathcal{O}^{\dagger}\mathcal{O})+\sum_{a=1}^{\mathrm{d}}\mathcal{O}_{aa}^2\sum_{e\neq f=1}^{\mathrm{d}}|\mathcal{O}_{ef}|^2\right]\\
        &=\frac{1}{\mathrm{d}_A^2 \mathrm{d}^2}\left[\mathrm{d}\sum_{a=1}^{\mathrm{d}}\mathcal{O}_{aa}^2+\sum_{a=1}^{\mathrm{d}}\mathcal{O}_{aa}^2\sum_{e\neq f=1}^{\mathrm{d}}|\mathcal{O}_{ef}|^2 +d\sum_{e\neq f=1}^{\mathrm{d}}|\mathcal{O}_{ef}|^2 - \mathrm{d}\sum_{e\neq f=1}^{\mathrm{d}}|\mathcal{O}_{ef}|^2\right]\\
        &=\frac{1}{\mathrm{d}_A^2 \mathrm{d}^2}\left[\mathrm{d}\Tr^{\delta E}(\mathcal{O}^{\dagger}\mathcal{O})+\sum_{a=1}^{\mathrm{d}}\mathcal{O}_{aa}^2\sum_{e\neq f=1}^{\mathrm{d}}|\mathcal{O}_{ef}|^2  - \mathrm{d}\sum_{e\neq f=1}^{\mathrm{d}}|\mathcal{O}_{ef}|^2\right]\\
        &=\frac{1}{\mathrm{d}_A^2}\left[1 + (\mathrm{d}-1)\sigma^2_{\text{diag}}\sigma^2_{\text{off-diag}}- (\mathrm{d}-1)\sigma^2_{\text{off-diag}}\right],
    \end{split}
\end{equation}
where we applied that $\Tr(\mathcal{O}^{\dagger}\mathcal{O})=\mathrm{d}$. Additionally, to introduce the fluctuations of the diagonal and off-diagonal elements, we use the result $\langle \mathcal{O}_{ab}\rangle =0$, given that the eigenvectors of the Hamiltonian are uncorrelated random vectors~\cite{ETH_review}, and we applied $\Tr(\mathcal{O}) =0$. Therefore, the fluctuations of the diagonal terms are simply given by $\sigma^2_{\text{diag}}=\langle \mathcal{O}_{aa}^2\rangle$, while the off-diagonal fluctuations are $\sigma^2_{\text{off-diag}}=\langle |\mathcal{O}_{ab}|^2\rangle$. Applying minus the logarithm to the final result, we get
\begin{equation}
    \Big\langle\overline{S_2(\rho_{\mathscr{A}}^{(\mathcal{O})}(t)\big)}\Big\rangle\approx 2\ln(\mathrm{d}_A)-\ln\left\{1 +(\mathrm{d}-1)\sigma_{\text{off-diag}}^2\left[\sigma^2_{\text{diag}}-1\right]\right\}.
\end{equation}

\begin{comment}
\subsubsection{${d_A=\sqrt{d}}$ and ${d=\text{O}(1),\, d\gg 1}$}

Through the magnitude analysis, we get that the leading order terms are those with weights indexed by $w_0$ and $w_2$. Rewriting the expression with the appropriate weight in the regime, we get

\begin{align}
    \Big\langle\overline{\Tr[\big(\rho^{(\mathcal{O})}_{\mathscr{A}}\big)^2]}\Big\rangle=\frac{2}{dd^2}\left(\sum_{a=1}^{d}\mathcal{O}_{aa}^4+\sum_{a=1}^{d}\mathcal{O}_{aa}^2\sum_{b=1}^{d}\mathcal{O}_{bb}^2\right),
\end{align}
which, applying minus the logarithm, provides us with the entropy
\begin{equation}
    \Big\langle\overline{S_2(\rho^{\mathcal{O}}(t))}\Big\rangle\approx \ln(d)-\ln\left[2\left(\langle \mathcal{O}^2_{aa}\rangle^2 + \frac{1}{d}\langle \mathcal{O}^4_{aa}\rangle\right)\right]
\end{equation}
\end{comment}

\subsubsection{${\mathrm{d}_A=\sqrt{\mathrm{d}}}$}

The leading order term is the one indexed with $w_4$, which is proportional to the fourth power of the trace of the operator over the window. Due to the observation that $\Tr(\mathcal{O}) =0$, we instead rely on the next order, given by terms $w_2$, $w_7$ and $w_{11}$. Summing up these contributions with the appropriate weights, we get
\begin{equation}
    \begin{split}
        \Big\langle\overline{\Tr[\big(\rho^{(\mathcal{O})}_{\mathscr{A}}\big)^2]}\Big\rangle&=\frac{1}{\mathrm{d}^3}\left(2\sum_{a=1}^{\mathrm{d}}\mathcal{O}_{aa}^2\sum_{a=1}^{\mathrm{d}}\mathcal{O}_{aa}^2+4\sum_{a=1}^{d}\mathcal{O}_{aa}^2\sum_{e\neq f=1}^{\mathrm{d}}|\mathcal{O}_{ef}|^2+\sum_{a\neq b=1}^{\mathrm{d}}|\mathcal{O}_{ab}|^2\sum_{c\neq d=1}^{\mathrm{d}}|\mathcal{O}_{cd}|^2\right)\\
        &=\frac{1}{\mathrm{d}^3}\left(2\sum_{a=1}^{\mathrm{d}}\mathcal{O}_{aa}^2\Tr^{\delta E}(\mathcal{O}^{\dagger}\mathcal{O})+2\sum_{a=1}^{\mathrm{d}}\mathcal{O}_{aa}^2\sum_{e\neq f=1}^{\mathrm{d}}|\mathcal{O}_{ef}|^2+\sum_{a\neq b=1}^{\mathrm{d}}|\mathcal{O}_{ab}|^2\sum_{c\neq d=1}^{\mathrm{d}}|\mathcal{O}_{cd}|^2\right)\\
        &=\frac{1}{\mathrm{d}^3}\left(2\sum_{a=1}^{\mathrm{d}}\mathcal{O}_{aa}^2\Tr^{\delta E}(\mathcal{O}^{\dagger}\mathcal{O})+\sum_{a=1}^{\mathrm{d}}\mathcal{O}_{aa}^2\sum_{e\neq f=1}^{\mathrm{d}}|\mathcal{O}_{ef}|^2+\sum_{a\neq b=1}^{\mathrm{d}}|\mathcal{O}_{ab}|^2\Tr^{\delta E}(\mathcal{O}^{\dagger}\mathcal{O})\right)\\
        &=\frac{1}{\mathrm{d}^3}\left\{\left[\Tr^{\delta E}(\mathcal{O}^{\dagger}\mathcal{O})\right]^2+\sum_{a=1}^{\mathrm{d}}\mathcal{O}_{aa}^2\left[\Tr^{\delta E}(\mathcal{O}^{\dagger}\mathcal{O})+\sum_{e\neq f=1}^{\mathrm{d}}|\mathcal{O}_{ef}|^2\right]\right\}\\
        &=\frac{1}{\mathrm{d}}\left\{1+\frac{1}{\mathrm{d}}\sum_{a=1}^{\mathrm{d}}\mathcal{O}_{aa}^2+\frac{1}{\mathrm{d}^2}\sum_{a=1}^{\mathrm{d}}\mathcal{O}_{aa}^2\sum_{e\neq f=1}^{\mathrm{d}}|\mathcal{O}_{ef}|^2\right\}\\
        &=\frac{1}{\mathrm{d}}\left\{1+\sigma^2_{\text{diag}}\left[1+(\mathrm{d}-1)\sigma^2_{\text{off-diag}}\right]\right\},
    \end{split}
\end{equation}
leading to the final result
\begin{equation}
    \Big\langle\overline{S_2(\rho_{\mathscr{A}}^{\mathcal{O}}(t))}\Big\rangle\approx \ln(\mathrm{d})-\ln\left\{1 + \sigma_{\text{diag}}^2\left[1+(\mathrm{d}-1)\sigma_{\text{off-diag}}^2\right]\right\}.
\end{equation}

\subsubsection{Numerical analysis of the weights}

The observed dip in the main text is consistent with the behavior of the leading order analytical weights $w_n$ obtained after Haar averaging. For all the regimes studied in the approximations, the leading order weights contributing are a combination of the terms $w_2$, $w_7$, and $w_{11}$. Table~\ref{tab:regimes_approximations} presents the observed combinations for the two regimes. Figure~\ref{fig:weights_distribution_10} presents how each of these $w_n$ weights behaves as a function of $n_A$. For small $\mathrm{d}_A$ we get only Page-like volume law contributions, weights $2$ and $7$ in Figure~\ref{fig:weights_distribution_10}. Even considering a linear combination of these two terms the behavior is always a regular volume law. This is pictorially represented in the inset on the left panel of Figure~\ref{fig:weights_distribution_10}. Restricting to the region for which the dip occurs, $\mathrm{d}_A=\sqrt{\mathrm{d}}$, we get contributions coming from $w_2$, $w_7$, and $w_{11}$, having only the last one a behavior that contrasts with the linear increase of the volume law, decaying linearly instead. When these three terms are summed, we get a regular volume law. However, if we change the linear combination, giving a greater contribution to the decaying weight $w_{11}$, the dip appears naturally even for a combination $w_2+w_7+3.5\times w_{11}$, getting more and more pronounced as the contribution coming from $w_{11}$ grows, see right inset on Figure~\ref{fig:weights_distribution_10} for a pictorial representation of the behavior. Therefore, we can argue that the overall behavior of the LOE comes from a combination of both the randomness of the eigenstates decomposed on a bipartite basis (which generates the different weights $w_n$) and the ETH behavior of the operator (which filters magnitude orders and the most influential weights).

\begin{table}[htb]
	\caption{Leading order weights for the complete LOE at each approximation regime depending on $\mathrm{d}_A$ and $\mathrm{d}$.}\label{tab:regimes_approximations}%
	\centering
	\setlength{\tabcolsep}{12pt}
	
	\begin{tabular}{cc}
		\toprule
		   $\mathrm{d}_A = O(1),\, \mathrm{d} \gg 1$ & $d_A \approx d$ \\
		\midrule 
		 $2, 7$ & $2, 7, 11$ \\
		\bottomrule
	\end{tabular}
\end{table}

\begin{figure*}[htb]
    \centering
    \includegraphics[width=\columnwidth]{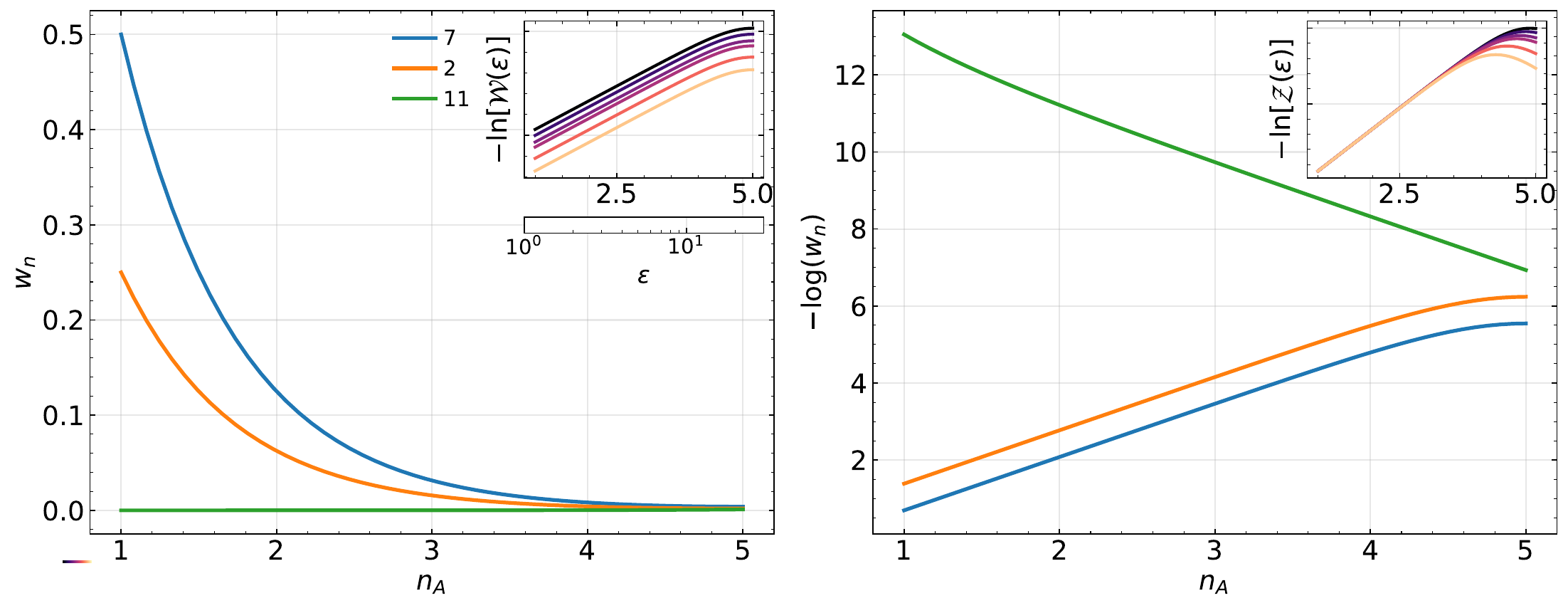}
    \caption{Weights distribution for the leading order contributions of the terms appearing in the analytical formula related to the LOE computation, named $2$, $7$ and $11$. Insets: minus the logarithm of linear combinations of weights $\mathcal{W}(\epsilon) \equiv w_7+\epsilon\cross w_2$ (left inset), representing the regime $\mathrm{d}_A = \text{O}(1)$, and $\mathcal{Z}(\epsilon) \equiv w_2+w_7+\epsilon\cross w_{11}$ (right inset), representing $\mathrm{d}_A \approx \mathrm{d}$, for different $\epsilon$ values. The set of $\epsilon$ values illustrated is $(1,\,2,\, 3.5,\, 5,\, 10,\, 20)$, identified according to the color map below the left inset.}
    \label{fig:weights_distribution_10}
\end{figure*}

\section{\label{app:additional_numerics}Additional numerical data}

In this appendix we present the numerical data mentioned and discussed throughout the main text as complementary results without further discussions. The first result we introduce is the comparison between numerical results for the average eigenstate entanglement for the $20\%$ of states in the middle of the spectrum for the maximally chaotic and integrable ($J =1.0$ and $h_x=1.1$, other parameters set to $0$) set of parameters of the model (Equation~\eqref{eq:mixed_field_Ising})~\cite{chaos_distributions_entanglement_khemani}, Figure~\ref{fig:eigenstate_entanglement}. Together, data for uniformly sampled vectors are shown and the analytical calculations considering state designs (traced lines). It is possible to see that the increase as a function of $n_A$ is different when comparing the integrable and chaotic sets of parameters, being the chaotic case very close to the random state computations. Additionally, the correspondence between the random state average and eigenstates of the chaotic model is less faithful for higher values of $\alpha$ and at subsystem sizes closer to $L/2$. For additional and more in depth discussion, check references~\cite{volume_law_eigenstates} and~\cite{renyi_entropy_chaotic_eigenstates}.

\begin{figure}[htb]
    \centering
    \includegraphics[width=0.5\columnwidth]{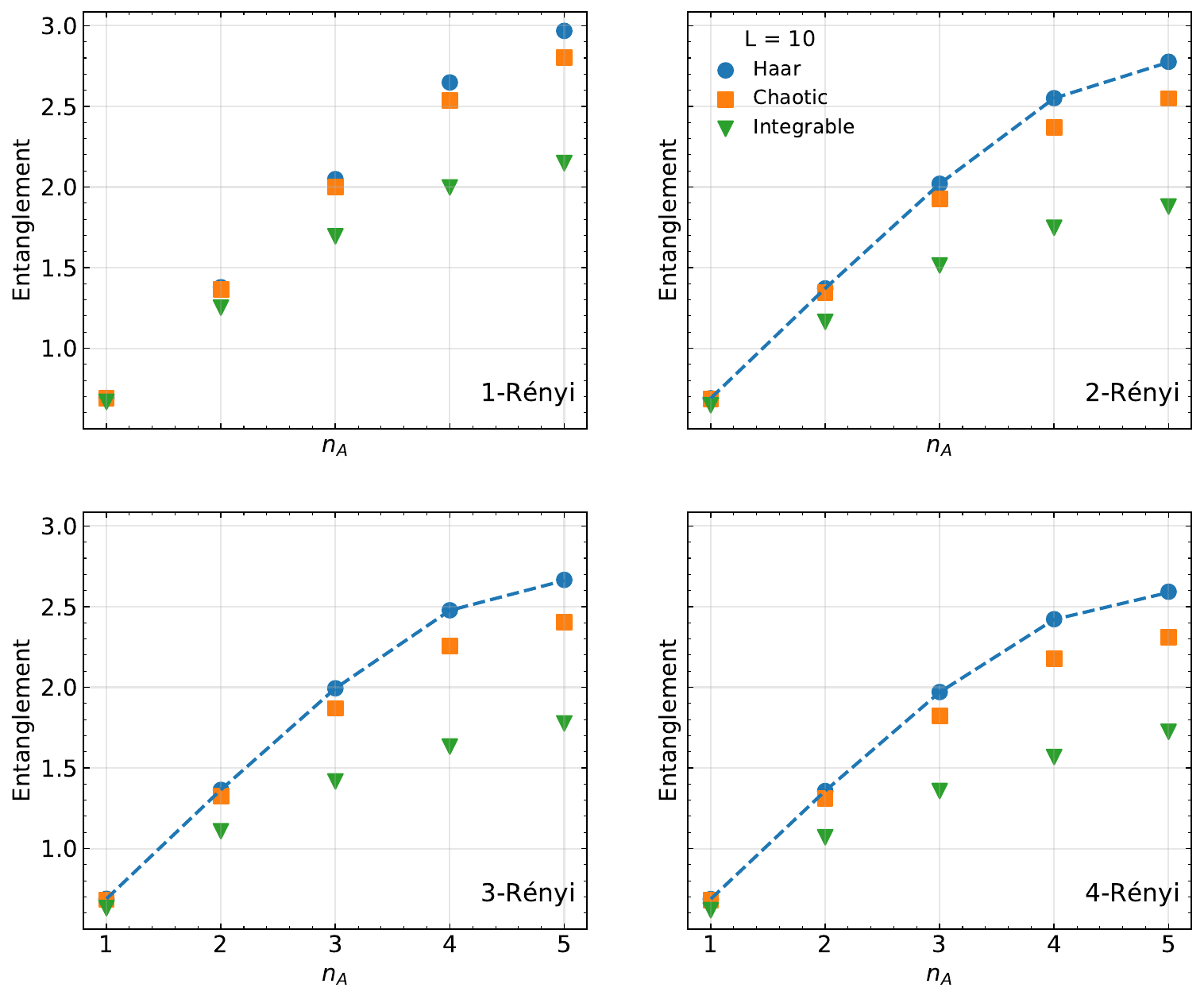}
    \caption{Average entanglement for eigenstates of a $10$ spins mixed field Ising model, considering $\alpha-$Rényi entropies from $1$ to $4$. The dots represent simulation data for Haar random vectors (blue circles), eigenstates of the maximally chaotic Hamiltonian (orange squares), and the integrable Hamiltonian (green upside down triangles). Blue traced lines are analytical calculations using random tensor-networks and Weingarten calculus. The averages for the Hamiltonians were performed with the eigenvectors corresponding to $20\%$ of the eigenvalues around the center of the spectrum.}
    \label{fig:eigenstate_entanglement}
\end{figure}

Figures~\ref{fig:12_spins_F_evolution} and~\ref{fig:error_12_spins_F_evolution} present the evolution of the values of $F(\mathcal{O})$ for different energy windows as a function of the subsystem size and the error as a function of the window dimension for $12$. We observe a similar behavior when compared to the $10$ spins case, where the only important reference values are regarding to the portion of the complete system size $L$ considered, i.e., the errors will be of the same order for different $L$ when the subsystem size considered is $L/n$. This feature can be seen also in the comparison between different values of $L$ (therefore different dimensions $\mathrm{d}$), but with fixed window dimension $d$ in Figures~\ref{fig:evolution_different_n_spins_F_evolution} and~\ref{fig:error_different_n_spins_F_evolution}. The behavior of the evolution is similar as a function of $n_A$ for both the values and error. By comparing the errors as a function of the number of spins $L$, we see that the value of error $10^{-1}$ (middle panel), for instance, occurs when $n_A=L/4$, which means $2$ for $L=10$, $3$ for 12, and so on. Figures~\ref{fig:log_F_10_spins},~\ref{fig:log_F_12_spins}, and~\ref{fig:log_F_different_n_spins} show the behavior of $-\log(F)$, displaying a characteristic volume law behavior at all $n_A$ values.

\begin{figure}[htb]
    \centering
    \includegraphics[width=0.5\columnwidth]{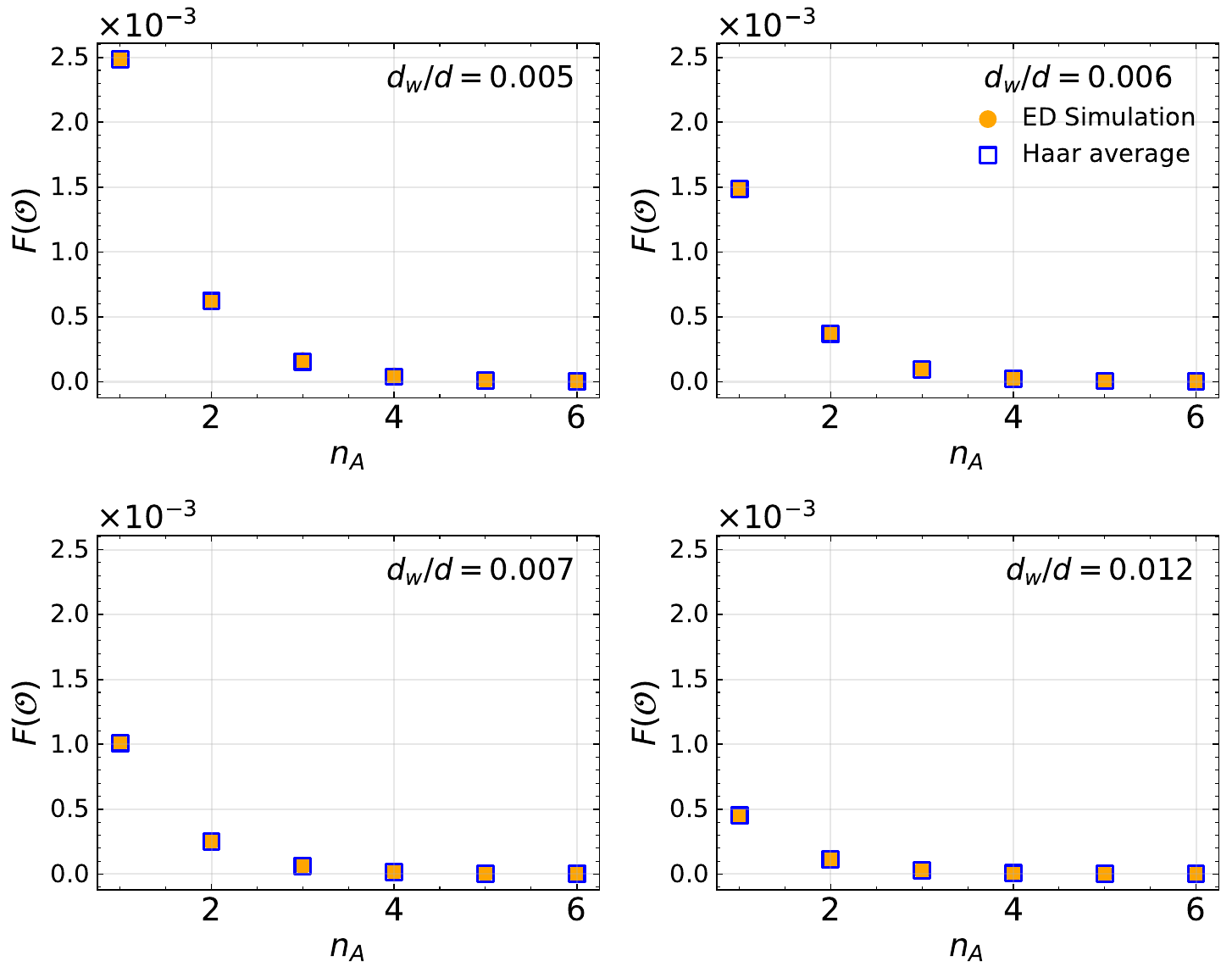}
    \caption{Evolution of the $F(\mathcal{O})$ (Equation~\eqref{eq:F_explicit_form}) term depending only on diagonal elements of the operator in the energy basis as a function of the size of subsystem A. The shown values are for total $L=12$, and each panel corresponds to a different size of the energy shell.}
    \label{fig:12_spins_F_evolution}
\end{figure}

\begin{figure}[htb]
    \centering
    \includegraphics[width=0.5\columnwidth]{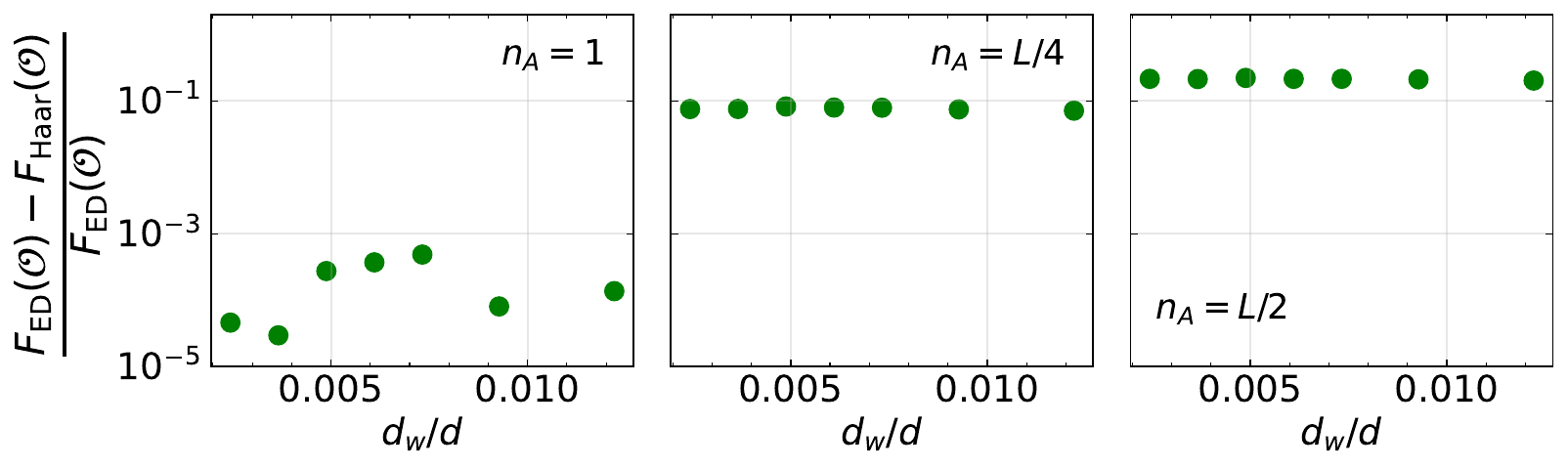}
    \caption{Error function comparing the ED simulations with Haar random averaged values of term $F(\mathcal{O})$ (Equation~\eqref{eq:F_explicit_form}) as a function of the rescaled dimension of the energy shell (number of eigenstates included divided by the total dimension of the Hilbert space) for $L=12$. Each panel shows the values at different bipartition sizes, including system $A$ with $1$, $L/4$ and $L/2$ spins. At this region, the number of states has an approximately linear correspondence to the energy window size, therefore the dependence is the same.}
    \label{fig:error_12_spins_F_evolution}
\end{figure}

\begin{figure}[htb]
    \centering
    \includegraphics[width=0.5\columnwidth]{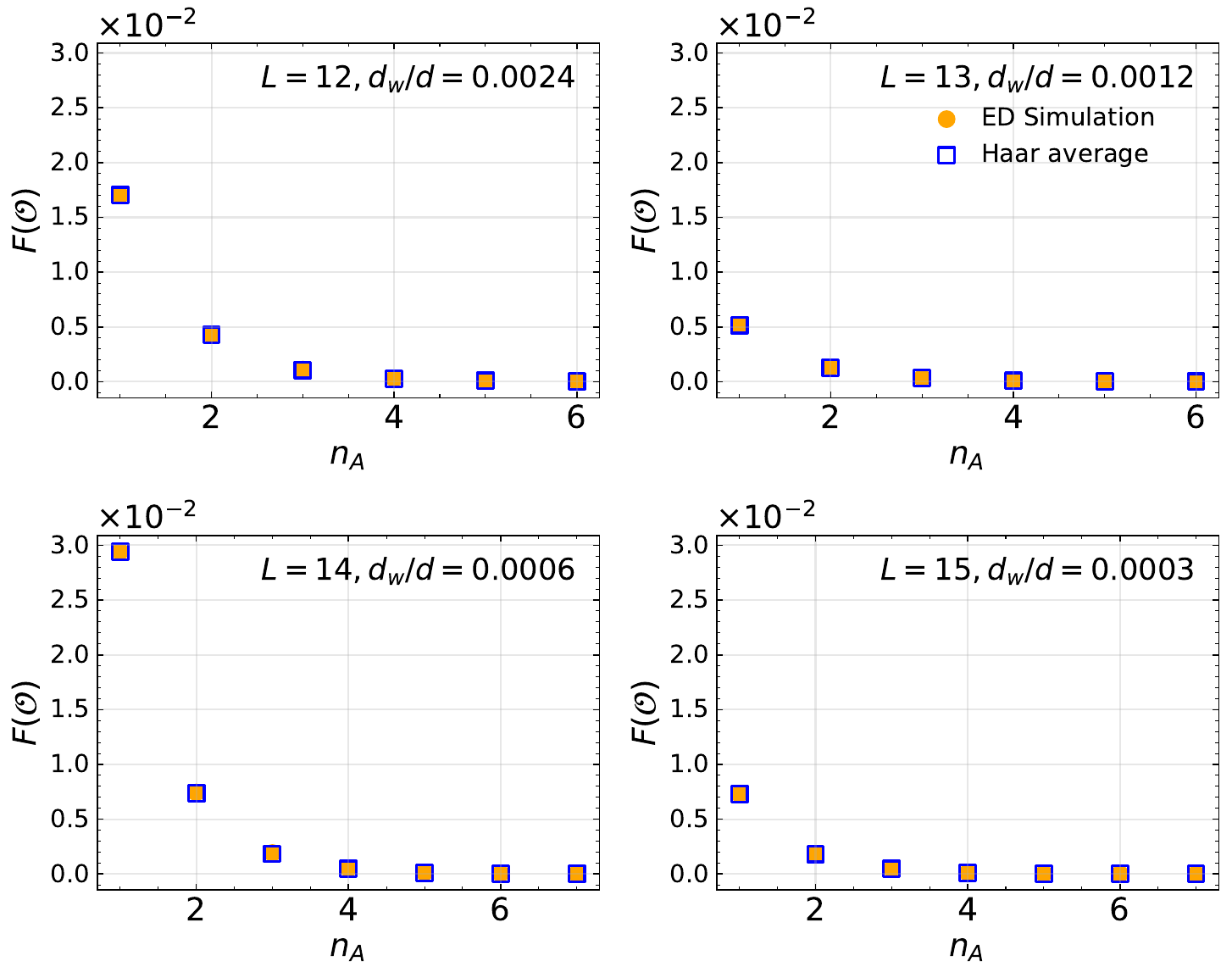}
    \caption{Evolution of the $F(\mathcal{O})$ (Equation~\eqref{eq:F_explicit_form}) term depending only on diagonal elements of the operator in the energy basis as a function of the size of subsystem $\mathscr{A}$. The shown values are for window dimension $\mathrm{d}_w=10$, and each panel corresponds to a different total number of spins $L$.}
    \label{fig:evolution_different_n_spins_F_evolution}
\end{figure}

\begin{figure}[htb]
    \centering
    \includegraphics[width=0.5\columnwidth]{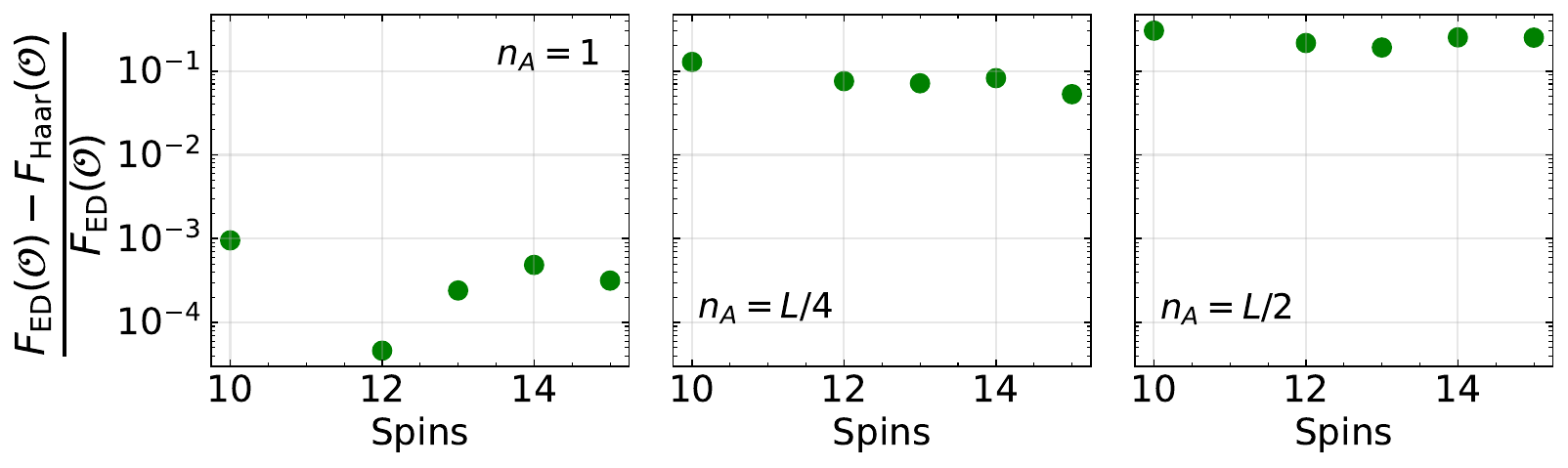}
    \caption{Error function comparing the ED simulations with Haar random averaged values of term $F(\mathcal{O})$ (Equation~\eqref{eq:F_explicit_form}) as a function of the dimension of the Hilbert space for window size $\mathrm{d}_w=10$. Each panel shows the values at different bipartition sizes, including system A with $1$, $L/4$ and $L/2$ spins.}
    \label{fig:error_different_n_spins_F_evolution}
\end{figure}

\begin{figure}[htb]
    \centering
    \includegraphics[width=0.5\columnwidth]{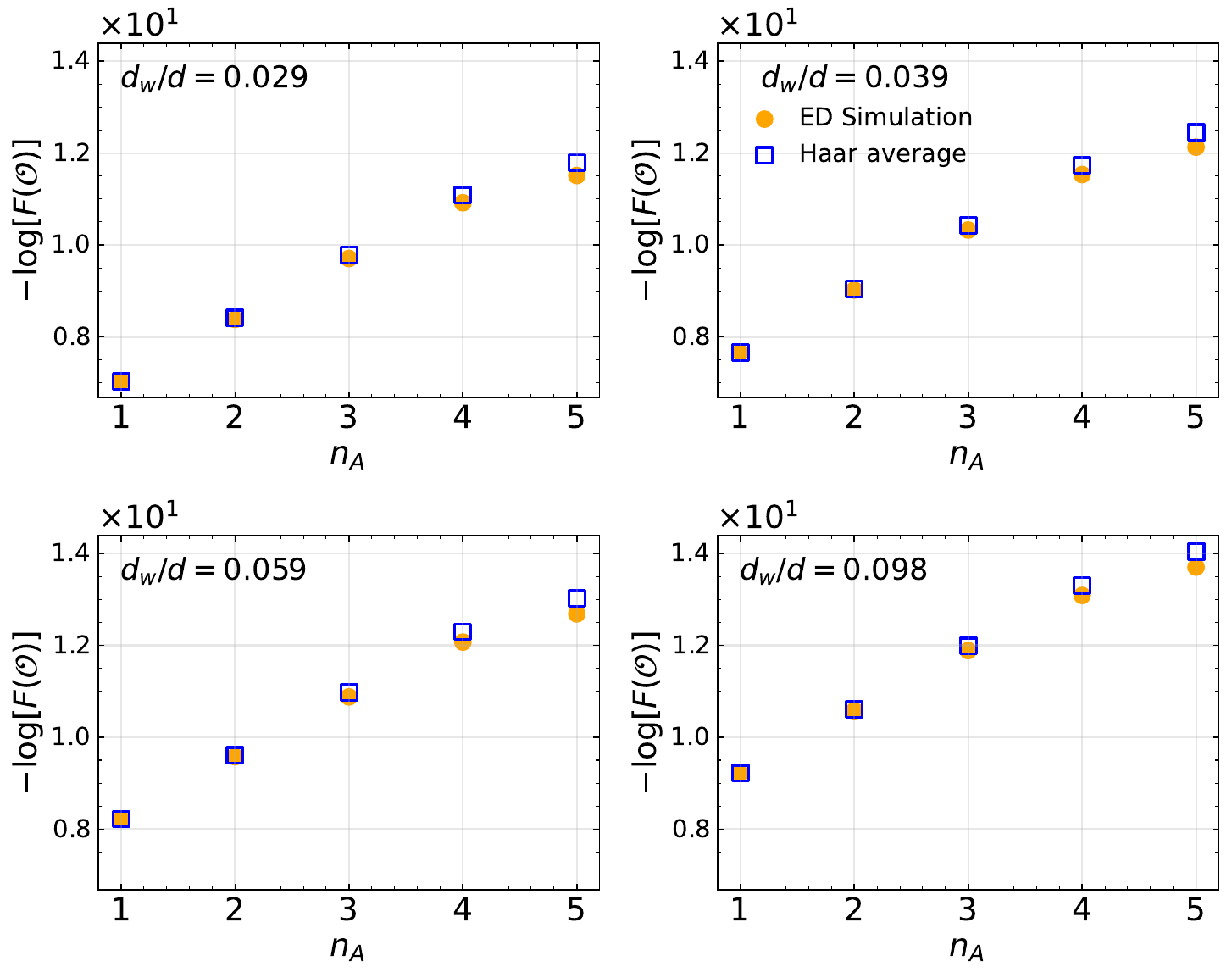}
    \caption{Evolution of $-\ln[F(\mathcal{O})]$ (Equation~\eqref{eq:F_explicit_form}) depending only on diagonal elements of the operator in the energy basis as a function of the size of subsystem A. The shown values are for total $L=10$, and each panel corresponds to a different size of the energy shell.}
    \label{fig:log_F_10_spins}
\end{figure}

\begin{figure}[htb]
    \centering
    \includegraphics[width=0.5\columnwidth]{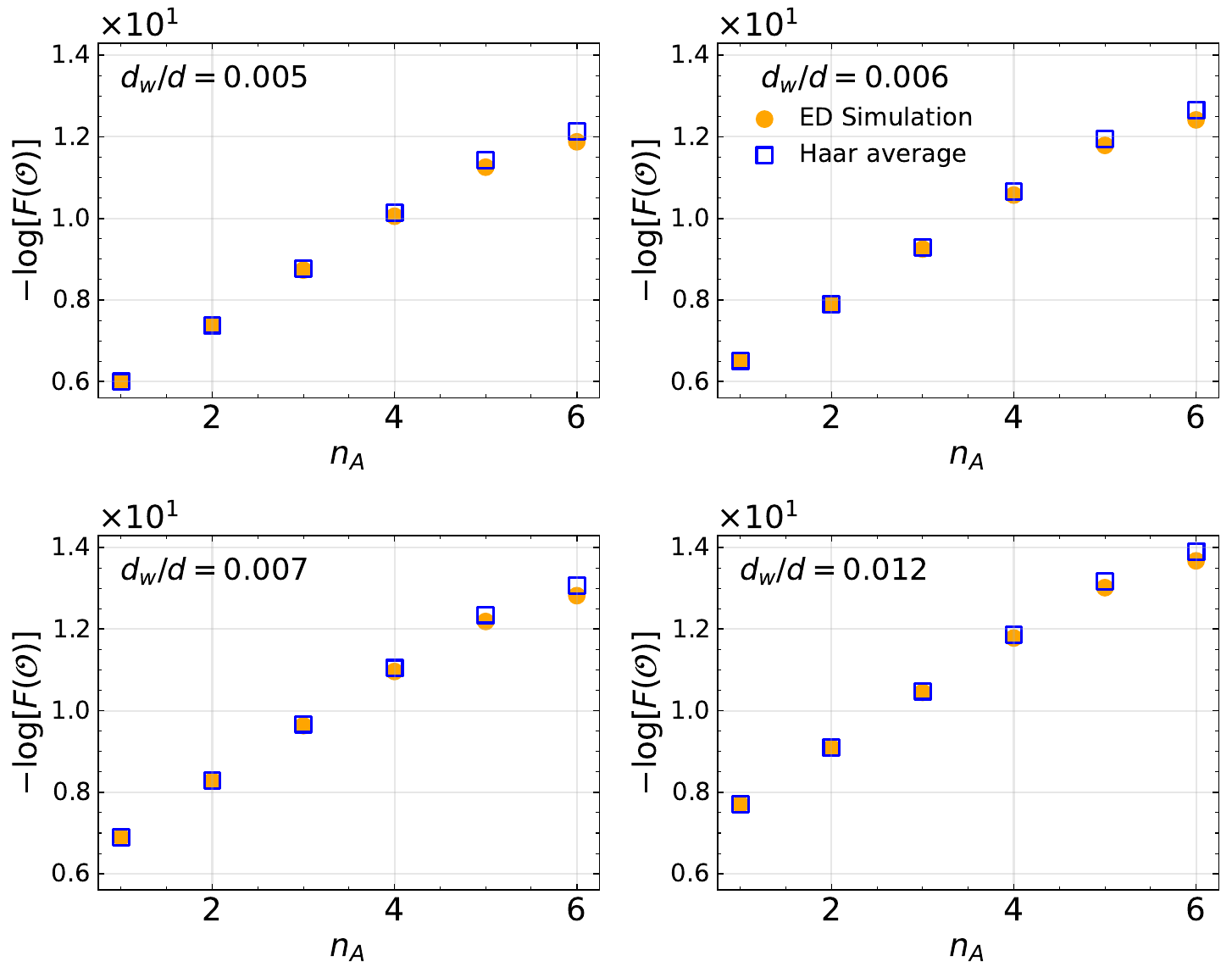}
    \caption{Evolution of $-\ln[F(\mathcal{O})]$ (Equation~\eqref{eq:F_explicit_form}) depending only on diagonal elements of the operator in the energy basis as a function of the size of subsystem A. The shown values are for total $L=12$, and each panel corresponds to a different size of the energy shell.}
    \label{fig:log_F_12_spins}
\end{figure}

\begin{figure}[htb]
    \centering
    \includegraphics[width=0.5\columnwidth]{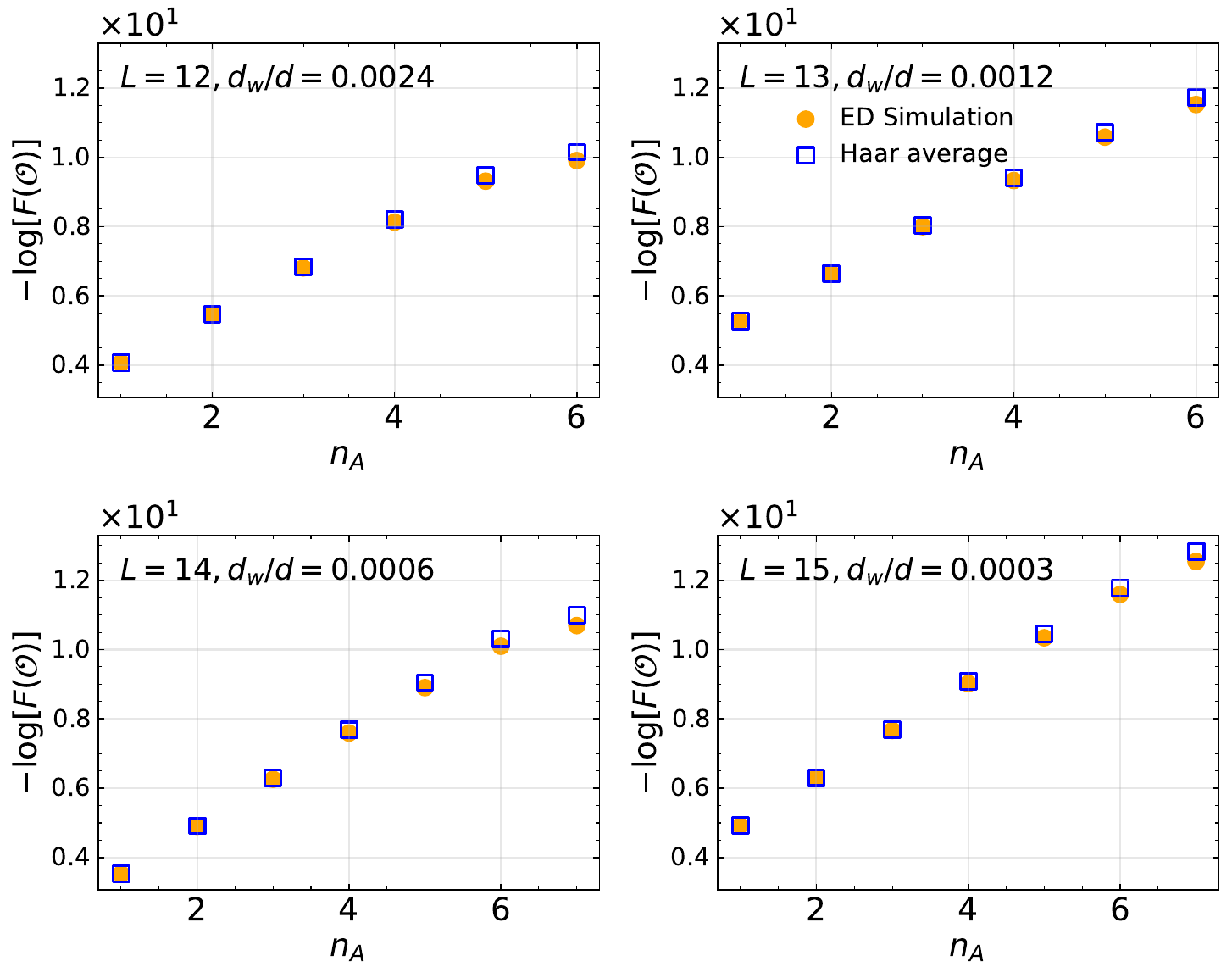}
    \caption{Evolution of $-\ln[F(\mathcal{O})]$ (Equation~\eqref{eq:F_explicit_form}) depending only on diagonal elements of the operator in the energy basis as a function of the size of subsystem A. The shown values are for window dimension $\mathrm{d}_w=10$, and each panel corresponds to a different total number of spins $L$.}
    \label{fig:log_F_different_n_spins}
\end{figure}

In order to avoid redundancies, we present only the values of $-\log(G)$, instead of showing also the plots of $G$. We see a break in the volume law for both $10$ and $12$ spins (Figures~\ref{fig:log_G_10_spins} and~\ref{fig:log_G_12_spins}), where an approximately linear increase appears, followed by a dip and decay of the function. We can see that this behavior appears for smaller values of $n_A$ for $10$ spins, with maximum value when $n_A = 2$, appearing at the bigger value $n_A=4$ for $12$ spins. This behavior corroborates the discussion presented in the main text with Figure~\ref{fig:log_offdiag_different_n_spins}, showing that the same behavior is observed even for bigger energy windows with $L=12$. The same discussion of the previous paragraph regarding the error is also valid in this case, comparing values of $G$ when $n_A = L/4$ (Figures~\ref{fig:error_G_12_spins} and~\ref{fig:error_G_different_n_spins}).

\begin{figure}[htb]
    \centering
    \includegraphics[width=0.5\columnwidth]{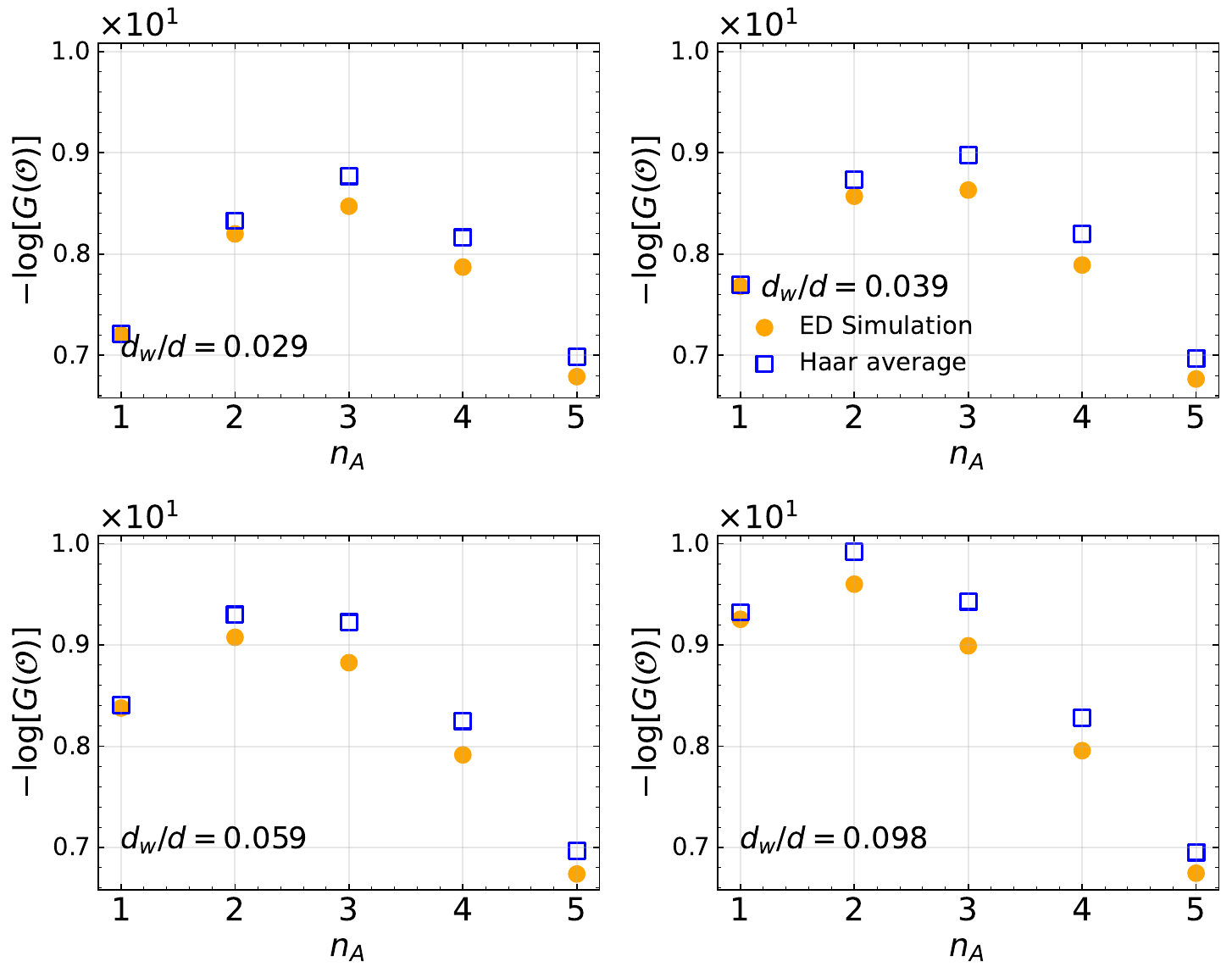}
    \caption{Evolution of $-\ln[G(\mathcal{O})]$ (Equation~\eqref{eq:G_explicit_form}) depending only on off-diagonal elements of the operator in the energy basis as a function of the size of subsystem A. The shown values are for total $L=10$, and each panel corresponds to a different size of the energy shell.}
    \label{fig:log_G_10_spins}
\end{figure}

\begin{figure}[htb]
    \centering
    \includegraphics[width=0.5\columnwidth]{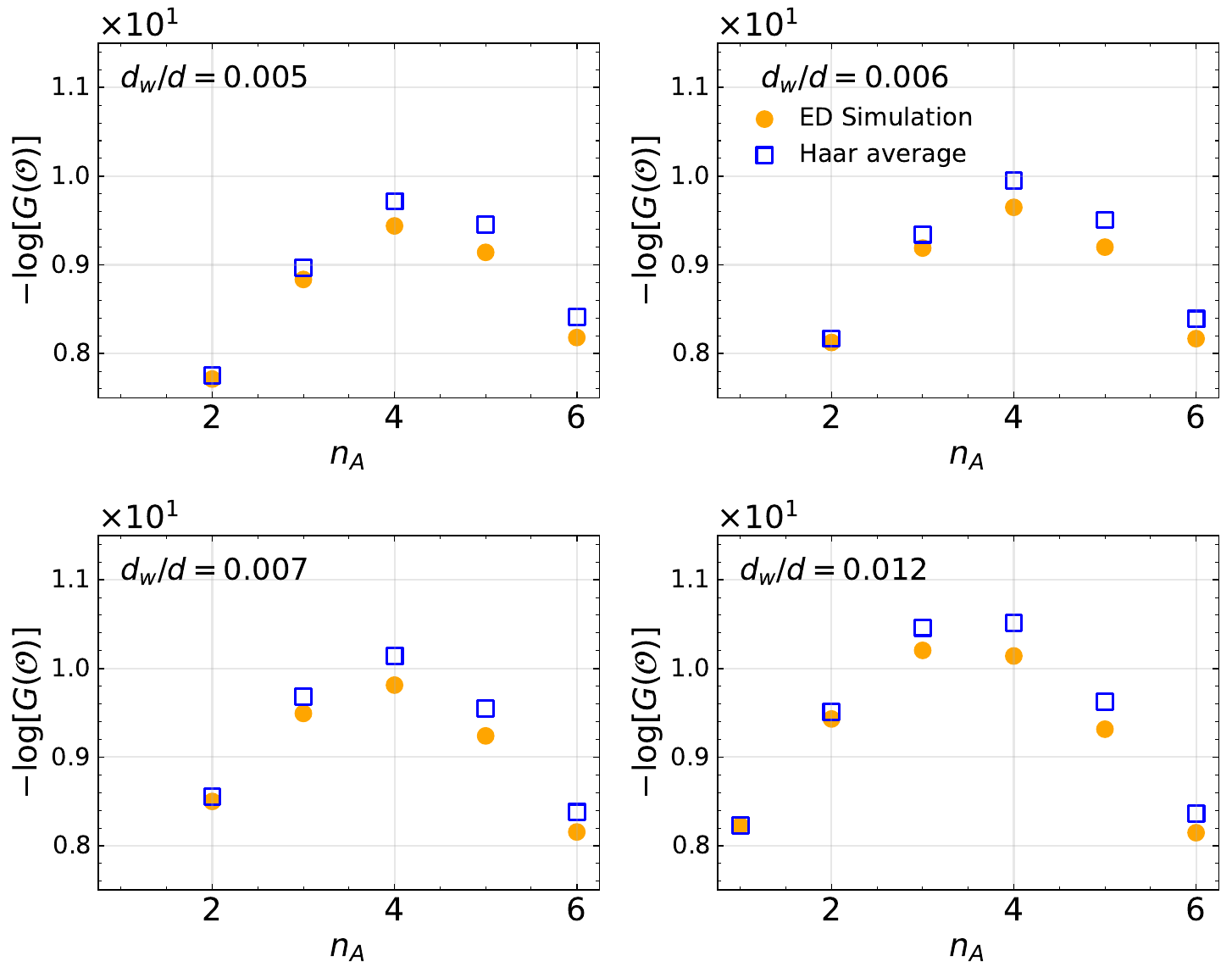}
    \caption{Evolution of $-\ln[G(\mathcal{O})]$ (Equation~\eqref{eq:G_explicit_form}) depending only on off-diagonal elements of the operator in the energy basis as a function of the size of subsystem A. The shown values are for total $L=12$, and each panel corresponds to a different size of the energy shell.}
    \label{fig:log_G_12_spins}
\end{figure}

\begin{figure}[htb]
    \centering
    \includegraphics[width=0.5\columnwidth]{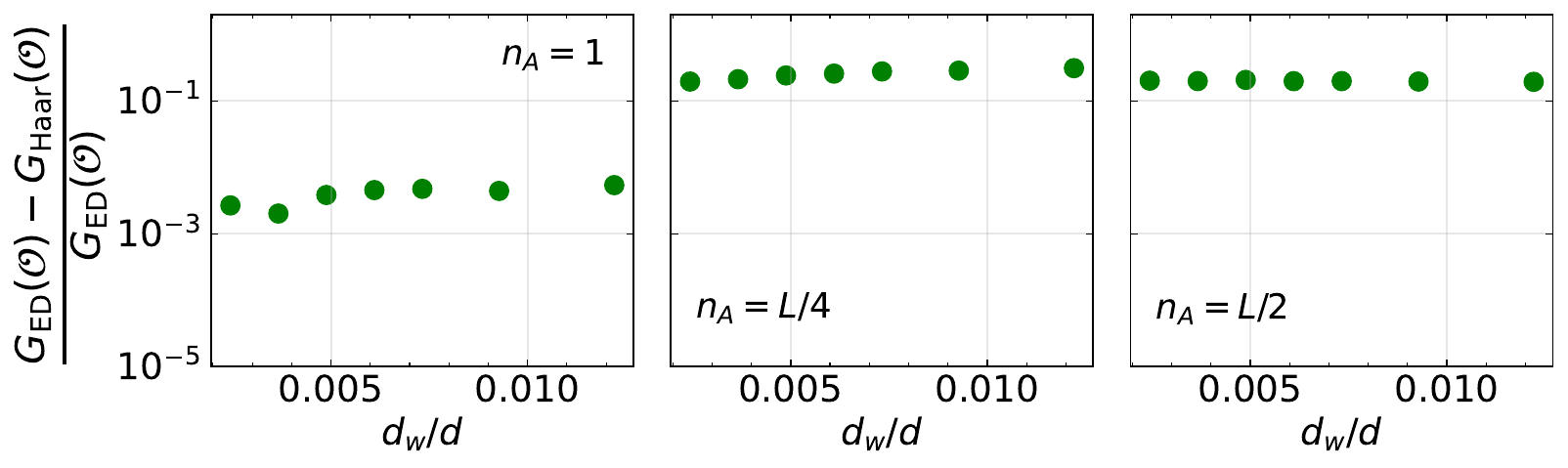}
    \caption{Error function comparing the ED simulations with Haar random averaged values of term $G(\mathcal{O})$ (Equation~\eqref{eq:G_explicit_form}) as a function of the rescaled dimension of the energy shell (number of eigenstates included divided by the total dimension of the Hilbert space) for $L=12$. Each panel shows the values at different bipartition sizes, including system A with $1$, $L/4$ and $L/2$ spins. At this region, the number of states has an approximately linear correspondence to the energy window size, therefore the dependence is the same.}
    \label{fig:error_G_12_spins}
\end{figure}

\begin{figure}[htb]
    \centering
    \includegraphics[width=0.5\columnwidth]{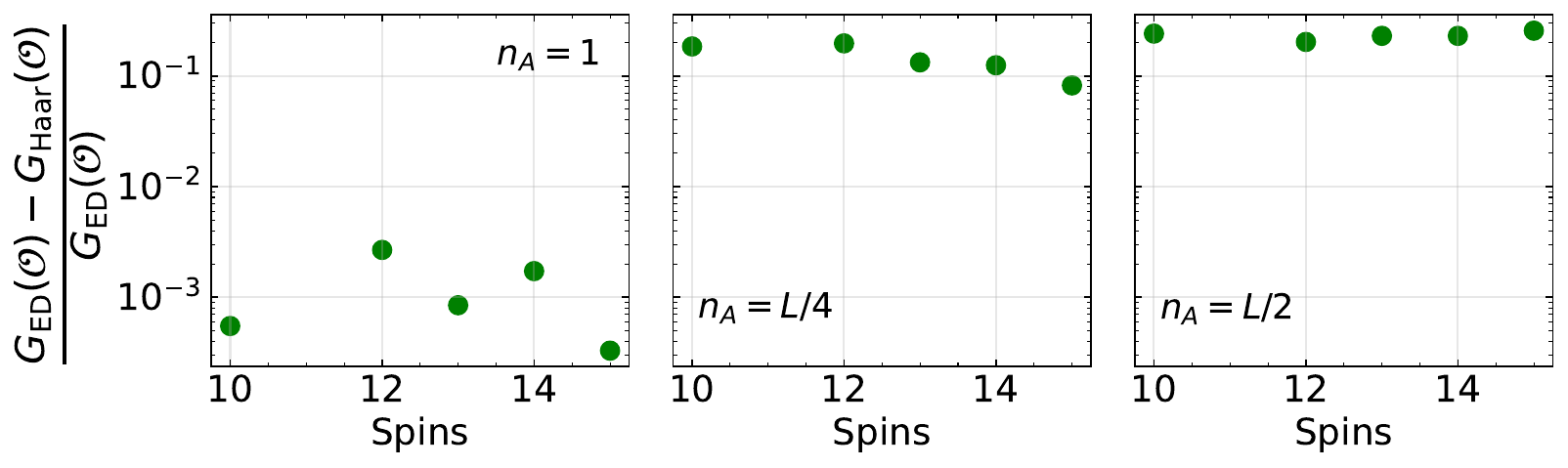}
    \caption{Error function comparing the ED simulations with Haar random averaged values of term $G(\mathcal{O})$ (Equation~\eqref{eq:G_explicit_form}) as a function of the dimension of the Hilbert space for window size $\mathrm{d}_w=10$. Each panel shows the values at different bipartition sizes, including system A with $1$, $L/4$ and $L/2$ spins.}
    \label{fig:error_G_different_n_spins}
\end{figure}

Finally, Figure ~\ref{fig:fluctuations_ETH} presents the fluctuations of the diagonal and off-diagonal elements of the operator $\sigma_x$ at the center of the chain in the basis of the chaotic MFIM. We notice a decay of the fluctuations as a function of the global dimension of the system, which justifies the approximations and discussions presented through the magnitude analysis of ETH. This result is actually one of the most important consequences of ETH, leading to the convergence of late-time average values of observables to the statistical mechanics predictions with very small fluctuations for big system sizes~\cite{eth_Deutsch1991, eth_Srednicki1994, eth_Srednicki1999, eth_Rigol2008}. We highlight that only the off-diagonal elements decay to very small values, while the diagonal elements decay with the dimension to values different from zero, given that the fluctuations will never be close to zero for it due to the order $\text{O}(1)$ term that always contributes. Figure~\ref{fig:diagonal_terms_ETH} illustrates the reduction of the fluctuations of $\mathcal{O}_{nn}$ by increasing the dimension, still preserving finite $\text{O}(1)$ fluctuation values.

\begin{figure}[htb]
    \centering
    \includegraphics[width=0.5\columnwidth]{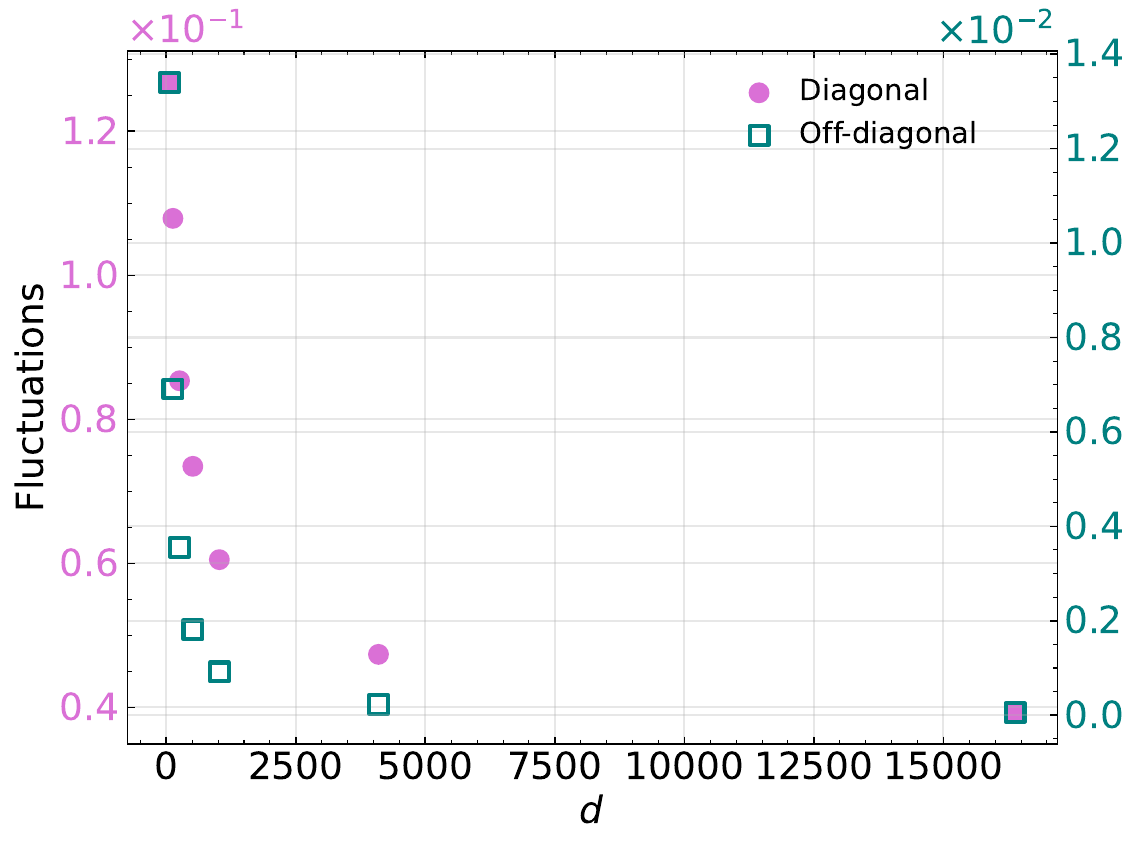}
    \caption{Fluctuations of the matrix elements of the operator $\sigma_x$ at the center of the chain (at $L//2$) in the energy eigenbasis of the maximally chaotic mixed field Ising model~\eqref{eq:mixed_field_Ising} as a function of the dimension of the Hilbert space. Filled circles represent the diagonal elements fluctuations (Equation~\eqref{eq:diagonal_average}), while open squares represent the off-diagonal values (Equation~\eqref{eq:off_diagonal_average}). Each point was calculated for a system of $L$ spins, from $L=6$ to $L=14$.}
    \label{fig:fluctuations_ETH}
\end{figure}

\end{document}